\documentclass[a4paper,12pt]{article}

\usepackage{jheppub}
\usepackage[english]{babel}
\pdfoutput=1
\usepackage[utf8]{inputenc}
\usepackage[T1]{fontenc}
\usepackage{amsmath}
\usepackage{float}
\usepackage{graphicx}
\usepackage{comment}
\usepackage[dvipsnames]{xcolor}
\usepackage{color}
\usepackage{epsfig}
\usepackage{dcolumn}
\usepackage{hyperref}
\usepackage{bm}
\usepackage{amssymb}
\usepackage{slashed}
\usepackage[caption=false]{subfig}
\usepackage{epstopdf}
\usepackage{setspace}
\usepackage[normalem]{ulem}	
\usepackage{multicol}
\usepackage{mathrsfs}
\pdfoutput=1

\newcommand{\di}{{\rm d}}
\newcommand{\ie}{{\it i.e.}}

\usepackage{tikz,pgfplots}
\usetikzlibrary{arrows,shapes,trees,calc,positioning,patterns}

\definecolor{corlinks}{RGB}{0,0,128}
\definecolor{cormenu}{RGB}{0,0,128}
\definecolor{corurl}{RGB}{0,0,128}

\definecolor{colRed0}{rgb}{0.85, 0.05, 0.12}
\definecolor{colRed1}{rgb}{0.92, 0.1, 0.05}
\definecolor{colRed2}{rgb}{0.95, 0.35, 0.05}
\definecolor{colYellow1}{rgb}{1., 0.82, 0.}
\definecolor{colBlue1}{rgb}{0.0, 0., 0.4}
\definecolor{colBlue2}{rgb}{0.1, 0.3, 0.9}
\definecolor{colBlue3}{rgb}{0.15, 0.4, 0.75}
\definecolor{colBlue4}{rgb}{0.3, 0.8, 0.93}
\definecolor{colGreen0}{rgb}{0.0, 0.15, 0.05}
\definecolor{colGreen1}{rgb}{0.0, 0.35, 0.1}
\definecolor{colGreen2}{rgb}{0.1, 0.65, 0.2}
\definecolor{colGreen3}{rgb}{0.3, 0.85, 0.5}
\definecolor{colBrown1}{rgb}{0.3, 0.18, 0.12}
\definecolor{colBrown2}{rgb}{0.5, 0.3, 0.20}
\definecolor{colViolet1}{rgb}{0.4, 0.18, 0.42}
\definecolor{colViolet2}{rgb}{0.5, 0.3, 0.70}

\hypersetup{
colorlinks=true,
urlcolor=corlinks,
linkcolor=corlinks,
menucolor=cormenu,
citecolor=corlinks,
pdfborder= 0 0 0
}

\DeclareRobustCommand{\Eq}[1]{Eq.~(\ref{#1})}
\DeclareRobustCommand{\Fig}[1]{Fig.~\ref{#1}}
\DeclareRobustCommand{\Ref}[1]{Ref.~\cite{#1}}
\DeclareRobustCommand{\Sec}[1]{Sec.~\ref{#1}}

\definecolor{cor2}{rgb}{0,0,0}
\definecolor{cor1}{rgb}{0,0,0}
\definecolor{darkpink}{rgb}{0.8,0,0.4}

\makeatother

\newcommand{\ew}{\textnormal{\tiny EW}}

\newcommand{\MPl}{M_\text{Pl}}

\newcommand{\Dtrel}{\Delta t_\text{rel}}

\newcommand{\eV}{\,\text{eV}}
\newcommand{\keV}{\,\text{keV}}
\newcommand{\MeV}{\,\text{MeV}}
\newcommand{\GeV}{\,\text{GeV}}

\newcommand{\s}{\,\text{s}}

\newcommand{\D}{\mathcal{D}}
\newcommand{\tr}{\textrm{Tr}}

\newcommand{\Yphimis}{Y^\phi_\textrm{mis}}
\newcommand{\Yphith}{Y^\phi_\textrm{th}}
\newcommand{\Tpp}{T_\textrm{pp}}

\newcommand{\HD}{\Phi_H} %Higgs doublet

\allowdisplaybreaks %Equations on more pages

\title{Higgs relaxation after inflation}

\date{\today}

\author[a]{Nayara Fonseca,}
\author[a]{Enrico Morgante,}
\author[a,b]{G\'eraldine Servant}

\affiliation[a\,]{DESY, Notkestrasse 85, 22607 Hamburg, Germany}
\affiliation[b\,]{II. Institute of Theoretical Physics, Univ. Hamburg, D-22761 Hamburg, Germany}

\abstract{
We show that the mechanism of cosmological relaxation of the electroweak scale can take place  independently of the inflation mechanism, thus relieving burdens from the original relaxion proposal. What eventually stops the (fast-rolling) relaxion field during its cosmological evolution is the production of particles  whose mass is controlled by the Higgs vacuum expectation value. We first show that Higgs particle production does not work for that purpose as the Higgs field does not track the minimum of its potential in the regime where Higgs particles get efficiently produced through their coupling to the relaxion. We then focus on gauge boson production.
We provide a detailed analysis of the scanning and stopping mechanism
and determine the parameter space for which the relaxion mechanism can take place after inflation, while being compatible with cosmological constraints, such as the relaxion  dark matter overabundance and Big Bang Nucleosynthesis. We find that the cutoff scale can be as high as two hundreds of  TeV.
In this approach, the relaxion sector is responsible for reheating the visible sector. The stopping barriers of the periodic potential are large and Higgs-independent, facilitating model-building. 
The allowed relaxion mass ranges from  200 MeV up to the weak scale. In this scenario, the relaxion field excursion is subplanckian, and is thus many orders of magnitude smaller than in the original relaxion proposal.
}
\begin{document}

\begin{flushright} 
DESY 18-069
\end{flushright}

\maketitle
\clearpage

\section{Introduction}

Recently, a new mechanism has been proposed to generate a naturally small electroweak scale without relying on new symmetries at the electroweak scale~\cite{Graham:2015cka}.
It exploits the coupling of the Higgs boson to an axion-like field
which induces a long cosmological evolution of the Higgs mass parameter. This is the so-called {\it relaxion mechanism}. 
While  classically rolling down its potential, the relaxion field scans down the Higgs mass parameter, starting  from a value of the order of the cutoff scale, until a stopping mechanism comes into play precisely when the Higgs mass parameter approaches zero. This would naturally explain the smallness of the Higgs mass parameter compared to the cutoff scale of the theory.
In this context, there may be no expectation for the existence of new particles at the TeV scale. Instead, one predicts at least one very light and very weakly interacting particle. 

This very change of paradigm has therefore far-reaching implications for strategies to search for new physics linked to the understanding of the weak scale
and has consequently triggered a large literature: On general model building concerns~\cite{Espinosa:2015eda,Gupta:2015uea,Abel:2015rkm, Choi:2015aem, Ibanez:2015fcv, Hebecker:2015zss, McAllister:2016vzi}, on attempts to do this without inflation~\cite{Hardy:2015laa, Hook:2016mqo}, on issues related to inflation~\cite{Patil:2015oxa, Jaeckel:2015txa, Marzola:2015dia, DiChiara:2015euo, Evans:2017bjs, Tangarife:2017rgl} and reheating~\cite{Choi:2016kke}, on UV completions involving supersymmetry~\cite{Batell:2015fma, Evans:2016htp,Evans:2017bjs}, composite Higgs~\cite{Batell:2017kho,Antipin:2015jia,Agugliaro:2016clv}, two-Higgs-doublet models \cite{Lalak:2016mbv}, a mirror copy of the Standard Model~\cite{Matsedonskyi:2015xta}, a Nelson-Barr model~\cite{Davidi:2017gir}, clockwork axions~\cite{Choi:2015fiu, Kaplan:2015fuy, Giudice:2016yja}, warped extra dimensions~\cite{Fonseca:2017crh} or other constructions with multiple axions~\cite{Fonseca:2016eoo},
on phenomenological aspects and experimental signatures~\cite{Kobayashi:2016bue, Choi:2016luu, Flacke:2016szy, Beauchesne:2017ukw}, and on alternative implementations of the mechanism that do not require any barriers~\cite{Huang:2016dhp, Matsedonskyi:2017rkq}.

In the original proposal \cite{Graham:2015cka} (GKR), the system starts in the symmetric electroweak  phase and stopping barriers in the periodic relaxion potential with period $f$ get generated only once the Higgs mass parameter turns tachyonic.
Such mechanism relies on Higgs-dependent barriers as well as a long period of inflation to guarantee the slow  classical evolution of the relaxion. 
The slope of the potential along which the axion-like field is rolling and its coupling to the Higgs field are generated due to a small effective breaking of a shift symmetry. 
This can be reconciled with the pseudo-Nambu-Goldstone (PNGB) boson nature of the relaxion  if the slope arises from a second oscillatory potential with a period much larger than $f$ in the so-called clockwork axion framework~\cite{Choi:2015fiu, Kaplan:2015fuy, Giudice:2016yja}, in $N$-site constructions~\cite{Fonseca:2016eoo} or in warped extra-dimensional ones~\cite{Fonseca:2017crh}.\footnote{
A completely different approach for a UV completion was discussed in string compactifications.
It was shown in \cite{McAllister:2016vzi} that UV completions of the relaxion in string theory realizations via axion monodromy are strongly constrained.
The field excursion corresponds to a physical charge carried by branes or fluxes which 
backreacts on the ten-dimensional configuration, and can suppress
the barriers generated by strong gauge dynamics. This leads to a ``runaway'' relaxion, thus ruining  the stopping mechanism.}

If the relaxion is the QCD axion (GKR1),  the Peccei-Quinn solution to the strong CP problem can be preserved only if  new dynamics is introduced at the end of inflation. Besides, the cutoff scale cannot be pushed higher than 30 TeV while the coupling between the Higgs and the relaxion has to be smaller than $10^{-30}$ (see \cite{Nelson:2017cfv, Jeong:2017gdy} for an updated discussion on this model).
If instead the relaxion comes from a new strongly interacting sector (GKR2), the cutoff scale can be pushed up to $10^8$ GeV. However, this requires new EW scale fermions, generating a coincidence problem.
This problem was solved via a double-scanner mechanism in \cite{Espinosa:2015eda} where the barrier height depends on an additional extra field. In this context, initial conditions are very different from GKR1 and GKR2 as the barrier's height starts large until it gets cancelled by the additional scanner field.
Overall, the maximal cutoff scale that can be achieved  by the relaxion mechanism is $\sim 10^{8}$ GeV.
At this scale, it is assumed that some other mechanism kicks in to protect the Higgs mass against the Planck scale, either via supersymmetry, compositeness, or other mechanisms.

A successful implementation  of the original relaxion proposal requires a low scale of inflation associated with non-trivial inflation model building. This has tarnished the appeal of the relaxion mechanism.
We are therefore interested to consider instead the possibility that the friction acting on the relaxion is not provided by an inflation era but instead by particle production during the field classical evolution. This would enable to decouple the relaxion mechanism from inflation and thus no further constraint on the number of e-folds or inflation scale would be imposed.
A very interesting framework along these lines was proposed in  \cite{Hook:2016mqo}, where the source of particle production comes from a Chern-Simons coupling between the relaxion and  the Standard Model gauge fields. The back-reaction mechanism is then provided by electroweak gauge boson production which is triggered when the Higgs vacuum expectation value (VEV) approaches zero. A motivation of the authors was to avoid superplanckian field excursions (see also~\cite{You:2017kah, Son:2018avk}). While the conditions for a successful friction mechanism were derived, a full derivation of all phenomenological constraints and a precise picture of the cosmological history in this context was lacking. This is what we do in this work.
In addition, we show explicitly why  the relaxion coupling to the Higgs cannot lead to a friction term from Higgs particle production. Relying on the relaxion coupling to  a   combination of SM gauge fields,  we suggest that the relaxion may be responsible for the reheating of the universe, thus connecting reheating and the Higgs sector.    We also stress that the relaxion is heavier than in the original GKR proposal and its mass can be as large as  $\mathcal{O}(100)$ GeV.
 In \Ref{Hook:2016mqo}, the impact of particle production on the relaxion mechanism was discussed  before inflation, during inflation and at the end of inflation. In our setup, we focus on the situation where one can  ignore the inflaton. The relaxion is dominating the energy density of the universe during relaxation, although it is not slow-rolling.

In this setup, the main improvements compared to the `standard relaxion' models are:
\begin{itemize}
\item a weak scale relaxation mechanism independent from inflation (no need for a gigantic number of e-folds $\mathcal{N}_e$ nor a  small Hubble rate during inflation  $H_I$);
\item sub-Planckian field excursions for the relaxion;
\item the barriers of the relaxion periodic potential are independent from the Higgs vacuum expectation value;
\end{itemize}

The plan of the paper is the following.
In Section \ref{sec:conditions1}, we discuss the general conditions for realizing the relaxion mechanism after inflation. In Section \ref{sec:stop_pp}, we discuss the conditions for using particle production as friction instead of inflation. We consider first Higgs particle production and then gauge boson production. 
In Section \ref{sec:photons}, we present the induced relaxion couplings to photons and fermions. Section 
\ref{sec:requirements} lists all requirements and summarizes the result of the combination in terms of constraints on the cutoff scale and relaxion coupling to the Higgs. 
The relaxion properties are presented in Section \ref{sec:pheno}. We then consider in Section \ref{sec:cosmology} the phenomenological, cosmological (relic abundance and Big Bang nucleosynthesis)  and astrophysical constraints   and determine the parameter space where a successful implementation is realised.
We conclude in Section \ref{sec:conclusion}.
The equations of motion for the Higgs, relaxion and the gauge bosons are reproduced in Appendix \ref{sec:appendix}, with a display of their numerical solutions.

%%%%%%%%%%%%%%%%%%%%%%%%%%%%%%%%%%%%%%%%%%%%%%%%%%%%%%%%%%%%%%%%%%%%%%%%%%%%%
\section{General conditions for relaxation after inflation}
\label{sec:conditions1}

The scalar potential for the Higgs $h$ and relaxion $\phi$ fields reads:
\begin{equation}
V(\phi, h) = \Lambda^4 - g \Lambda^3 \phi +\frac{1}{2} \left(-\Lambda^2 + g'\Lambda \phi\right)h^2+ \frac{\lambda}{4} h^4 + \Lambda_b^4\cos\left(\frac{\phi}{f'}\right)\,,
\label{eq:relaxion_potential}
\end{equation}
where $\Lambda$ is the cutoff scale up to which we want to solve the hierarchy problem using the relaxion. 
 The relaxion $\phi$ is an axion-like field with decay constant $f'$.
The dimensionless couplings $g$ and $g'$  are assumed to be spurions that quantify the explicit breaking of the axion shift symmetry, and $\Lambda_b$ is the scale at which  the $\phi$ periodic potential is generated. The term $\Lambda^4$  cancels the final value of the cosmological constant and corresponds to the usual tuning of the cosmological constant.

We want the scanning of the Higgs mass parameter  to occur when the inflaton is a subdominant component of the energy of the universe so as to decouple the relaxation scenario from inflation. 
For that, a crucial difference with respect to the  original relaxion scenario~\cite{Graham:2015cka} is that we start in the broken electroweak phase, where the Higgs mass parameter in the Higgs potential is large and negative \cite{Hook:2016mqo}.
Another important difference is that that the amplitude $\Lambda_b^4$  of the cosine potential is constant and does not depend on the Higgs vacuum expectation value.\footnote{The existence of large  barriers  was also present in the double scanning mechanism of the CHAIN model presented in \cite{Espinosa:2015eda}. } We require $g \gtrsim g'/(16 \pi^2)$ for the stability of the potential (closing the Higgs loop from the  third term in~\Eq{eq:relaxion_potential} generates  the slope $\sim g\, \Lambda^3 \phi$). In our numerical analysis, we will take $g=g'$.

The cosmological history is the following. After inflation ends, the universe is reheated. As soon as the temperature drops below $\sim \Lambda$, the shift symmetry that protects the relaxion potential breaks and $\phi$ starts rolling. The scanning of the Higgs mass parameter starts.  
The initial condition for $\phi$ is such that the Higgs field has a large negative mass term, which corresponds to  
\begin{equation}\label{eq: Higgs negative mass}
-\mu_h^2 \equiv -\Lambda^2 + g'\Lambda \phi \ll 0 \ .
\end{equation}
 The electroweak symmetry is broken and all gauge bosons coupled to the Higgs have large masses.
The friction is negligible and the relaxion rolls down its potential with some high speed, overshooting the large barriers. The Higgs mass parameter is thus scanned as the VEV of the relaxion is increasing.
The relaxion evolves towards large positive values until it approaches the critical point 
\begin{equation}
\phi_c \equiv\Lambda/g'
\end{equation}
  where the Higgs mass term is zero. 
 As the electroweak gauge bosons become light enough, they can be produced exponentially through their coupling to the relaxion. As we will discuss  in Sec.\,\ref{sec:stop_pp}, this particle production is so efficient that it quickly slows down the relaxion, which has no longer enough kinetic energy to overshoot the barriers. 
 The relaxion is therefore stopped right before it reaches the critical point  $\phi_c$,  in such a way that the final Higgs VEV is small compared to its initial value  set by $\Lambda$.  
The use of particle production in relaxion models  appeared previously in  \cite{Hook:2016mqo} (see also \cite{Choi:2016kke, You:2017kah, Matsedonskyi:2017rkq}). 

\begin{figure}
\begin{center}
\includegraphics[width=.7\textwidth]{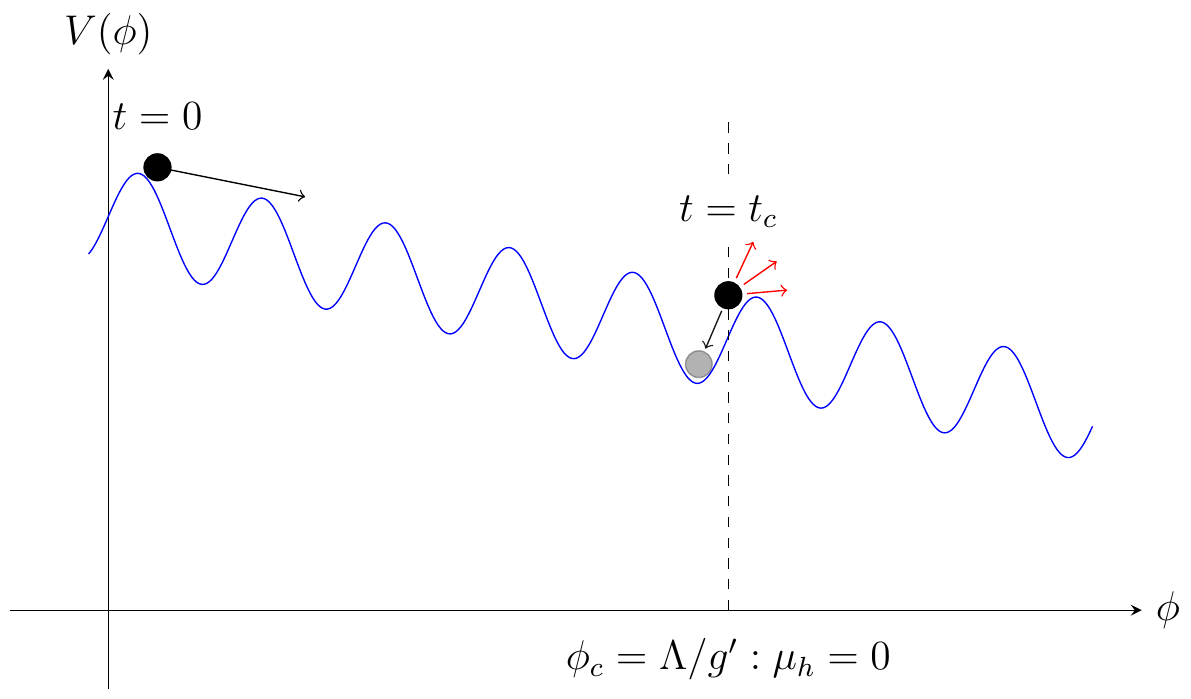}
\end{center}
\caption{Sketch of the relaxion field evolution.  }
\label{fig:relaxion evolution}
\end{figure}

 In the sketch below we show the hierarchy of  scales ($f', M_I > \Lambda, \Lambda_b > v_\ew$), where the solid line indicates when the scanning starts and $M_I$ is the inflation scale. 
\begin{center}
\begin{tikzpicture}
\draw[<-,dashed,thick] (-3,0)--(0,0);
\draw[thick] (0,0) -- (4,0);
%\draw (.7,-.1)--(.7,.1);
\draw (3.3,-.1)--(3.3,.1);
\node at (-1.5,-.5) {$ f', M_I$};
\node at (.7,-.5) {$\Lambda,\Lambda_b$};
\node at (3.3,-.5) {$v_\ew$};
\draw[->] (-.2,.4)--(1.2,.4);
\node at (.5,.6) {time};
\end{tikzpicture}
\end{center}
A sketch of the relaxion potential and its evolution is shown in Fig.~\ref{fig:relaxion evolution}. 

\subsection{Relaxation in a non-inflationary phase and the reheating of the universe} 
\label{sec:noinflation}

To realize the relaxation mechanism after the end of inflation, it is crucial to discuss how the temperature affects the scenario described above.  
As we shall see in the following, there are two possibilities for the reheating: (i) the inflaton sector  reheats the SM degrees of freedom, and (ii) the inflaton reheats a hidden sector decoupled from the SM and later the relaxion reheats the visible sector. In what follows we will discuss these two possibilities, and we will show that the first case can  be realized only in the region of the parameter space corresponding to the smallest values of the coupling $g'$  in \Eq{eq:relaxion_potential}.

\begin{figure}
\begin{center}
\includegraphics[width=\textwidth]{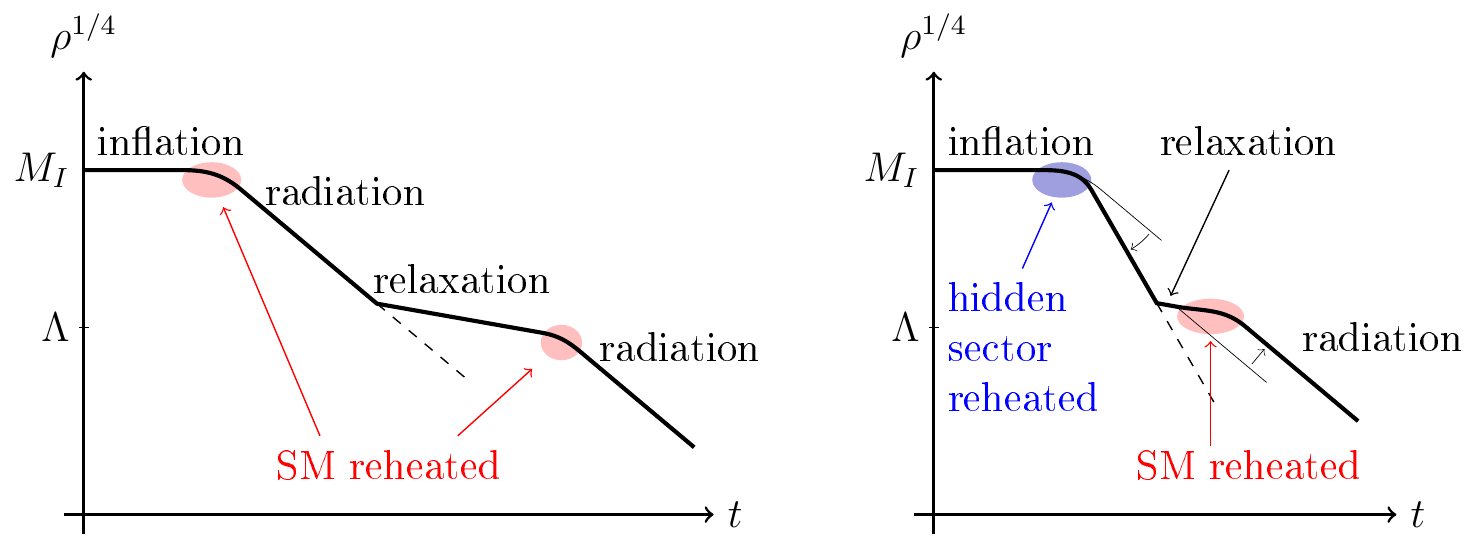}
\caption{
Sketch of the evolution of the energy density of the universe under two distinct assumptions about the first reheating stage. \textit{Left:} The inflaton reheats the visible sector, and relaxation starts when its temperature drops below $\Lambda$. The period of relaxation is characterised by a short stage of inflation lasting for a few e-folds as needed to suppress the thermal mass of the Higgs.  A second reheating takes place after relaxation. \textit{Right:}~The inflaton transfers its energy to a hidden sector, and the visible one is reheated after relaxation. In this case, the relaxation phase can be short, and some mechanism to dilute the dark radiation is needed. For example, a period of kinetic energy domination in the dark sector or a prolonged reheating phase with a matter-like equation of state can be sufficient to suppress the dark component with respect to the visible one.}
\label{fig:energy evolution}
\end{center}
\end{figure}

First, let us assume that, at the end of the inflationary phase, the energy density of the inflaton field is transferred to the SM sector, initiating the radiation era.
Relaxation starts when the temperature drops below the cutoff $\Lambda$ of the theory and the potential $V(\phi)$ is generated. The relaxion then dominates the energy density until it is stopped and its energy is converted into radiation. 
A sketch of how the energy density evolves  is shown in the left panel of~\Fig{fig:energy evolution}.
During the relaxation era, the equation of state changes along the evolution, with $w = p/\rho < 0$ and close to $w=-1$ (cosmological constant) for the lowest values of the coupling $g'$.

An important concern comes from the fact that, if the SM is reheated to a too high temperature, the negative mass-squared of the Higgs field is turned positive by a thermal mass term $\sim y_t^2 T^2$. This could spoil the relaxation mechanism, since the field $\phi$ would stop in the wrong position as soon as $\mu_h^2+y_t^2 \, T^2=0$, where $y_t\sim 1$ is the top Yukawa.
To consider this issue more carefully, we have to compare the time scales of relaxation with that of  the cooling of the universe. 
Relaxation starts when the temperature drops below $T\sim\Lambda$. Initially, the squared mass term $\mu_h^2$ and the Higgs thermal mass are both of order $\Lambda$, and we have to assume that the former is larger than the latter. As relaxation goes on, both terms will decrease. In order for the mechanism not to be spoiled by thermal effects, it is necessary that the condition
\begin{equation}\label{eq:tempmass}
|\mu_h|\gtrsim T
\end{equation}
holds during the whole process. 
The validity of condition (\ref{eq:tempmass}) in terms of the parameters $\Lambda$ and $g'$ can be understood as follows.   The temperature at the end of relaxation must be smaller than the electroweak scale  $T_{\rm{end}} \lesssim v_\ew$ and   $T_{\rm{ini}} \sim \Lambda$, then we have
\begin{equation} \label{eq:efolds1}
 \frac{a_{\rm{ini}}}{a_{\rm{end}}} \lesssim \frac{v_\ew}{\Lambda},
\end{equation}
where  $a_{\rm{ini}} \equiv a(t= t_{\rm{ini}})$ and $a_{\rm{end}} \equiv a(t= t_{\rm{end}})$ and we used that the temperature scales with the inverse of the scale factor.  On the other hand,   $a_{\rm{ini}}/a_{\rm{end}}  = \exp\left(-\int H dt\right) \equiv  e^{-\mathcal{N}}$,  within $\mathcal{N}$ being the number of e-folds, therefore condition (\ref{eq:efolds1}) gives
\begin{equation}\label{eq:Nefoldslarge}
\mathcal{N} \gtrsim \log\left(\frac{\Lambda}{v_\ew}\right),
\end{equation}
implying that the roll-down phase must last for at least a few e-folds. In the approximation of a constant Hubble rate $H \sim \Lambda^2/(\sqrt{3}\MPl)$ we get
\begin{equation}\label{eq:T<v}
\mathcal{N} \sim H \Dtrel \sim  \frac{\Lambda}{\sqrt{3}g'\MPl} \gtrsim \log\left(\frac{\Lambda}{v_\ew}\right),
\end{equation}
where we used  that
\begin{equation}
\Dtrel \sim \frac{\Delta\phi}{\dot\phi} = \frac{\Lambda/g'}{\Lambda^2} = \frac{1}{g'\Lambda} \,,
\label{eq:Dtrel}
\end{equation}
which is obtained in the approximation of negligible Hubble friction. Additionally, the number of e-folds cannot be too large, in order not to wash out the perturbations generated during inflation, and therefore we impose that $\mathcal{N}\lesssim 20$.%
\footnote{If the relaxion drives a long period of inflation, it is difficult to match the curvature perturbations that are generated in this phase with the COBE normalization~\cite{Tangarife:2017rgl}.
If this second inflationary period driven by the relaxion is shorter, some of the modes that had reentered the horizon after inflation could exit again and reenter after the end of relaxation, possibly imprinting observable features in the CMB power spectrum.
}
In this way one gets
\begin{equation}\label{eq:noinflation}
\frac{\Lambda}{20\sqrt{3}\MPl} \lesssim g' \lesssim \frac{\Lambda}{\sqrt{3}\MPl\log(\Lambda/v_\ew)}\,,
\end{equation}
leaving only a small window of parameter space open (imposing a more stringent bound on the duration of a secondary inflationary stage will reduce the region accordingly). Note that the lower bound on $g'$ in (\ref{eq:noinflation}) automatically avoids very large (super-Planckian) field excursions.

\begin{figure}
\centering
\includegraphics[width=.83\textwidth]{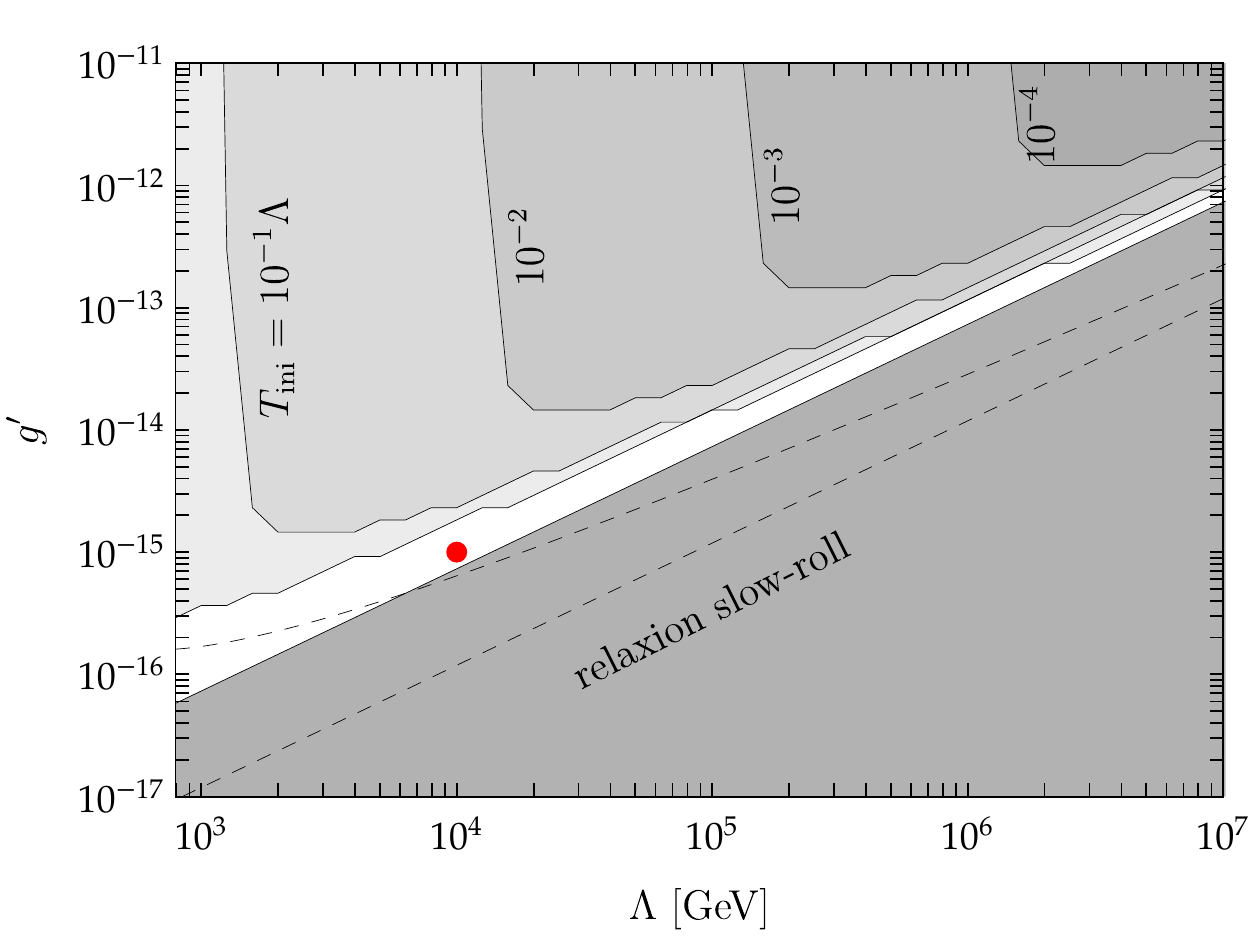}\\
\includegraphics[width=.415\textwidth]{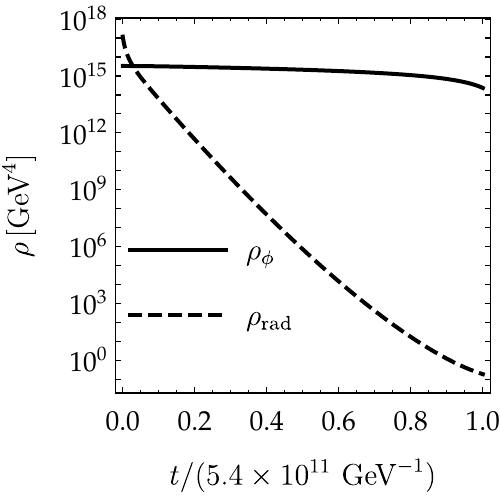}
\includegraphics[width=.415\textwidth]{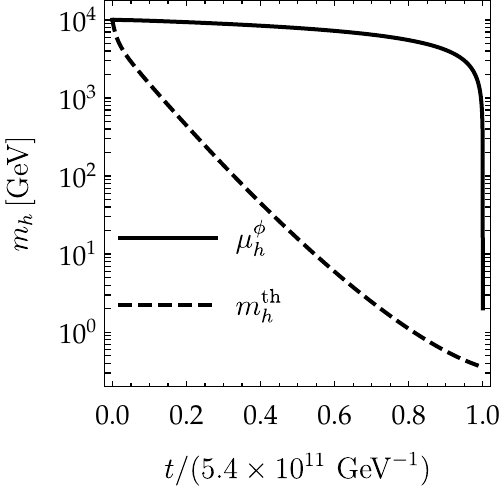}
\caption{  \textit{Top:} In the white region, the temperature $T$ of the visible universe (SM) is suppressed with respect to the EW scale and the relaxation lasts less than $20$ e-folds. It is obtained by solving numerically the relaxion equation of motion in a universe where the energy is initially equipartioned between the relaxion and radiation, and the initial temperature is $T_{\rm ini}=\Lambda$. The dashed lines indicate the corresponding region obtained using the approximation  \Eq{eq:noinflation}. The different shades of gray show how the parameter space opens up if  $T_{\rm ini}$ is assumed to be a fraction $10^{-1},10^{-2},10^{-3},10^{-4}$ of the cutoff scale $\Lambda$.
\textit{Bottom left:} Time evolutions of the relaxion energy density and radiation energy density  for the benchmark point $\Lambda=10^4\GeV$, $g'=10^{-15}$, marked in red in the top panel.  \textit{Bottom right:}  Time evolutions of the Higgs mass term and the Higgs thermal mass for the same benchmark point.}
\label{fig:reheating SM allowed g}
\end{figure}

To check the validity of this naive estimate, we solved numerically the equation of motion for the relaxion field rolling down the linear potential, in an universe where the energy is initially equipartitioned between $\phi$ and radiation.
The top panel of Fig.~\ref{fig:reheating SM allowed g} shows the bounds in \Eq{eq:noinflation} and the comparison with the corresponding bounds computed numerically by solving the equations of motion. The difference in the two estimates is due to the assumption of constant Hubble rate in the analytical one.
The other two lower panels show the evolution of the relaxion  and radiation component of the energy density, and their contribution to Higgs mass parameter, for a benchmark point at $\Lambda=10^4\GeV$ and $g'=10^{-15}$, that corresponds to a relaxation that lasts for $\sim 10$ e-folds. We see that the energy density is dominated by the relaxion, and that the thermal mass term is always subdominant with respect to the relaxion one, thus not spoiling the mechanism.

The bottomline of this discussion is that, if the SM is reheated to a temperature larger than the electroweak scale, there is little room for the relaxion mechanism to take place after reheating, in the radiation era. If instead the  reheating temperature of the universe is below the electroweak scale, the scanning can be finished during the radiation era without spoiling the mechanism. However, since the scale of inflation is larger than $\Lambda$, the relaxation process would start during the last e-folds of inflation or during the reheating phase. 
In this case, either relaxation takes place when the universe is still inflaton dominated~\cite{Hook:2016mqo}, and the reheating phase has not yet started, or one has to worry about the maximal temperature of the SM plasma during reheating, which can exceed the EW scale.

\medskip

On the other hand, one can assume that a large fraction of the energy which is initially stored in the inflaton field is transferred to a hidden sector gas, with no interaction with the SM, and that the temperature of the SM is much smaller than $\Lambda$ at the time when relaxation starts. In this case, we can consider that the relaxion mechanism starts while the universe is in a radiation-dominated era and  the temperature of the visible sector $T$ is much smaller than the temperature of the universe, so that 
the bound presented in Eq.~(\ref{eq:T<v}) is evaded for $T_\textrm{ini}\lesssim |\mu_h|$.\,\footnote{The same could be obtained if the reheating temperature is larger than $v_\ew$, but a second field locks the relaxion at its initial position until the temperature has dropped, with a mechanism similar to~\cite{Espinosa:2015eda, Matsedonskyi:2017rkq, Fonseca:2017crh}. While one can envisage a model of this kind, we are not going to discuss this possibility further.}
The different shadings in the first panel of Fig.~\ref{fig:reheating SM allowed g} show how the parameter space opens up if the temperature of the SM plasma at the beginning of the relaxation phase is taken to be a fraction $10^{-1},10^{-2},10^{-3},10^{-4}$ of the cutoff $\Lambda$.

In the following, we are going to assume for simplicity that the SM temperature after reheating is negligible, and all the energy is transferred to a sector decoupled from the SM.  This implies that the upper bound on $g'$ in \Eq{eq:noinflation} disappears and in  the following   we are simply going to assume the lower bound
\begin{equation}
g'\gtrsim 0.2\frac{\Lambda}{\MPl}, 
\label{eq:lowerboundongp}
\end{equation}
which results from the numerical solution, and expresses the condition of a short-enough period of relaxion-driven inflation. 
Note that in this case Eq.~(\ref{eq:Nefoldslarge}) does not apply, and therefore the energy stored in the dark sector is not diluted during relaxation. In principle, this could be at odds with bounds on dark radiation. In order to avoid this, one can invoke for example a period of matter domination in the visible sector during the second reheating phase or a period of kination domination in the hidden sector after relaxation. This can efficiently dilute away the dark radiation (see right panel of Fig.~\ref{fig:energy evolution}).  Another possibility is to assume that the hidden sector decays into the SM model after the reheating phase and before the BBN epoch.

In this scenario the role of particle production is two-fold: It stops the relaxion evolution at the right place and it reheats the visible sector after relaxation.
In addition, $\phi$ can be responsible for the generation of the primordial curvature perturbations. If the Hubble rate ($H_I$) 60 e-folds before the end of inflation is larger than the scale $\Lambda$, then the field $\phi$ has to be regarded as a free field, which has quantum fluctuations governed by $H_I$. Then, the relaxion acts as a curvaton field, generating a sufficient amount of curvature perturbation, that is transferred to the SM during the particle production phase.
A detailed investigation of this aspect requires a dedicated study that we leave for future work.

\subsection{Relaxion initial velocity}

\begin{figure}
\begin{center}
\includegraphics[width=.55\textwidth]{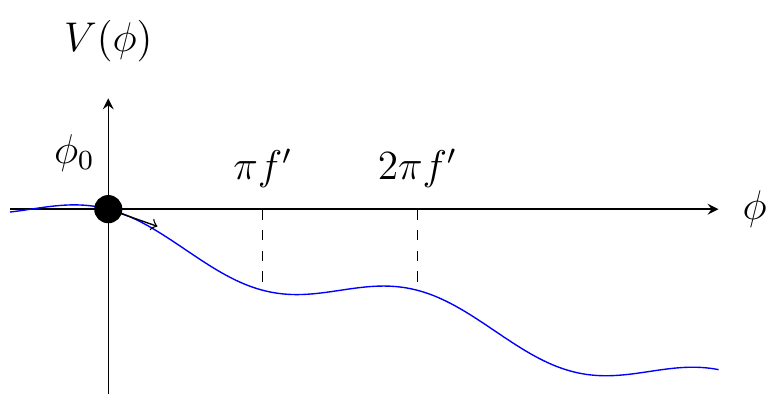}
\end{center}
\caption{ Sketch of the relaxion potential assuming a coincidence of scales that may avoid the requirement of initial large velocity. Here for convenience we set $V(\phi_0)=0$ and  $\phi_0=0$. }
\label{fig:byeye}
\end{figure}

The  initial velocity of the relaxion has to be
large enough to overcome the barriers in the periodic potential.
One could think that an initially small velocity is allowed for a generic initial position of $\phi$, if the height of the barriers is small enough compared with the average slope of the potential, as it is shown Fig.\,\ref{fig:byeye}.
In the following we estimate under which condition   the assumption of large initial velocity is not necessary, meaning that the field is in an  `easy rolling' regime.
If the field starts rolling from a generic point $\phi_0$, that without loss of generality we assume to be in the interval $0-2\pi f'$, it will be able to overcome the first barrier at $2\pi f'$ and consequently start the rolling phase only if $V(\phi_0)>V(2\pi f')$.
To quantify the requirement that this condition is generic, we impose that it holds for all values of $\phi_0$ between $0$ and $\pi f'$, implying $V(\pi f')\geq V(2\pi f')$, where
\begin{align}
 V(\pi f')  =& - g\,\Lambda^3\pi f'  + \Lambda_b^4 \cos\left(\frac{\pi f'}{f'}\right) \\
V(2 \pi f')  =& - g\,\Lambda^32 \pi f'  + \Lambda_b^4 \cos\left(\frac{2\pi f'}{f'}\right)
\end{align}
In order for the relaxion to roll down easily, one obtains
\begin{equation} \label{eq:scales}
\Lambda_b^4 \lesssim \pi g\,\Lambda^3 f'.
\end{equation}
However, once  particle production friction slows down the field, the constant periodic potential  needs to cancel the slope, stopping the field evolution, which implies that the height of the barrier  should be at least 
\begin{equation}
\Lambda_b^4 \gtrsim g\, \Lambda^3 f' \ .
\end{equation}
 Therefore, in order to avoid an initial large velocity, one could in principle saturate this bound  by assuming a coincidence of scales $\Lambda_b^4 \sim g\, \Lambda^3 f'$. Once we simultaneously consider all the requirements for successful particle production (see Sec.~\ref{sec:stop_pp}),  this condition cannot be satisfied.
At the end of the scanning process, the relaxion acquires a speed $\Lambda^2$ from the slope, and since $\phi$ has to be able to pass the barriers we have $\Lambda_b \lesssim \Lambda$, which would imply $\Lambda \sim g  f'$ (where we saturated the condition $\Lambda_b \sim \Lambda$ as such scales are very close to each other given the resulting parameter space in Sec.~\ref{sec:stop_pp}). Unfortunately, this is in conflict with the condition on the precision of the Higgs mass scanning that we will discuss in Sec.~\ref{sec:stop_pp}, which requires Eq.(\ref{eq:4}), and which would imply $\Lambda^2\lesssim m_h^2/(2\pi)$.
We therefore disagree with the statement in \cite{Hook:2016mqo,Son:2018avk} that the relaxion can start at rest by assuming a coincidence of scales $\Lambda_b^4 \sim g\, \Lambda^3 f'$.

As a result, we have to assume that the initial velocity satisfies
\begin{equation}
\label{eq:importantinequality}
\dot\phi\gtrsim \Lambda_b^2. 
\end{equation}
Realizing such a velocity as an outcome of a previous inflationary period is relatively simple. For example, one can introduce a coupling of the relaxion to the inflaton field, suppressed by a small coupling $\tilde g$, such that during inflation the relaxion obtains an effective slope $-\tilde g M_I^3 \phi$. Then imposing that the slow-roll velocity of the relaxion is larger than $\Lambda_b^2$ we obtain $\tilde g \gtrsim \Lambda_b^2 / (M_I \MPl)$, which would allow for a rather high inflationary scale while having a small coupling $\tilde g$. The smallness of the coupling $\tilde g$ is important to guarantee that the shift symmetry is only softly broken, and in addition to avoid the reheating of the SM after inflation.\footnote{Assuming a coupling $\tilde g M_I^2 \sigma \phi$, where $\sigma$ is the inflaton, we estimate the decay width of the inflaton into SM fields through the mixing with the relaxion as $\Gamma \sim \tilde g^2 M_I^3/f^2$, where $f$ is the scale controlling the coupling of the relaxion to gauge bosons that will be introduced in Sec.~\ref{sec:gauge}. To avoid reheating to the SM, one needs $\Gamma/H<1$, which can be obtained by taking a small coupling $\tilde g$, while still satisfying the inequality above.}
 Furthermore, a precise statement about the  initial velocity requires to know the relaxion interaction with the hidden sector and also the UV physics that generate these barriers and how fast they appear. From now, we will assume \Eq{eq:importantinequality}.

A potential worry is if this velocity can be sustained during the relaxation phase, or whether the Hubble friction generated by the relaxion itself could reduce this velocity, in particular since the relaxion has to pass a very large number of barriers.
We estimate the condition for this not to happen in the following way. Let us consider only the first barrier of the potential.
The effect of Hubble friction is very small and we can use the approximate energy conservation to compute the velocity gain when the field goes from the first to the second peak
\begin{equation}
\Delta \dot\phi_g = \sqrt{\dot\phi^2 + 4\pi f' g \Lambda^3} - \dot\phi \,.
\end{equation}
This has to be compared with the velocity loss due to Hubble friction, which  is obtained as $\Delta\dot\phi_H = \int 3H\dot\phi dt \approx 6\pi f' H$. Imposing that $\Delta\dot\phi_g > \Delta\dot\phi_H$, one obtains
\begin{equation}\label{eq:domino}
f' < \frac{g \Lambda^3}{9\pi H^2} - \frac{\dot\phi}{3\pi H} \approx \frac{g \MPl^2}{9\pi\Lambda} \,,
\end{equation}
where the second term was neglected in the last step and to get the last equality we used $\dot\phi\approx\Lambda^2$ and $H\approx\Lambda^2/\MPl$. Given \Eq{eq:lowerboundongp}, condition~(\ref{eq:domino}) is trivially satisfied for sub-Planckian values of $f'$, which as we are going to see in Sec.~\ref{sec:requirements} is always the case in our setup.

\subsection{Higgs field following its minimum } \label{sec:Higgs_tracking}

In the case where the relaxation mechanism starts in the broken EW phase, we have to make sure that the Higgs field follows the minimum of its potential, to guarantee that the stopping mechanism is triggered when the Higgs VEV is small.
Considering the potential in \Eq{eq:relaxion_potential}, the Higgs field has a minimum given by  
\begin{equation} 
v = \frac{1}{\sqrt{\lambda}} (\Lambda^2 - g'\Lambda\,\phi)^{1/2}\,.
 \label{eq:higgsvev}
\end{equation}
If the Higgs field does not efficiently track the minimum of the potential,  \textit{i.e.}  if it has a much larger value than the  small value at the potential minimum near the critical value  $(\Lambda^2 - g'\Lambda\,\phi_c) \approx 0$,  then the relaxion mechanism is spoiled.

\begin{figure}
   \centering
  \begin{minipage}{0.4\textwidth}
\includegraphics[width=\textwidth]{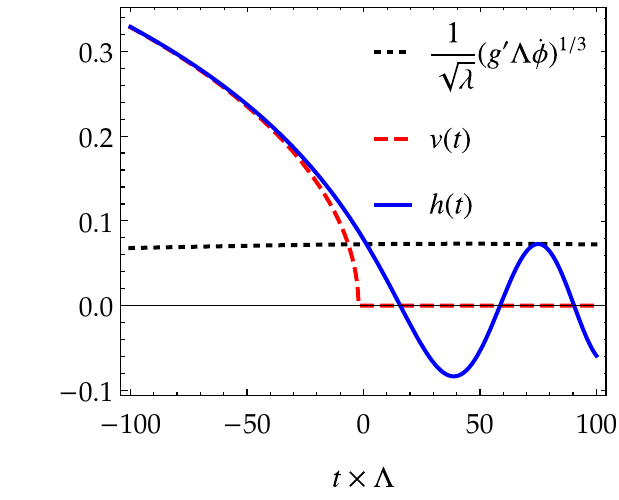}
\end{minipage}
\begin{minipage}{0.39\textwidth}
\includegraphics[width=\textwidth]{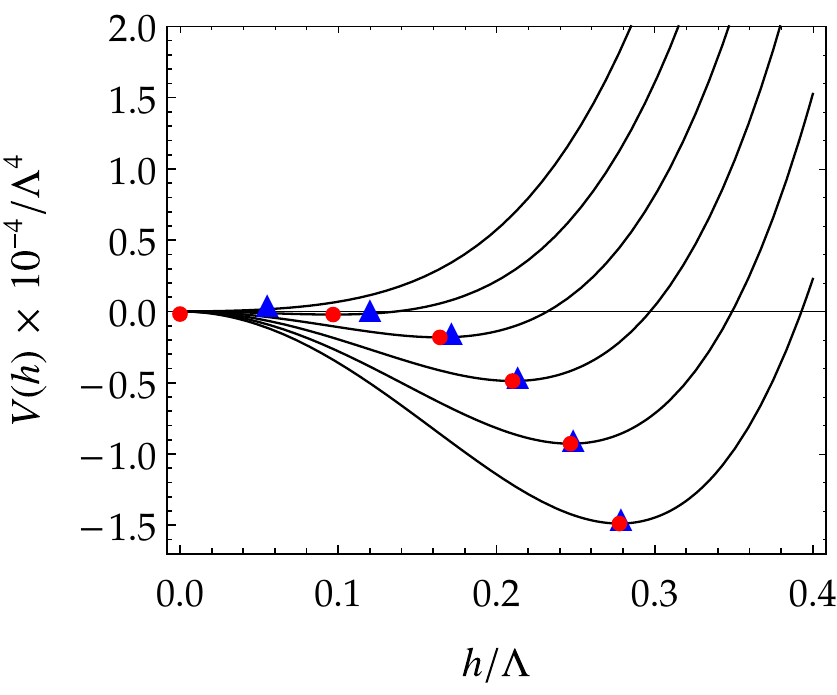}
\end{minipage}
   
  \begin{minipage}{0.4\textwidth}
\includegraphics[width=\textwidth]{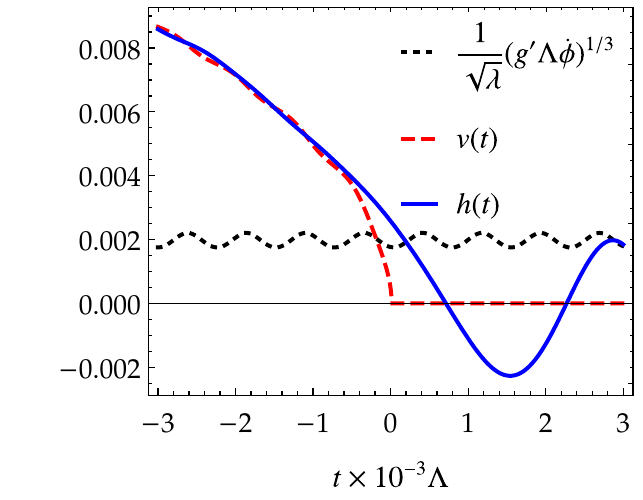}
\end{minipage}
\begin{minipage}{0.37\textwidth}
\includegraphics[width=\textwidth]{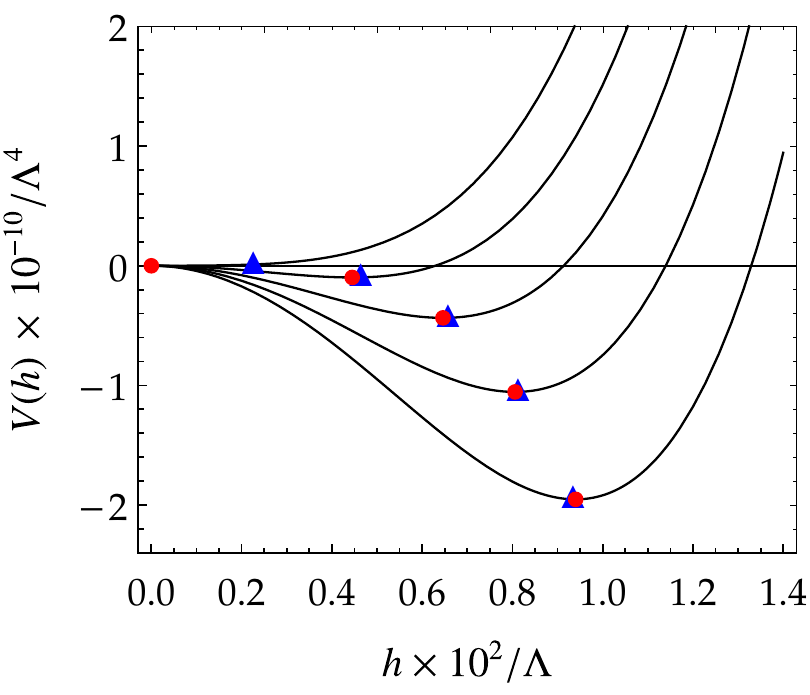}
\end{minipage}
\caption{\label{fig:tracking}   \emph{Left}: Higgs field $h(t)$ (blue) and its value at the minimum of the potential $ v(t)$  (red) as a function of time $t$  in units
 of $\Lambda^{-1}$. The black curve indicates when the tracking stops from the approximation in (\ref{eq:trackingvew})  \emph{Right}: Higgs potential where the blue points represent the  field value and the red ones the value at the minimum. $\Lambda =10^4~\text{GeV}, f'=10^6~\text{GeV}~ \text{and}~ \Lambda_b=7\times 10^3~\text{GeV}$.
\emph{Top}: $g'= 10^{-4}$, the Higgs is not
 efficiently tracking the minimum  of its potential  (the tracking stops
 when the value at the minimum is still too large,  $v \sim 0.1 \,\Lambda$).  \emph{Bottom}:  $g'= 3\times  10^{-9}$, 
  the Higgs is   efficiently following its minimum 
(the tracking stops when the value at the minimum is already below the electroweak scale, 
 $v \sim 0.003 \, \Lambda$).  }
\end{figure}

The Higgs efficiently follows the minimum of its potential as the mass is being scanned if the VEV evolves adiabatically, \ie
\begin{equation} \label{eq:tracking1}
\frac{\dot{v}}{v^2} \lesssim 1.
\end{equation}
 Note that $m_h^2 = 2 \lambda\, v^2$ is the Higgs mass-squared. Therefore, assuming that the evolution starts with the Higgs field at the minimum, the relation  (\ref{eq:tracking1}) tells us that if the mass is  large,  the field is kept at the minimum of the correspondingly deep well in the potential during the scanning process. 
In the regime  where the Higgs follows its minimum, we can write  \Eq{eq:tracking1} in terms of the Higgs field,
\begin{equation} \label{eq:tracking2}
h \gtrsim \frac{1}{\sqrt{\lambda}} (g' \Lambda\, \dot{\phi})^{1/3},
\end{equation}
where we assumed constant velocity $\dot{\phi}$.
The success of the mechanism requires that \Eq{eq:tracking2} can only be violated  when the Higgs field value is  below the electroweak scale, then we impose 
\begin{equation} \label{eq:trackingvew}
h \sim \frac{1}{\sqrt{\lambda}} (g' \Lambda\, \dot{\phi})^{1/3} \lesssim v_\ew\,,
\end{equation}
At the end of the evolution, the velocity is expected to be 
\begin{equation}
\dot{\phi}\sim\mathcal{O}(\Lambda^2) \ , 
\end{equation}
so the bound in \Eq{eq:trackingvew} shows that  we can make the Higgs follow  the minimum of the potential as
close as we want by decreasing  $g'$ and/or decreasing the cutoff scale $\Lambda$. 
As we shall see in Sec.~\ref{sec:requirements}, this is an important constraint in our parameter space.
One can obtain the same result given in \Eq{eq:trackingvew} by studying how  efficiently the field  tracks the minimum using the expansion of the Higgs around the tracking solution as in \Ref{Hook:2016mqo}.
In Fig.~\ref{fig:tracking},  we compare the actual evolution of $h$ and $v$ with the approximation given by \Eq{eq:trackingvew}.  These solutions  were obtained by solving numerically the classical equations of motion for the fields. 
In the plots on the top,  the Higgs does not track the minimum close enough to the critical point.  In the plots on the  bottom,  the Higgs is   efficiently following its minimum.

\subsection{Baryogenesis}

While we are interested to decouple the relaxion mechanism from inflation, it is important to note at this point that according to \Fig{fig:energy evolution}, while relaxation will happen after the reheating stage in which the inflaton energy density is transferred to an invisible sector, the relaxion energy density eventually takes over  and a second reheating stage will follow at the end of relaxation when the energy density of the relaxion is transferred to the SM.
The entropy injected in the plasma will then dilute the baryon asymmetry produced in earlier phases with a factor $ (v_\ew/\Lambda)^3$. This affects in particular a possible scenario in which baryogenesis takes place at high temperature before relaxation and with a very large Higgs VEV.
Our relaxion mechanism therefore calls for some alternative baryogenesis mechanism taking place at the end of or after the relaxion mechanism. Electroweak baryogenesis is not option given that a first-order electroweak phase transition requires new physics at the electroweak scale, which is at odds with the relaxion principle. Other mechanisms will rely on non-vanishing $B-L$ processes. It might also be possible that the relaxion itself generates the baryon asymmetry through the $\phi F\widetilde{F}$ coupling  with a mechanism similar to \cite{Kusenko:2014uta, Jimenez:2017cdr}.  As this work is being completed,  a recent proposal appeared in \cite{Son:2018avk} that is constrained by the bounds we derive in the following sections.

%%%%%%%%%%%%%%%%%%%%%%%%%%%%%%%%%%%%%%%%%%%%%%%%%%%%%%%%%%%%%%%%%%%%%%%%%%%%%
\section{General conditions  for relaxation  through particle production}\label{sec:stop_pp}
%%%%%%%%%%%%%%%%%%%%%%%%%%%%%%%%%%%%%%%%%%%%%%%%%%%%%%%%%%%%%%%%%%%%%%%%%%%%%

Particle production was successfully used to trap moduli fields at enhanced symmetry points~\cite{Kofman:2004yc} or to generate slow-roll inflation with a non-flat potential~\cite{Green:2009ds,Pearce:2016qtn,Anber:2009ua}.
This mechanism was  applied to the relaxion scenario  in~\cite{Hook:2016mqo}, exploiting the higher-dimensional anomalous coupling of the relaxion to a combination of electroweak gauge fields.
Here we will discuss this possibility in detail.

Before doing so,  we start in the next subsection by considering the much more minimal possibility of Higgs particle production. This would be very appealing as it would not require any additional ingredient, exploiting the already existing Higgs-relaxion coupling.
Unfortunately, it will turn out that there is a fundamental obstruction related to the requirement that the Higgs tracks the minimum of its potential. We will then turn our attention to the model in which the relaxion couples to the Chern-Simons term of the SM massive gauge bosons, as considered in~\cite{Hook:2016mqo} and derive the constraints that this mechanism poses on the parameter space.
In Sec.~\ref{sec:cosmology}, we will discuss the further cosmological bounds on this model.

\subsection{Friction from Higgs particle production}

Let us   consider   the case of  production of Higgs particles through the coupling to the relaxion.  We decompose the Higgs field $h$  in a classical background field and a quantum fluctuation,
\begin{equation}
h = h_0 + \chi \ .
\end{equation}
 The set of equations of motion  for the model in (\ref{eq:relaxion_potential}) can be found in  App.~\ref{app:Higgs_pp}.   One can study how a single Fourier mode $\chi_{\vec{k}}$ evolves by looking at the linearized equation of motion,
\begin{equation}
\label{eq:brokenchi}
\left(\partial^2 + k^2 + g' \Lambda \phi - \Lambda^2 + 3\lambda\, h_0^2 \right)\chi_{\vec{k}} = 0\,.
\end{equation}
Efficient particle production requires the adiabatic condition to be violated, \emph{i.e.} 
\begin{equation} \label{eq:efficient_pp}
\left |\frac{~\dot{\omega}_k~}{\omega_k^2}\right|\gg 1,
\end{equation}
 where 
 \begin{equation}
 \omega_k = \sqrt{k^2 + (g' \phi \Lambda -\Lambda^2 +3\,\lambda\, h_0^2)}
 \end{equation}
  is the frequency of the quantum field $\chi_{\vec{k}}$.   
 If one naively assumes that the field $h_0$ always efficiently tracks  its minimum,  \Eq{eq:efficient_pp} tells us that, for low momentum $k$, particle creation is efficient when $ (g' \Lambda\, \phi_c  - \Lambda^2) \approx 0$. This implies that $ v \approx 0  \leftrightarrow \omega_k \approx 0 $, where $v$ is the Higgs minimum in \Eq{eq:higgsvev}. This  behaviour would be exactly what we are looking for, meaning that particle production friction is large for small $v$, then the relaxion could get trapped when the Higgs mass is small. More concretely, for low momentum $k$,  \Eq{eq:efficient_pp} is satisfied if
\begin{equation} \label{eq:eff_pp}
 \dot{\phi} \gtrsim 2 \sqrt{2} \,  (g' \Lambda)^{1/2}  \left|\phi - \frac{\Lambda}{g'}\right|^{3/2}.
\end{equation}
On the other hand, from \Eq{eq:tracking2}, we know that the condition for the Higgs field to track the minimum of its potential is $h_0 \gtrsim (g' \Lambda \dot{\phi})^{1/3}/\sqrt{\lambda} $, leading to
\begin{equation}\label{eq:trackingeff}
\dot{\phi} \lesssim (g' \Lambda)^{1/2} \left|\phi - \frac{\Lambda}{g'}\right|^{3/2}.
\end{equation}
Therefore, in the regime the approximations we use are valid, conditions (\ref{eq:eff_pp}) and (\ref{eq:trackingeff}) show that one cannot simultaneously have efficient particle production and  the Higgs field following the minimum of its potential. This result may be expected as for the  potential in \Eq{eq:relaxion_potential} the condition (\ref{eq:tracking1}) is basically the opposite of (\ref{eq:efficient_pp}). In fact,  for the benchmark points in Fig.~\ref{fig:tracking}, we find that the maximum  of  $|\dot{\omega}_k/\omega_k^2|$ are   $|\dot{\omega}_k/\omega_k^2|_{\rm{max}}\sim 0.6 $ and $|\dot{\omega}_k/\omega_k^2|_{\rm{max}}\sim 0.2$, respectively for the top and bottom cases in Fig.~\ref{fig:tracking}, showing that the adiabaticity condition is not violated in those cases.

\subsection{Stopping the relaxion with gauge bosons production} \label{sec:gauge}

We are now going to discuss the possibility that the relaxion is coupled to a massive vector field that becomes tachyonic when the VEV of the Higgs is sufficiently small. When the tachyonic instability occurs, the field grows exponentially, at the expense of the kinetic energy of the relaxion field, which decreases until the point $\dot{\phi} \sim\dot\phi_{\textrm{stop}}\lesssim\Lambda_b^2$ when the relaxion is no more able to overcome the barriers.
Here we review the main ingredients and the constraints that we must impose for the successful implementation of the mechanism.

\medskip

Given a generic vector field $V_\mu$, its equation of motion presents a tachyonic instability if the field is coupled to the relaxion via a term
\begin{equation}
\frac{\phi}{4 \mathcal{F}} V_{\mu\nu}\widetilde V^{\mu\nu}\,,
\label{eq:FFtilde}
\end{equation}
where $\widetilde{V}^{\mu\nu} = \epsilon^{\mu\nu\rho\sigma}V_{\rho\sigma}$ (notice that this differs from the usual convention by a factor of $1/2$). In our case, the field $V_\mu$ should be a massive SM vector, as we want its mass to be related to the Higgs VEV. It is crucial that the Lagrangian does not contain a term similar to \Eq{eq:FFtilde} for the photon, as otherwise the tachyonic instability would be present during all the evolution, independently of the smallness of the Higgs mass.
One can write the following Lagrangian, invariant under $\textrm{SU}(2) \times\textrm{U}(1)$ \cite{Hook:2016mqo}:
\begin{align}
\mathcal{L} = & \frac{1}{2}\partial_\mu \phi \partial^\mu \phi +
(\D_\mu \HD)^\dagger \D^\mu \HD -
\frac{1}{2}\tr [W_{\mu\nu} W^{\mu\nu}] - \frac{1}{4} B_{\mu\nu} B^{\mu\nu} \nonumber \\
& - \frac{\phi}{4\mathcal{F}} \left(g_2^2 W^a_{\mu\nu}\widetilde{W}^{a\,\mu\nu} - g_1^2 B_{\mu\nu}\widetilde{B}^{\mu\nu} \right) - V(\phi, \HD^\dagger \HD)
\label{eq:lagrangian vectors}
\end{align}
where $g_1$ and $g_2$ are the $\textrm{U}(1)$ and $\textrm{SU}(2)$ coupling constants, respectively, and $\D_\mu$ is the usual covariant derivative. The   coupling to gauge bosons above  prevents the relaxion to couple to the photon's $F \widetilde F$ term and has to be protected by a symmetry in a UV model containing the SM group. An example of such a UV completion is  discussed in  \cite{Hook:2016mqo}, where the
 coupling structure $\phi\,(\theta_W W\widetilde W + \theta_B B\widetilde B)$ with $\theta_W = - \theta_B$ is fixed if the SM is embedded in a left-right symmetric model $\textrm{SU}(2)_L\times\textrm{SU}(2)_R \times\textrm{U}(1)_{B-L}$ (see e.g \cite{Pati:1974yy,Mohapatra:1974gc,Mohapatra:1974hk, Senjanovic:1975rk}). In this case, a global symmetry  forces the coupling structure  as in \Eq{eq:lagrangian vectors} and then forbids the relaxion coupling to  $ F \widetilde F$.
 The coupling structure  which prevents a coupling of the relaxion to the photon also appears in non-minimal composite Higgs models with coset $SO(6)/SO(5)$ where the interaction of the additional singlet PNGB arising via anomalies is of the form (\ref{eq:lagrangian vectors}) with no coupling to the photon~\cite{Cacciapaglia:2014uja, Gripaios:2016mmi, Molinaro:2017mwb, Chala:2017sjk}.\footnote{This motivates the investigation of  UV completions for the relaxation mechanism with particle production in the context of composite Higgs models \cite{Bertuzzo:2018}.}

After EW symmetry breaking, we can rewrite the relevant part of \Eq{eq:lagrangian vectors} in terms of the mass eigenstates $A_\mu$, $Z_\mu, W^\pm_\mu$
\begin{align}\label{eq:lagbroken}
\mathcal{L} \supset\, & m_W^2(h)W^-_\mu W^{+\,\mu} + \frac{1}{2}m_Z(h)^2Z_\mu Z^\mu \nonumber \\
& - \frac{\phi}{\mathcal{F}} \epsilon^{\mu\nu\rho\sigma} \left(
2 g_2^2 \partial_\mu W^-_\nu \partial_\rho W^+_\sigma +
(g_2^2-g_1^2) \partial_\mu Z_\nu \partial_\rho Z_\sigma
- 2g_1 g_2 \partial_\mu Z_\nu \partial_\rho A_\sigma
\right)\,,
\end{align}
where the masses of the gauge bosons are $m_W(h) = g_2 h/2$ and $m_Z(h) = \sqrt{g_2^2+g_1^2}h/2$.
The contribution of the $WW$ and of the $ZA$ terms can be safely neglected when discussing the evolution of the relaxion and the particle production phase. Indeed, we expect the $WW$ contribution to be suppressed by thermal effects, due to its non-abelian nature~\cite{Linde:1980ts, Gross:1980br, Espinosa:1992kf}. The photon, instead, does not present a tachyonic instability in its equation of motion, and we expect its behaviour to be oscillatory up to  the critical point when the $Z$ field starts growing. After this point, one of the two components of the field will grow following the evolution of the $Z$, while the other will decrease. Since this behaviour will only start after the critical point, we expect the photon to contribute at most with a $\mathcal{O}(1)$ factor to our results. From now on we just consider the tachyonic instability from the $Z\widetilde{Z}$ term.

 It is important to stress that  shift symmetry breaking terms like $\sim g'\Lambda\, \phi\, h^2$  in \Eq{eq:relaxion_potential} do not respect the global symmetry which forces the coupling structure in \Eq{eq:lagrangian vectors}, so an anomalous interaction  with photons is generated through the small mixing with the Higgs. Even more important are the contributions to the $\phi F\widetilde F$ coupling coming from a $W$ loop through the interaction in \Eq{eq:lagbroken}~\cite{Bauer:2017ris, Craig:2018kne}. 
 These contributions are non-zero once the symmetry protecting the PNGB mass is broken and they disappear in the massless limit ($m_\phi^2 \rightarrow 0$). 
 There are several reasons why an anomalous coupling with photons can be dangerous. First, if photon production is efficient, the corresponding  friction term, which is always active, should be included in the relaxion evolution. Second, these photons may generate a temperature  that is large enough to deconfine the strong sector which is supposed to stop the relaxion field evolution through the generation of the potential barriers. Finally, if such temperature is large ($ T_\gamma \gtrsim \Lambda$), we may end up scanning the Higgs thermal  mass instead  of the vacuum mass parameter $\mu_h^2$.   In Sec.~\ref{sec:photons} we  specify the relaxion effective coupling to photons and discuss the condition one should impose to guarantee that this coupling is enough suppressed.

\medskip

A massive vector field $V_\mu$ can be decomposed in three independent parts: two transverse components and one longitudinal component. The term $V_{\mu\nu}\widetilde{V}^{\mu\nu}$ does not contain the longitudinal one, whose equation of motion is therefore the usual Klein-Gordon equation with a positive mass term, plus an additional term that encodes the variation of the mass.
Because this equation does not predict a tachyonic growth, we can neglect the longitudinal mode in our description.
As shown in App.~\ref{app:gaugebosons}, absorbing the gauge couplings in the definition of $f$,
\begin{equation} \label{eq:f convention}
\frac{1}{f} = \frac{(g_2^2 -g_1^2)}{\mathcal{F}},
\end{equation}
the equations of motion for the relaxion, the Higgs, and the transverse modes of the vector fields are:
\begin{align}
\ddot\phi - g\Lambda^3 + g'\Lambda h^2 + \frac{\Lambda_b^4}{f'}\sin\frac{\phi}{f'} + \frac{1}{4f}\langle V\widetilde{V} \rangle &= 0 \label{eq:phieomvector} \\
\ddot h + (g'\Lambda\phi-\Lambda^2) h + \lambda h^3 - \frac{1}{2} g_V^2\langle V_\mu V^\mu\rangle h &=0 \label{eq:heomvector} \\
\ddot V_\pm +(k^2 + m_V^2 \mp k\frac{\dot\phi}{f})V_\pm &= 0 \label{eq:Aeomvector}
\end{align}
where
\begin{equation}
m_V^2 = g_V^2 h^2,
\end{equation}
 and we have neglected the spatial fluctuations of $h$ and $\phi$. 
 We also define 
 \begin{equation}
 \omega^2_{k\pm}=k^2 + m_V^2 \mp k\frac{\dot\phi}{f}.
  \label{eq:zero T omega}
 \end{equation}
 Notice that, for simplicity, we are working in Minkowski space. This simplification is well justified, since the particle production process is very fast, as we will see shortly, and cosmic expansion can be neglected.
The quantities in brackets should be interpreted as the expectation values of the corresponding quantum operators  (see the derivation in App.~\ref{app:gaugebosons}):
\begin{equation}
 \label{eq:fftilde2}
\langle V\widetilde{V} \rangle = \frac{1}{4\pi^2}\int\di k \, k^3 \frac{\partial}{\partial t} (|V_+|^2-|V_-|^2).
\end{equation}

When the Higgs VEV (and consequently the mass of the gauge bosons) decreases, \Eq{eq:Aeomvector} exhibits a tachyonic instability for the $V_+$ polarization, in some range of $k$.
The first mode $k_*$  that becomes tachyonic is the one for which $ \omega^2_{k\pm}$ is minimum, \ie 
\begin{equation}
0 = \frac{\di}{\di k}\left.(k^2 + m_V^2 - k\frac{\dot\phi}{f})\right|_{k=k_*} = 2k_* - \frac{\dot\phi}{f} \Longrightarrow k_* =\frac{\dot\phi}{2f}
\end{equation}
Plugging this into the equation of motion we get
\begin{equation} \label{eq:Atachyion}
\ddot V_+ +( m_V^2 - \frac{\dot\phi^2}{4f^2})V_+ = 0
\end{equation}
that becomes tachyonic for\footnote{ Note that the tachyonic instability should turn on before the relaxion reaches  the critical value $\Lambda/g'$ which is when the EW symmetry is recovered. To simplify the notation, we do not make a distinction between  $\Lambda/g'$ and the point where the first mode becomes tachyonic (given in (\ref{eq:Atachyion}))  as these points are very close to each other compared to the evolution range.}
\begin{equation}
\dot\phi\gtrsim \dot\phi_c = 2 f m_V \ .
\end{equation}
 We call $t_c$  the time when this condition turns true.
Initially, the velocity is large enough that the field jumps over the barriers. 
At the time $t=t_c$, the terms $\langle V\widetilde{V} \rangle$ and $\langle VV\rangle$ in the equations of motion grow exponentially. Particle production starts.
The term $\langle V\widetilde{V} \rangle / 4f$ comes to dominate the equation of motion of $\phi$, which  slows down and is captured in the potential wells.%
\footnote{Notice that the intuition of $\langle V\widetilde{V} \rangle$ acting as a friction term which is valid in inflationary models as in~\cite{Anber:2009ua} does not apply here, since it is derived from the assumption of a constant slow-roll velocity.}
The parameters of the model must be chosen in such a way that, at $t_c$, the relaxion field is as close as possible to the critical value $\Lambda/g'$, thus generating the hierarchy between the cutoff $\Lambda$ and the electroweak scale.
Numerical solutions to the system of equations (\ref{eq:phieomvector},
\ref{eq:heomvector}, \ref{eq:Aeomvector}) are shown for illustration in Fig.~\ref{fig:numerics} in App.\,\ref{sec:appendix}. The plotted time evolution is limited to a range where numerics is under control (shortly after $t_c$, fields are subject to a large oscillatory behaviour). This is enough to see that the relaxion slows down and  $\phi$ stops growing, as soon as 
\begin{equation}
h(t) < \dot{\phi}/(2 g_V f) \ ,
\end{equation}
 which is when some $k$ modes becomes tachyonic ($\omega^2_{k+}<0$),  stabilising the Higgs mass parameter,  while the temporary additional contribution to the Higgs mass parameter from the gauge field production grows.

\begin{figure}
\begin{center}
\includegraphics[width=.85\textwidth]{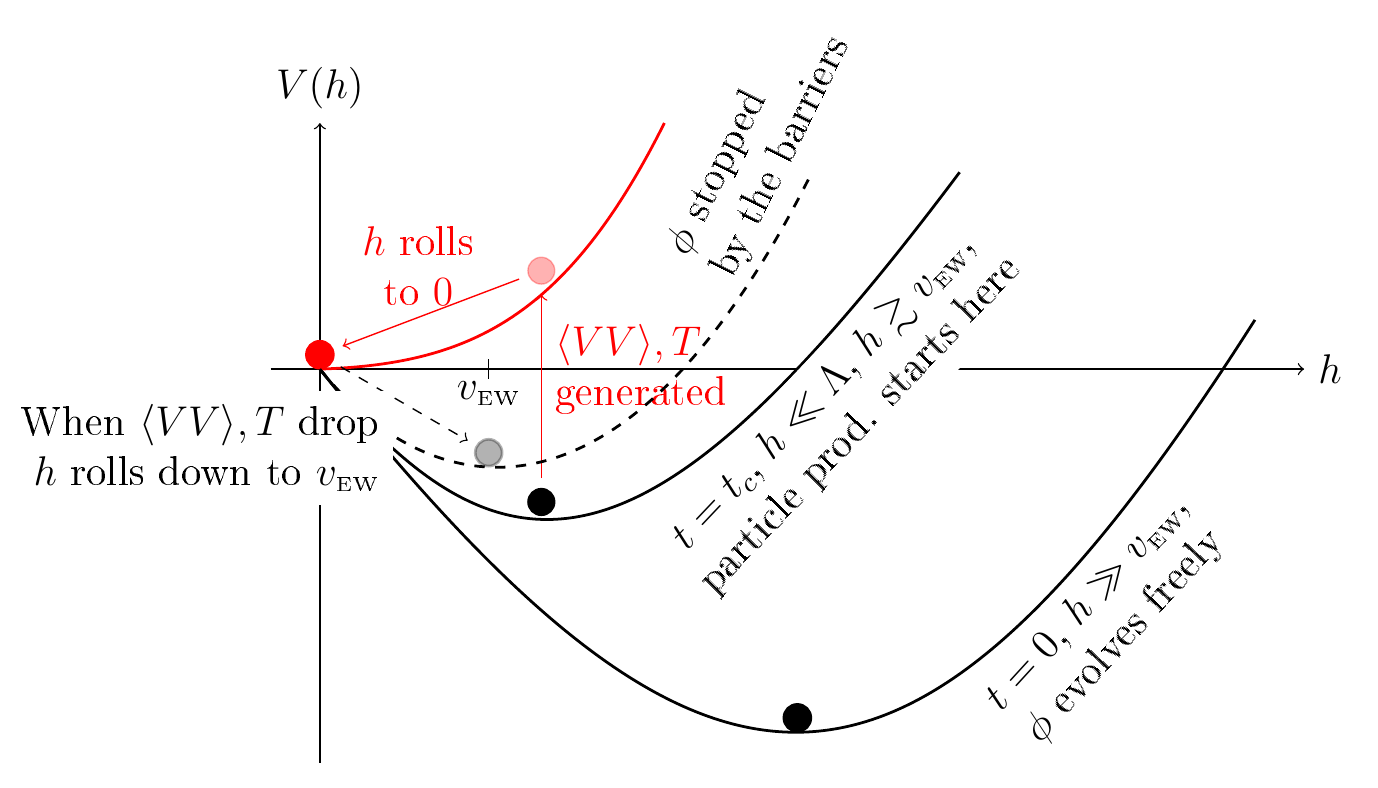}
\end{center}
\caption{Evolution of the Higgs field during cosmological relaxation intertwined with gauge boson production.}
\label{fig:Higgs evolution}
\end{figure}

A crucial role is played by the evolution of the Higgs field right after particle production starts. Two main effects should be considered here.
First, as the $V_+$ field grows, the $\langle VV\rangle$ term in \Eq{eq:heomvector} induces a positive mass term for the Higgs field, temporarily restoring the electroweak symmetry. The field $h$ rapidly rolls to zero, and so does the mass of the vector boson, making the tachyonic growth even faster.
Another effect, which on the other hand is not included in the system (\ref{eq:phieomvector},
\ref{eq:heomvector}, \ref{eq:Aeomvector}), is due to the temperature.
The produced particles are expected to thermalise, generating an additional thermal mass term for the Higgs, which adds to the one mentioned above.
After temperature has dropped, the Higgs relaxes to the minimum of the $T=0$ potential, which is now given by $v_\ew$. Such evolution of the Higgs field is summarised in Fig.~\ref{fig:Higgs evolution}.

The thermalisation process and the computation of the reheating temperature are a topic of study in their own that goes beyond the scope of this paper. On the other hand, the time scale for particle production in the presence of a thermal bath can be estimated as follows \cite{Hook:2016mqo}.
The generation of a thermal mass  for the Higgs is not the only effect of thermalisation.  The  production of gauge bosons will be affected by the presence of the thermal plasma, and the dispersion relation for the tachyonic mode ($V_+$) is modified into
\begin{equation}\label{eq:dispersion relation}
%\omega^2 = \omega^2_{k\mp} + \Pi[\omega,k]
\omega^2 = k^2 + m_V^2 - k\frac{\dot\phi}{f}
+ \Pi[\omega,k]
\end{equation}
where, in a hard thermal loop (\ie\,  high temperature) limit \cite{Bellac:2011kqa},
\begin{equation}\label{eq:dispersion relation 2}
\Pi[\omega,k] = m_D^2\frac{\omega}{k}\left(\frac{\omega}{k}+\frac{1}{2}\left(1-\frac{\omega^2}{k^2}\right)\log\frac{\omega+k}{\omega-k} \right)\,.
\end{equation}
Here $m_D^2 = g_\ew^2 T^2/6$ is the Debye mass of the plasma. In  pure QED, it  comes from evaluating an electron loop and it is proportional to the coupling $e^2$. Doing the same calculation but including all SM fermions (all assumed to be light in this phase), one gets a factor $g_1^2 \times ((-1)^2\times 3 + (2/3)^2\times 3\times 3 + (-1/3)^2 \times 3 \times 3) = (32/9)g_1^2$, where $g_1$ is the SM hypercharge coupling. This result should be multiplied by $\sin^2\theta_W\approx0.23$ to project the $Z$ onto its abelian component. Taking the value of the SM coupling $g_1\approx0.5$ we get $g_\ew^2 \approx 0.2$.

For imaginary frequency $\omega=i\Omega$ the function $\Pi[\omega,k]$ is positive, which already shows that the tachyonic instability is damped by the thermal bath.
As we discussed above, the instability first develops for $k\sim\dot\phi/(2f)$, and its timescale $\Omega$ is initially small. For $\Omega / k \to 0$ we can expand \Eq{eq:dispersion relation 2} obtaining
\begin{equation}
\Pi[\Omega,k] \approx \frac{\pi}{2}\frac{|\Omega|}{k}m_D^2.
\end{equation}
Plugging this back in \Eq{eq:dispersion relation} and neglecting the bare mass $m_V$ and $\mathcal{O}(\Omega^2)$ terms we obtain that $\Omega$ is maximized for $k = 2\dot\phi/ (3 f)$, 
\begin{equation}\label{eq:Omega thermal}
\Omega_\textrm{max} \approx \frac{8}{27\pi}\frac{\dot \phi^3}{f^3 m_D^2} = \frac{16}{9\pi g_\ew^2} \frac{\dot \phi^3}{T^2 f^3}\,.
\end{equation}
Equation~(\ref{eq:Omega thermal}) gives an estimate of the typical timescale for the exponential growth of the Fourier modes of the vector field $V_+$ in the presence of a thermal bath,
\begin{equation}
\Delta t_{\rm{pp}} \sim \frac{9\pi g_\ew^2}{16} \frac{T^2 f^3}{\dot \phi^3} \,,
\end{equation}
which we are going to use in the following sections in quantifying the efficiency of the mechanism.

\section{Effective   coupling  with photons and fermions} \label{sec:photons}

In this section, we discuss the different contributions to the $ (\phi/f_\gamma)  F \widetilde F$  effective coupling, which are generated due to the relaxion-Higgs mixing in  \Eq{eq:relaxion_potential} or by the relaxion interaction with the electroweak gauge bosons  in \Eq{eq:lagbroken}.   
Note that the contribution originated from the mixing with the Higgs is  also present in other relaxion models. 
As we show in the following,  the effective coupling  $\phi F \widetilde F$ from the   mixing with the Higgs  is sufficiently suppressed, and consequently it is harmless to the particle production mechanism. On the other hand, the contribution resulting from the leading interaction with the SM gauge bosons in \Eq{eq:lagbroken} is not negligible  and can constrain  part of our parameter space.

First, let us discuss the coupling $(\phi/f_\gamma) F \widetilde F$ generated through the mixing with the Higgs (see Sec.~\ref{sec:relaxion mixing}) which decays into two photons with a CP-violating coupling that is generated at three loops. It  is suppressed by one power of the SM 
Jarlskog invariant $J=  \textrm{Im}[V_{ij}V_{kl}V^*_{il}V^*_{kj}]\sim 10^{-5}$  \cite{Patrignani:2016xqp}. Its size can be estimated as
\begin{equation}
\label{eq: mixing gamma gamma}
\frac{1}{f_\gamma} \sim \frac{g' \, J}{h} \, \frac{\alpha_W^2\alpha_\text{em}}{(4\pi)^6}
\end{equation}
where $h\sim\Lambda$ is the Higgs VEV during the relaxation process, $\alpha_{\textrm{em}}$ is the fine-structure constant, and $\alpha_W =\alpha_{\textrm{em}}/\sin^2\theta_W$ with $\theta_W$ being the SM weak angle. The particle production rate  (see Sec.\,\ref{sec:gauge}) should be smaller than the Hubble rate that gives the dilution  of these photons due to cosmic expansion. Given the smallness of the prefactor in Eq.~(\ref{eq: mixing gamma gamma}), this bound is trivially satisfied.\footnote{In addition, as the relaxion mixes with the Higgs, one can generate the term $ \sim \alpha_V/(4 \pi)g'(\Lambda/v_\ew) \phi/f F_{\mu \nu} F^{\mu \nu }$ using a  fermion loop, which contributes to the  vectors' kinetic term. Using  $\phi \sim \Lambda/g'$ and $f \sim \Lambda^2/(2 m_Z)$ (see \Eq{eq:mZprediction}), one can see that such contribution is  sub-dominant compared to the canonical kinetic term.}

In addition to the Higgs-relaxion mixing, the leading interaction with the electroweak gauge bosons  (see \Eq{eq:lagbroken}) also generates a coupling of the relaxion to SM fermions at one loop and to photons at one and two loops~\cite{Bauer:2017ris, Craig:2018kne}
\begin{equation}\label{eq:coupling fermions photons}
\frac{\partial_\mu \phi}{f_F} (\bar{\psi} \gamma^\mu \gamma_5 \psi)
\quad\text{and}\quad
\frac{\phi}{4f_\gamma} F \widetilde F
\end{equation}
where
\begin{equation}\label{eq:fermions}
\frac{1}{f_F} = \frac{3 \alpha_{\textrm{em}}^2}{4 \mathcal{F}}\left[  \frac{Y_{F_L}^2 +Y_{F_R}^2}{\cos^4\theta_W} -\frac{3}{4 \sin^4\theta_W}
\right]\log\frac{\Lambda^2}{m_W^2},
\end{equation}
and
\begin{equation} \label{eq:fgamma}
\frac{1}{f_\gamma} =  \frac{2 \alpha_{\rm em}}{ \pi \sin^2\theta_W \mathcal{F}} B_2\left(x_W\right)+\sum_F \frac{N_c^F Q_F^2}{2 \pi^2 f_F}   B_1\left( x_F\right),
\end{equation}
where $N_c^F$ and $Q_F$  are respectively  the color multiplicity  and the electric charge of the fermion $F$ with mass $m_F$, and the $x_i$ is defined as $x_i \equiv 4 m_i^2/m_\phi$. The functions $B_{1,2}$ are written as follows:
\begin{eqnarray}
\begin{aligned}
    B_1(x) &= 1 - x [f(x)]^2\\
    B_2(x) &= 1 - (x - 1) [f(x)]^2
  \end{aligned}
 \qquad  \qquad f(x)=\begin{cases}
    \arcsin \frac{1}{\sqrt{x}} & x \geq 1\\
    \frac{\pi}{2} + \frac{i}{2} \log \frac{1 + \sqrt{1-x}}{1 - \sqrt{1-x}} & x < 1.
  \end{cases}
\end{eqnarray}
In the small mass limit, these functions  asymptotically tend to  $B_1(x_F)\rightarrow -m_\phi^2/(12 m_F^2)$ and $B_2(x_W)\rightarrow m_\phi^2/(6 m_W^2)$ as $m^2_\phi\rightarrow 0$, such that these  contributions to the relaxion-photon effective coupling   are  suppressed  by the spurion $m_\phi^2$ and so  are absent  in the massless limit. 
Equation~(\ref{eq:fermions}) assumes that the coupling to fermions vanishes in the UV, and the logarithm accounts for the 1-loop RG evolution from the UV cutoff $\Lambda$ down to the electroweak scale.  
Differently from the relaxion-photon coupling, the coupling to fermions is generated independently of the relaxion mass as it respects the  axion shift symmetry.

\begin{figure}
\begin{center}
\includegraphics[width=.7\textwidth]{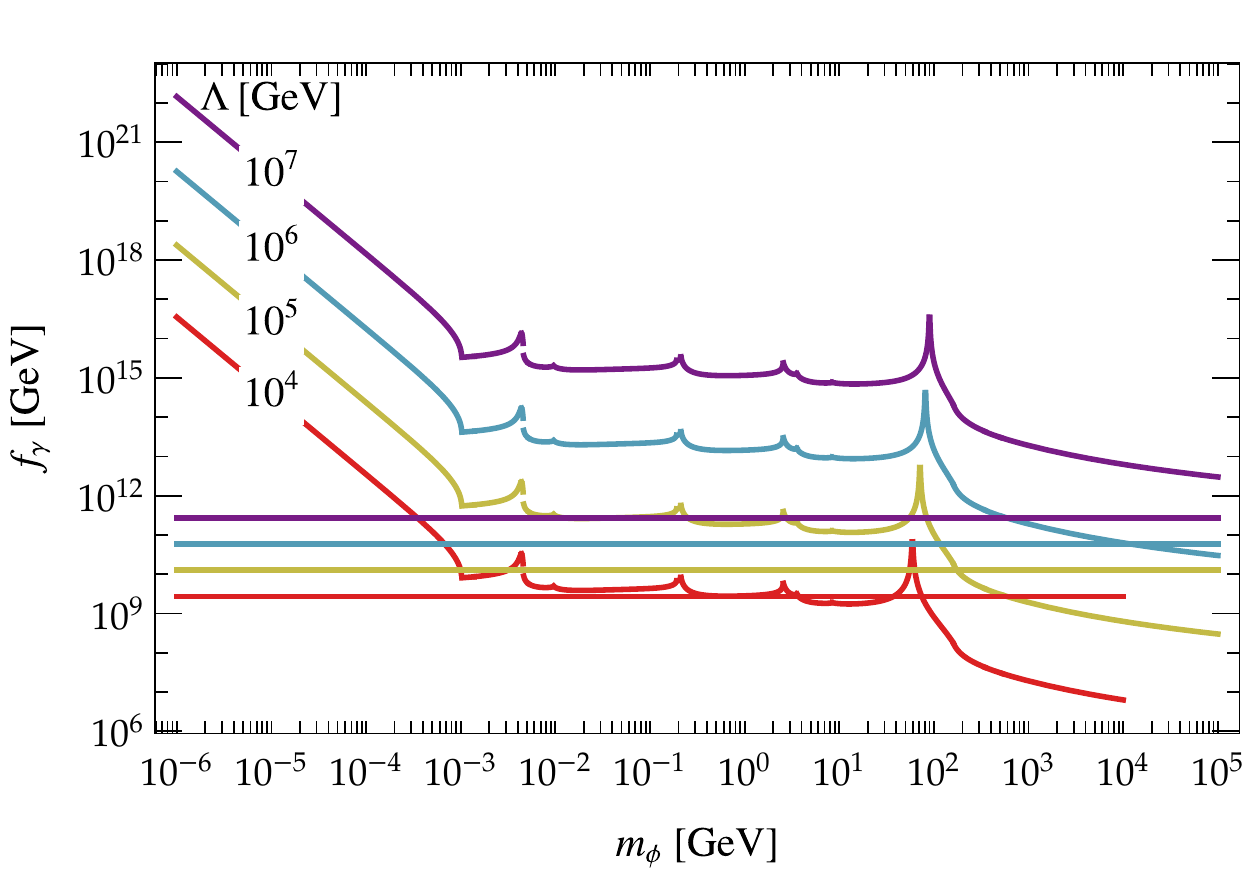}
\end{center}
\caption{Relaxion effective coupling to photons, ($\phi/f_\gamma) F \widetilde F$, given in \Eq{eq:fgamma} as a function of the relaxion mass for different cutoff scales from $\Lambda= 10^4~\textrm{GeV}$ (lowest, red) to $\Lambda =10^7~\textrm{GeV}$ (upmost, purple). The dilution condition (\Eq{eq:photon dilution}) is satisfied above the horizontal lines which set the bound for the different  the $\Lambda$  scales showed in the plot.}
\label{fig:fgamma}
\end{figure}

From the previous discussion, one  concludes  that  there is  an irreducible coupling to photons (\Eq{eq:fgamma}) once the axion shift symmetry is broken regardless of the absence of  an axion-photon coupling in the UV theory.  To guarantee that this coupling is sufficiently suppressed not to spoil the particle production mechanism, we have to impose that the timescale for photon production is longer than the Hubble time, 
\ie
\begin{equation} \label{eq:photon dilution}
\Delta t_{\gamma} > H^{-1},
\end{equation}
where  $\Delta t_{\gamma} \sim  T^2 f_\gamma^3/\dot{\phi}^3$ as discussed in Sec.\,\ref{sec:gauge} and $H\sim \Lambda^2/\MPl$. In Fig.\,\ref{fig:fgamma} we show the scale $f_\gamma$ as a function of the relaxion mass for different cutoff scales.
The horizontal lines set the condition in \Eq{eq:photon  dilution} such that above the lines the scale
$f_\gamma$   is large enough to suppress the coupling to photons. For instance, for $\Lambda =10^4~\textrm{GeV}$ (red line),  the condition in \Eq{eq:photon  dilution} is satisfied  if $m_\phi\lesssim \mathcal{O}(1)\,\textrm{GeV}$, and for $\Lambda =10^7~\textrm{GeV}$ (purple line)  the condition is always satisfied for the range showed in the plot. 

\section{Parameter space for  successful relaxation of the EW scale through gauge bosons production }
\label{sec:requirements}

In the following we list all  the constraints imposed to realize the relaxion idea 
using gauge bosons production as a stopping mechanism.

\begin{enumerate}
\item \textbf{Higgs field tracking the minimum of its potential:}  According to the discussion in Sec.~\ref{sec:Higgs_tracking}, the Higgs stops tracking its minimum  when $  h\sim (g' \Lambda \dot{\phi})^{1/3}/  \sqrt{\lambda}$ as in \Eq{eq:trackingvew}. We need to guarantee that this condition only breaks down when the Higgs field value is already below the electroweak scale, i.e. $h \lesssim v_{\ew}$.
This translates into
\begin{equation} \label{eq:tracking_conditions}
g' \lesssim \left(\frac{v_{\ew}  \, \sqrt{\lambda}}{\Lambda}\right)^{3},
\end{equation}
where we  neglect particle production and assume $\dot{\phi}\sim \Lambda^2$.

\item \textbf{Prediction for the electroweak scale:} To obtain  the correct electroweak scale, the dissipation should be important when the mass of the gauge boson is close to $m_Z$  
\begin{equation} \label{eq:phistop}
\dot\phi_c \sim 2\,  m_Z f,
\end{equation}
where $\dot{\phi}_c$ is the velocity when particle production becomes efficient (see the discussion in Sec.~\ref{sec:gauge}).
For the consistency of the effective theory we expect that $\dot{\phi}$ does not exceed 
 $\Lambda^2$. Assuming a non-vanishing initial  velocity, after traveling an entire field range $\Lambda/g'$, the velocity would be $\dot\phi \sim \Lambda^2$. Therefore, we can expect the final mass of the gauge bosons to be
\begin{equation}\label{eq:mZprediction}
m_Z \sim \frac{\Lambda^2}{2f} \,.
\end{equation}

On the other hand, during its rolling phase the field must be able to jump over the barriers, and therefore $ \dot{\phi} \gtrsim  \Lambda_b^2$. For this reason we  assume 
\begin{equation}
\Lambda_b \lesssim \Lambda \ .
\end{equation}

\item \textbf{Stopping condition:} Once particle creation slows down the relaxion, the constant barrier of the cosine potential ($\sim \Lambda_b^4 \cos{(\phi/f')}$) should be able to stop the field by cancelling the slope:
\begin{equation} \label{eq:stopping}
\Lambda_b^4 \gtrsim g\, \Lambda^3 f'.
\end{equation}

\item \textbf{Scanning precision:} the scanning of the Higgs mass should have enough precision so that we do not overshoot the electroweak scale, 
\begin{equation} \label{eq:scanning}
g' \Lambda\, \delta\phi =  g'\Lambda\, ( 2\pi f' )\lesssim m_h^2\,.
\end{equation}

\item \textbf{Efficient energy dissipation:} We need to impose that the kinetic energy that the relaxion looses due to particle production is larger than the one it gains by rolling down the potential slope,
\begin{equation}
\Delta K_\textrm{rolling}\, \lesssim\, \Delta K_\textrm{pp}.
\end{equation}
We assume that a $\mathcal{O}(1)$ fraction of the kinetic energy is dissipated away by particle creation, meaning that $\Delta K_\textrm{pp}\sim \dot\phi^2 /2$.
The energy gained by rolling can be estimated as $\Delta K_\textrm{rolling} \sim \frac{dK}{dt}  \Delta t_{\rm{pp}}$, where $dK/dt = -dV/dt \sim g \Lambda^3 \dot\phi$. To consider the most stringent bound, we  evaluate this condition for $\dot\phi^2/2 = \dot\phi_{\rm{stop}}^2/2 \sim\Lambda_b^4$, the maximum velocity the relaxion can have after it has been trapped. We get 
\begin{equation}\label{eq:condition energy dissipation}
\frac{9\pi g_\ew^2}{16} T^2 f^3 \lesssim  \frac{2 \Lambda_b^8}{g \Lambda^3} \,.
\end{equation}

Evaluating the condition at $\dot\phi_c$ would lead to a  similar bound on $\Lambda$, but would fail at constraining the scenario with $\Lambda_b\ll\Lambda$, which, as we will see, is excluded through Eq.~\ref{eq:condition energy dissipation}.

\item \textbf{Small variation of the Higgs mass:} When $\phi$ is loosing its kinetic energy, the Higgs mass parameter should not vary more than a fraction of the electroweak scale
during the time it takes for the relaxion velocity to become smaller than the barrier.
This can be satisfied if we impose
\begin{equation}\label{eq:f42}
\Delta m_h \sim \frac{\Delta m_h^2}{m_h} \sim  \frac{1}{m_h} g' \Lambda \, \dot\phi \, \Delta t_\textrm{pp} \lesssim m_h\,,
\end{equation} 
which, again, we evaluate at $\dot\phi = \dot{\phi}_{\rm{stop}}\sim\Lambda_b^2$ to derive the most stringent bound.

\item \textbf{Shift symmetry not restored:}
After the relaxion has been trapped, the temperature may be larger than the condensation scale of the cosine potential. In this case, the potential barriers would disappear, and the relaxion would start rolling again until the temperature is redshifted enough for the barriers to be generated again.
To avoid this scenario, we impose that
\begin{equation}
T < \Lambda_b \,.
\end{equation}
This condition only applies when the sector which generates the barriers is in thermal equilibrium with the SM. Assuming that the barriers are generated by some QCD-like gauge group coupled to the relaxion as $\phi G\widetilde G /f'$, we naively  estimate the rate for $gg\leftrightarrow ZZ$ interactions mediated by the relaxion as $\Gamma \sim T^5/(f^2f'^2)$, which must be larger than the Hubble rate $H\sim \Lambda^2/\MPl$.

\end{enumerate}

We now want to combine  all the above constraints. To display the allowed region of parameter space,  we make a  few simplifying assumptions. First, we assume $g=g'$, keeping in mind that from the perspective of a UV completion the terms proportional to $g'$ and $g$ in \Eq{eq:relaxion_potential}  should be generated in a similar way.
Secondly, we assume that a $\mathcal{O}(1)$ fraction of the relaxion kinetic energy is converted into radiation with temperature  given by
\begin{equation}\label{eq:Temperature}
\frac{\dot\phi_c^2}{2} \sim \frac{\pi^2}{30} g_*\, \Tpp^4 \,,
\end{equation}
where $g_*$ is the number of relativistic degrees of freedom in the SM at high temperature, and $\dot\phi_c \sim \Lambda^2$.
Finally, from \Eq{eq:mZprediction} we assume 
\begin{equation}
f = \Lambda^2/(2m_Z) \ .
\end{equation}
Under these assumptions, we are left with four free parameters: 
\begin{equation} \label{eq:parameters}
\{\Lambda, g', \Lambda_b, f'\},
\end{equation}
which are  constrained by the relations 1 to 7 together with the condition that the relaxion does not drive another period of inflation (see Eq.~(\ref{eq:lowerboundongp}) in Sec.\,\ref{sec:noinflation}) and that the produced photons are diluted by the cosmic expansion (see \Eq{eq:photon dilution} in Sec.\,\ref{sec:photons}). These constraints  can  be conveniently listed as:
\begin{align}
\label{eq:0}
g'\gtrsim\, & 0.2 \frac{\Lambda}{ \MPl} & \textrm{Avoid slow-roll}\\
\label{eq:1}
g' \lesssim & \left(\frac{v_{\ew}\sqrt{\lambda}}{\Lambda}\right)^{3} & \textrm{Higgs tracking the minimum}\\
\label{eq:photons}
f_\gamma^3 \gtrsim & \frac{9}{6\sqrt{5}} \frac{g_*^{1/2}}{g_\ew^2}\MPl \Lambda^2 & \textrm{Photons dilution} \\
%\Delta t_{\gamma} > & ~H^{-1} & \textrm{\new{Photons dilution}}\\
%
\label{eq:2}
\Lambda_b \lesssim\, & \Lambda & \phi\textrm{ initially rolls above the barriers}\\
\label{eq:3}
g' \lesssim\, & \frac{\Lambda_b^4}{\Lambda^3 f'} & \textrm{The barriers are high enough to stop }\phi\\
\label{eq:4}
g'\lesssim\, & \frac{m_h^2}{2 \pi f' \Lambda} & \textrm{Precision of the mass scanning}\\
\label{eq:5}
%\Lambda_b^8 \gtrsim\, & \frac{9 \sqrt{15}  g_\ew^2}{256 g_*^{1/2} } \frac{g' \Lambda^{11}}{m_Z^3} & \textrm{Efficient dissipation} \\
\Lambda_b^8 \gtrsim & 0.1 \frac{g_\ew^2}{g_*^{1/2} } \frac{g' \Lambda^{11}}{m_Z^3} & \textrm{Efficient dissipation} \\
\label{eq:6}
%\Lambda_b^4 \gtrsim\, & \frac{9 \sqrt{15}  g_\ew^2}{256 g_*^{1/2}} \frac{g' \Lambda^9}{m_h^2 m_Z^3} & \textrm{Small Higgs mass variation} \\
\Lambda_b^4 \gtrsim & 0.1 \frac{g_\ew^2}{g_*^{1/2}} \frac{g' \Lambda^9}{m_h^2 m_Z^3} & \textrm{Small Higgs mass variation} \\
\label{eq:8}
\Lambda_b\lesssim\, & f' & \textrm{Consistency of symmetry breaking pattern} \\
\label{eq:13}
\Lambda\lesssim\,   & f' & \textrm{EFT validity} \\
\label{eq:9}
%g' \lesssim & \frac{256 g_*^{1/2} }{9 \sqrt{15}  g_\ew^2} \frac{m_Z^3}{\Lambda^3} & \textrm{Combining~\ref{eq:2} and~\ref{eq:5}}\\
g' \lesssim\, & 7 \frac{g_*^{1/2} }{g_\ew^2} \frac{m_Z^3}{\Lambda^3} & \textrm{Combining~(\ref{eq:2}) and~(\ref{eq:5})}\\
\label{eq:10}
%g' \lesssim\, & \frac{256 g_*^{1/2} }{9 \sqrt{15}  g_\ew^2} \frac{m_h^2 m_Z^3}{\Lambda^5} & \textrm{Combining~\ref{eq:2} and~\ref{eq:6}} \\
g' \lesssim\, & 7 \frac{g_*^{1/2} }{g_\ew^2} \frac{m_h^2 m_Z^3}{\Lambda^5} & \textrm{Combining~(\ref{eq:2}) and~(\ref{eq:6})} \\
\label{eq:12}
%f'^2 \gtrsim\, & 180 \left(\frac{5}{3 \pi^{10} g_*^5}\right)^{1/4} \frac{ \MPl m_Z^2}{\Lambda } \lor \Lambda_b \gtrsim \left(\frac{15}{\pi^2 g_*}\right)^{1/4}\Lambda & \textrm{No symmetry restoration}
f'^2 \gtrsim\, & 12 \frac{ \MPl m_Z^2}{g_*^{5/4}\Lambda }~ \text{or}~
\Lambda_b \gtrsim \frac{\Lambda}{g_*^{1/4}} & \textrm{No symmetry restoration}
\end{align}

We used Eqs.~(\ref{eq:0}), (\ref{eq:1}), (\ref{eq:9}) and (\ref{eq:10}) to constrain the parameters $\Lambda$ and $g'$. Note that, while Eqs.~(\ref{eq:0}) and~(\ref{eq:1}) are generic, Eqs.~(\ref{eq:9}) and (\ref{eq:10}) depend on our assumptions on the behaviour of the gauge bosons after thermalisation, which is a difficult subject that would require a careful treatment beyond our simple description. Our choice was to maximize the strength of these constraints, therefore the reader should remember that they could in principle be relaxed.

\begin{figure}
\centering
\includegraphics[width=.65\textwidth]{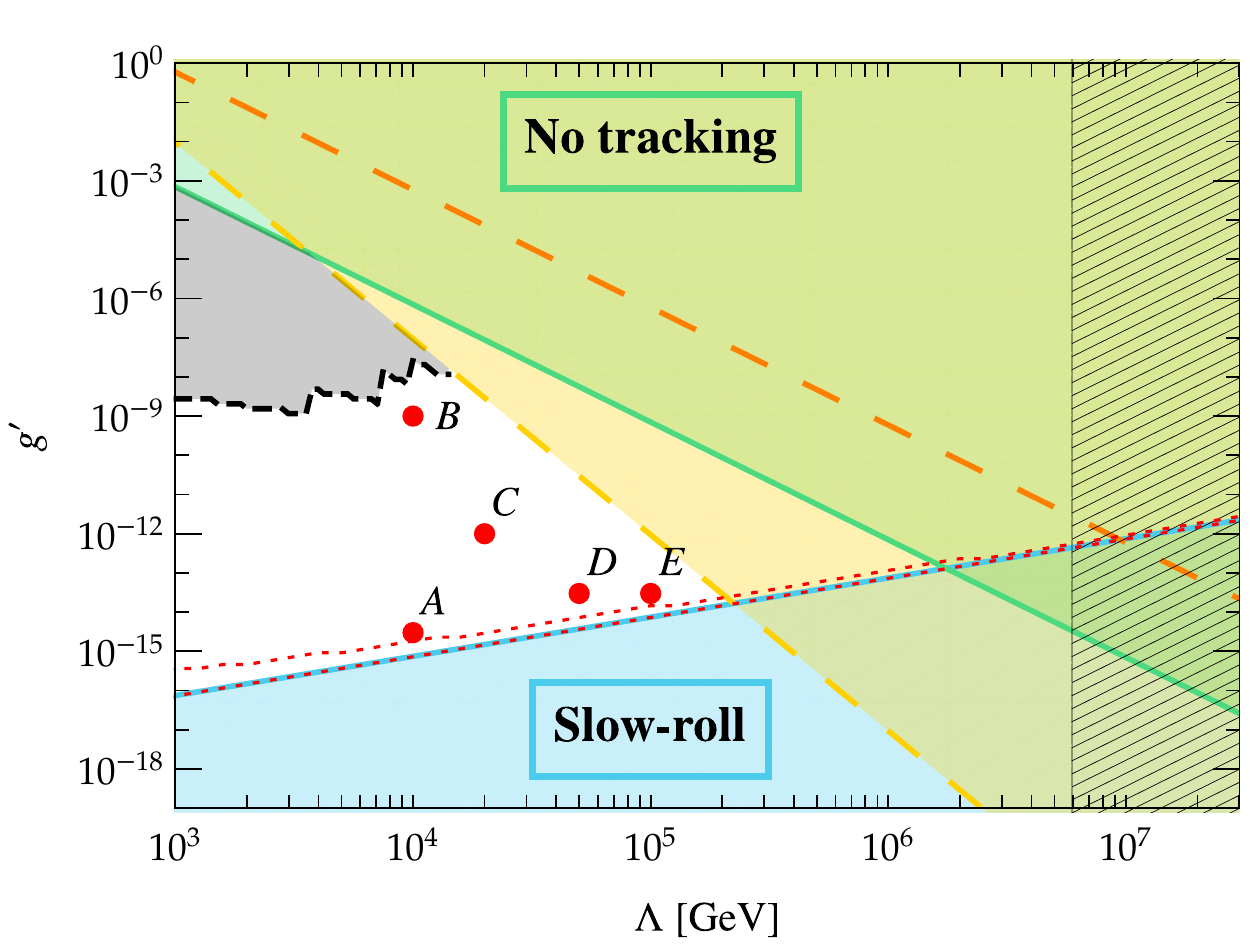}
\includegraphics[width=.34\textwidth]{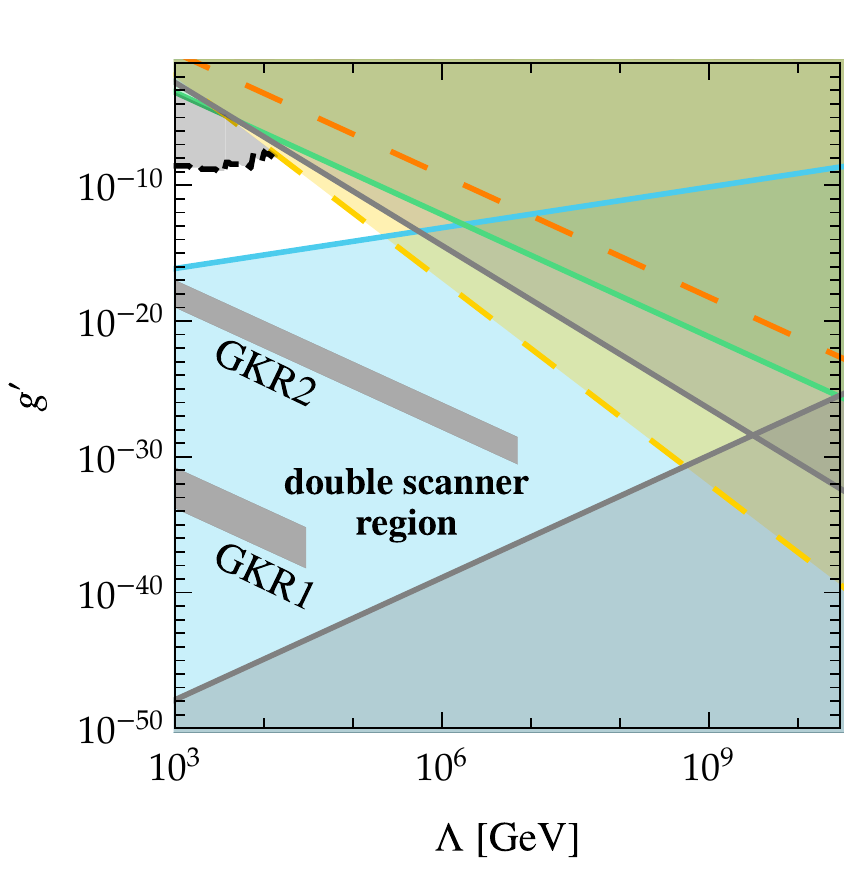}
%\begin{minipage}{\textwidth}
\begin{multicols*}{2}
\begin{itemize}
\item[{\tikz\fill[rounded corners = .5mm, colBlue4](0,0)rectangle(0.3,0.2);}] Slow-roll (Eq.~\ref{eq:0})

\item[{\tikz\fill[rounded corners = .5mm, colGreen3](0,0)rectangle(0.3,0.2);}] Untracked minimum (Eq.~\ref{eq:1})

\item[{\tikz\fill[rounded corners = .5mm, Black](0,0)rectangle(0.3,0.2);}] Unsuppressed $\phi F \widetilde F$ coupling \\ (Eq.~\ref{eq:photons})

\item[{\tikz\fill[rounded corners = .5mm, orange](0,0)rectangle(0.3,0.2);}] Small barriers $+$ efficient dissipation (Eq.~\ref{eq:9})

\item[{\tikz\fill[rounded corners = .5mm, colYellow1](0,0)rectangle(0.3,0.2);}] Small barriers $+$ small Higgs mass variation (Eq.~\ref{eq:10})
\end{itemize}
\end{multicols*}
%\end{minipage}

\caption{Summary of the available region of parameter space in the $g'-\Lambda$ plane consistent with the relaxion mechanism exploiting particle production instead of inflation.
Constraints come from Eqs.~(\ref{eq:0})--(\ref{eq:photons}), (\ref{eq:9}) and (\ref{eq:10}). The red dots correspond to the five benchmark cases of Fig.~\ref{fig:fpLambdac}. The dashed lines correspond to the bounds which rely on   conservative assumptions about the thermalisation process.
The region between the red dotted lines corresponds to the one in which the inflaton is allowed to reheat the SM before relaxation takes place (see Fig.~\ref{fig:reheating SM allowed g}).
The rectangular hashed region at large $\Lambda$ is the one in which the approximation of instantaneous particle production fails.
 Right plot: This smaller plot indicates in comparison the parameter space relevant for the two original relaxion models (GKR1 for the relaxion being the QCD axion and GKR2 for the model with strong  dynamics at the weak scale) proposed in \cite{Graham:2015cka}) as well as for the double-scanner mechanism of Ref.~\cite{Espinosa:2015eda}, which all use inflation as a source of friction.}
\label{fig:gpLambda}
\end{figure}

Despite all these constraints, interestingly, there is a sizeable region of parameter space that remains open.
In Fig.~\ref{fig:gpLambda} we show the bounds on $\Lambda$ and $g'$ that are drawn from the above inequalities.  Although the \Eq{eq:photons} depends on the relaxion mass ($m_\phi \sim \Lambda_b^2/f'$) which constrains the other plane $\Lambda_b-f'$, there is a region in the $g'-\Lambda$ plane that is excluded by this bound for the whole  mass range, which we also show in Fig.~\ref{fig:gpLambda}. The maximum cutoff that we can obtain in our model is 
\begin{equation}
\Lambda\sim3\times10^5\GeV \ ,
\end{equation}
 that can be extended up to $\Lambda\sim10^6\GeV$ in the case reheating is very inefficient.
In the right plot, we compare this open region with the parameter regions associated with the original relaxion models implemented during a long inflation era. The relaxion mechanism relying on gauge boson production as a source of friction is associated with much larger values of the coupling $g'$.

\begin{figure}
\centering
\includegraphics[width=.4\textwidth]{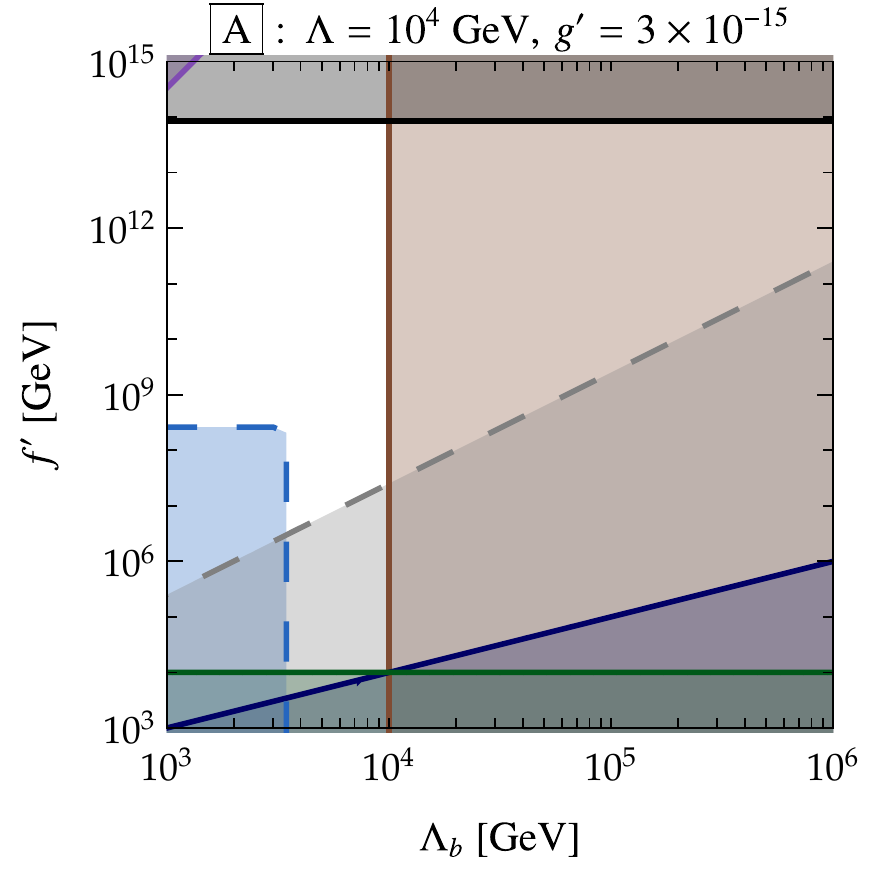}
\includegraphics[width=.4\textwidth]{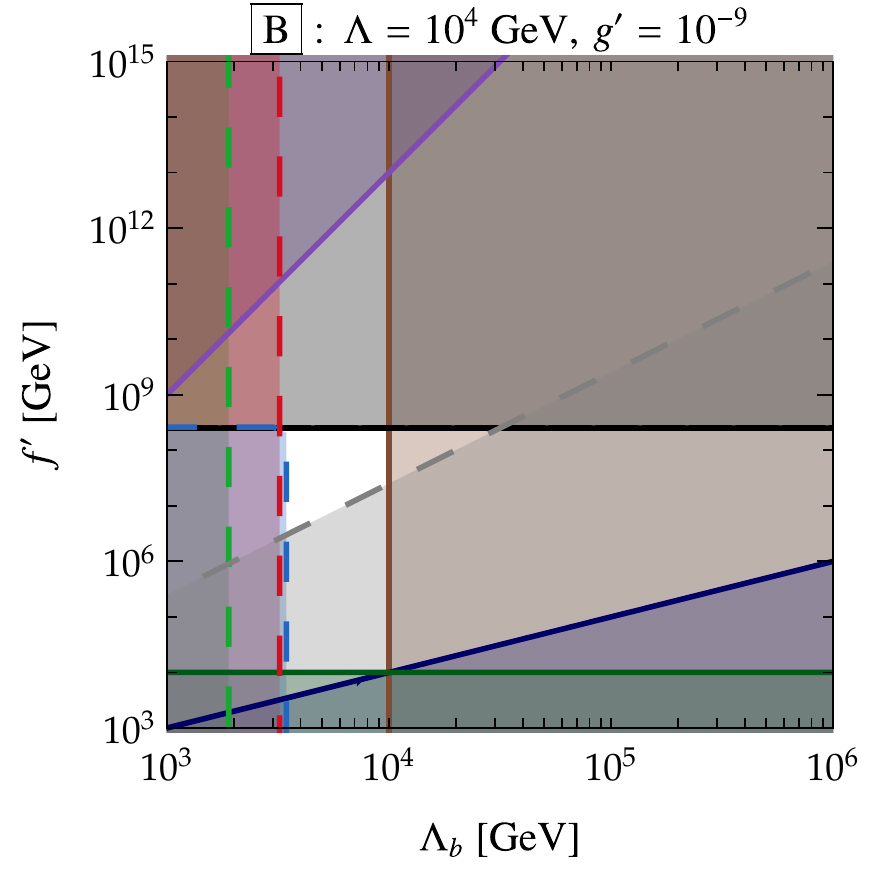}
\includegraphics[width=.4\textwidth]{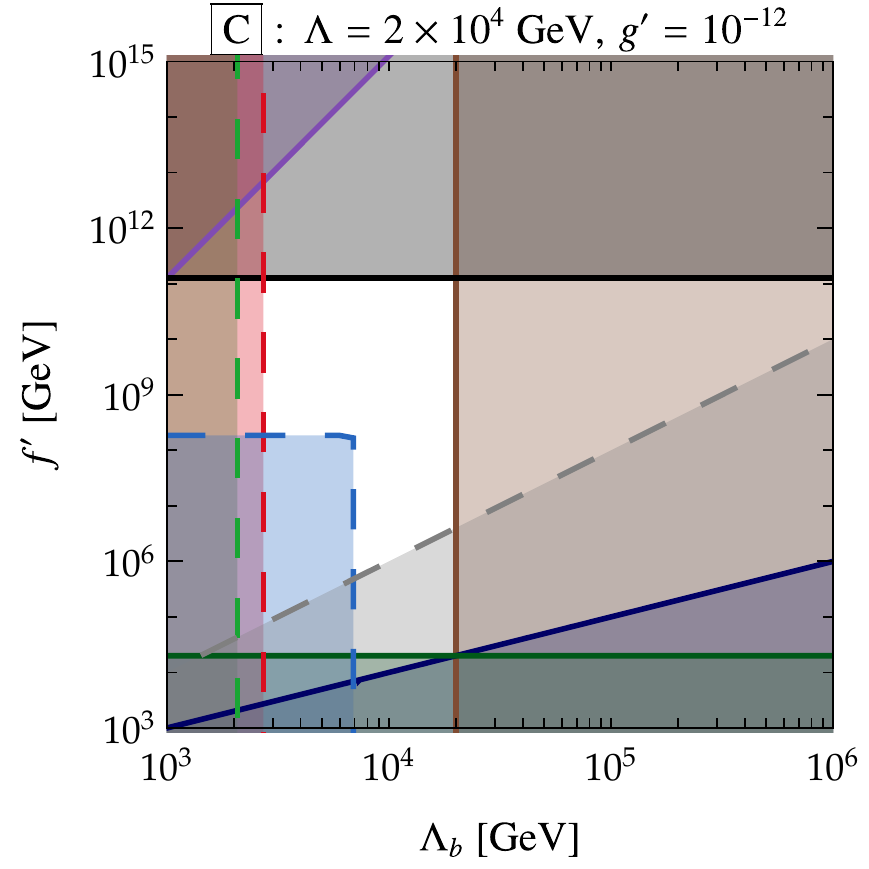}
\includegraphics[width=.4\textwidth]{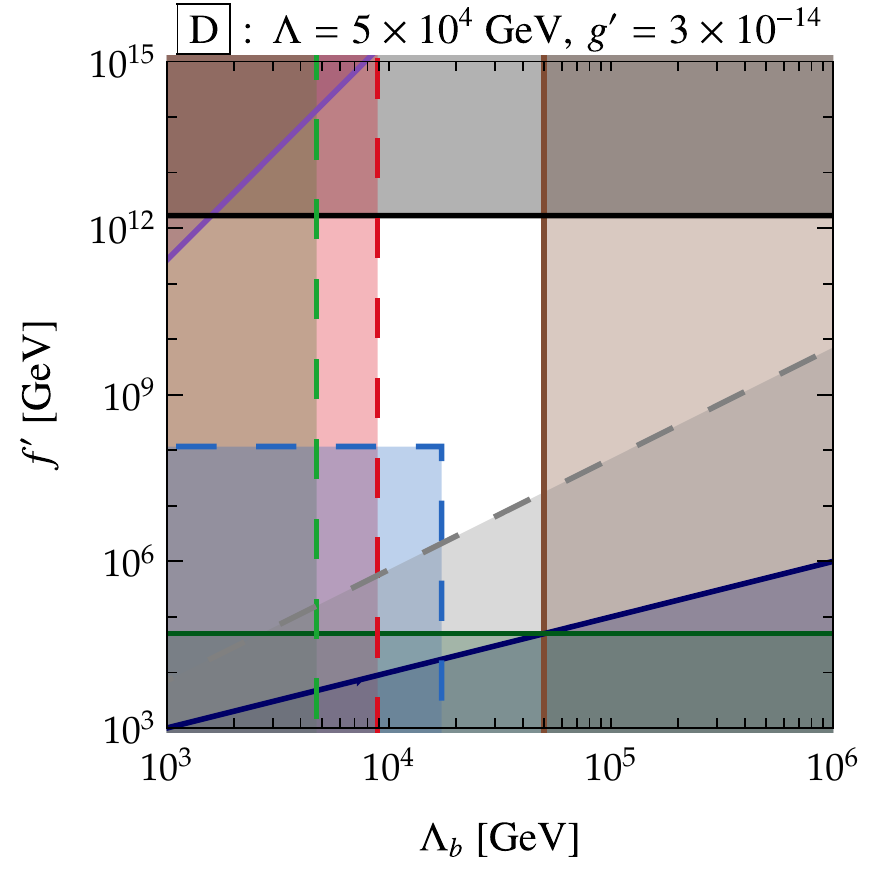}
\begin{minipage}{.4\textwidth}
\includegraphics[width=\textwidth]{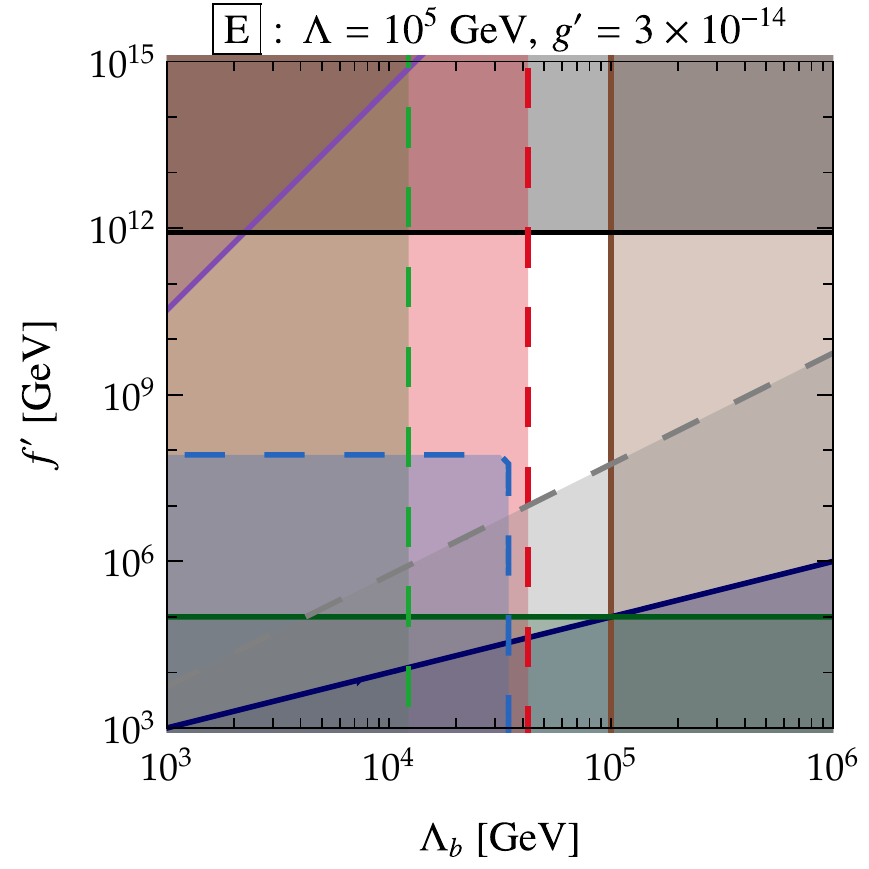}
\end{minipage}
\begin{minipage}{.4\textwidth}
\begin{multicols}{2}
\begin{itemize}
\item[{\tikz\fill[rounded corners = .5mm, Gray](0,0)rectangle(0.3,0.2);}] Eq.~\ref{eq:photons}

\item[{\tikz\fill[rounded corners = .5mm, colBrown2](0,0)rectangle(0.3,0.2);}] Eq.~\ref{eq:2}
\item[{\tikz\fill[rounded corners = .5mm, colViolet2](0,0)rectangle(0.3,0.2);}] Eq.~\ref{eq:3}
\item[{\tikz\fill[rounded corners = .5mm, black](0,0)rectangle(0.3,0.2);}] Eq.~\ref{eq:4}
\item[{\tikz\fill[rounded corners = .5mm, colGreen2](0,0)rectangle(0.3,0.2);}] Eq.~\ref{eq:5}
\item[{\tikz\fill[rounded corners = .5mm, colRed0](0,0)rectangle(0.3,0.2);}] Eq.~\ref{eq:6}
%
%\item[{\tikz\fill[rounded corners = .5mm, Orange](0,0)rectangle(0.3,0.2);}] Eq.~\ref{eq:7}
%
\item[{\tikz\fill[rounded corners = .5mm, colBlue1](0,0)rectangle(0.3,0.2);}] Eq.~\ref{eq:8}
\item[{\tikz\fill[rounded corners = .5mm, colGreen1](0,0)rectangle(0.3,0.2);}] Eq.~\ref{eq:13}
\item[{\tikz\fill[rounded corners = .5mm, colBlue3](0,0)rectangle(0.3,0.2);}] Eq.~\ref{eq:12}
%
%\item[{\tikz\fill[rounded corners = .5mm, colBlue3](0,0)rectangle(0.3,0.2);}] Eq.~\ref{eq:13}
\end{itemize}
\end{multicols}
\end{minipage}
\caption{Constraints on the $f'-\Lambda_b$ plane from Eqs.~(\ref{eq:photons})--(\ref{eq:13}),~(\ref{eq:12}) for the five benchmark cases listed in \Eq{eq:benchmarks}. The dashed lines correspond to the bounds which rely on   conservative  assumptions about the thermalisation process.}
\label{fig:fpLambdac}
\end{figure}

We also chose five benchmark points within the allowed region, and used Eqs.~(\ref{eq:photons})--(\ref{eq:13}) and (\ref{eq:12}) to constrain the remaining free parameters $\Lambda_b$ and $f'$ in each of these cases:
\begin{equation}
\begin{aligned}
\label{eq:benchmarks}
\text{scenario A:} \qquad & \Lambda = 10^4\GeV, \, &g'&=3\times10^{-15} \\
\text{scenario B:} \qquad & \Lambda = 10^4\GeV, \, &g'&= 10^{-9}\\
\text{scenario C:} \qquad & \Lambda = 2 \times 10^4\GeV, \, &g'&=10^{-12}  \\
\text{scenario D:} \qquad & \Lambda = 5 \times 10^4\GeV, \, &g'&=3\times10^{-14} \\
\text{scenario E:} \qquad & \Lambda = 10^5\GeV, \, &g'&=3\times10^{-14}.
\end{aligned}
\end{equation}
Figure \ref{fig:fpLambdac} shows the constraints on the $f'-\Lambda_b$ plane for the benchmark cases listed above. Again, conditions~(\ref{eq:5}), (\ref{eq:6}), and (\ref{eq:12})  depend on the details of the thermalisation process and should be considered as a pessimistic bound.
The results in Fig.~\ref{fig:fpLambdac} indicate that, in order for the mechanism to work, we need to require a coincidence of scales $\Lambda\sim\Lambda_b$, which may be reasonable if those two scales are generated by a common dynamics.

On top of these constraints, one should also check that particle production happens on a timescale shorter than a Hubble time, in order to justify our assumption of a Minkowski metric. During the whole relaxation process the relaxion dominates the energy density of the universe, with $H\sim \Lambda^2/\MPl$. Therefore we get the relation
\begin{equation}
\Delta t_\textrm{pp} \sim \frac{9\pi g_\ew^2}{16} \frac{T^2 f^3}{\dot \phi^3}  \lesssim \frac{\MPl}{\Lambda^2}
\end{equation}
which, for $\dot\phi\sim\Lambda^2$, is satisfied for $\Lambda\lesssim 7\times10^6 \GeV$. The region where the Minkowski approximation breaks down is shown in Fig.~\ref{fig:gpLambda} with a hashed area.

\begin{figure}
\centering
\includegraphics[width=.4\textwidth]{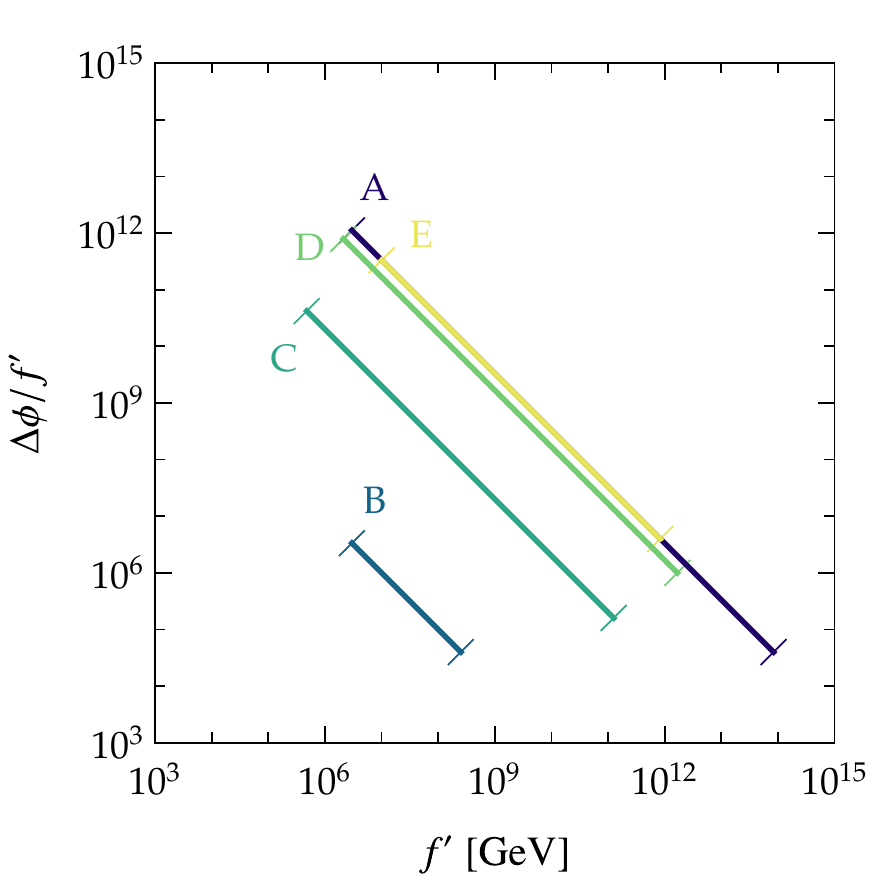}
\caption{Ratio of the total field excursion $\Delta\phi\sim\Lambda/g'$ over the decay constant $f'$, for the five benchmarks defined in Eq.~\ref{eq:benchmarks}. The lines are cut at the allowed values of $f'$. }
\label{fig: DeltaPhifp}
\end{figure}

In our setup,  the total field excursion over the decay constant $\Delta\phi/f'\sim \Lambda/(g'f')$ remains typically large, as shown in Fig.~\ref{fig: DeltaPhifp}, making a possible embedding of this model in string theory problematic, as discussed in~\cite{McAllister:2016vzi}.

One could wonder whether condition~(\ref{eq:10}) can be relaxed by assuming an inefficient thermalisation process. By naively setting $T=0$ and consider the zero temperature dispersion relation~\Eq{eq:zero T omega} when evaluating the time scale for particle production, the bounds indicated with dashed lines in Figs.~\ref{fig:gpLambda} and ~\ref{fig:fpLambdac} are relaxed, and the parameter space opens up to a value of the cutoff $\Lambda\approx 2\times 10^6\GeV$. The reason for this is that these bounds descend from upper limits on the time scale for particle production, which is shorter in the zero temperature case. This does not apply to the bound in \Eq{eq:photon dilution}, which, on the contrary, gives a lower bound on the same time scale. As a consequence, the black line in Fig.~\ref{fig:gpLambda} moves to lower values of the coupling, thus excluding all the parameter space that would be open by relaxing the other conditions. Anyway, this argument ignores the fact that, even out of thermal equilibrium, the dispersion relation~\Eq{eq:zero T omega} should be modified to account for finite density effects when a large number of particles is produced to dissipate the relaxion's kinetic energy. Because of this, we consider the results obtained under the assumption of thermal equilibrium (and consequently the use of the thermal dispersion relation~\Eq{eq:dispersion relation}) more reliable than the $T=0$ ones.

At this stage, we now need to consider phenomenological constraints and see whether they limit further the parameter space. For this, we determine the relaxion mass and lifetime in the next section.

%%%%%%%%%%%%%%%%%%%%%%%%%%%%%%%%%%%%%%%%%%%%%%%%%%%%%%%%%%%%%%%%%%%%%%%%%%%%%

%%%%%%%%%%%%%%%%%%%%%%%%%%%%%%%%%%%%%%%%%%%%%%%%%%%%%%%%%%%%%%%%%%%%%%%%%%%%%
\section{Relaxion properties}\label{sec:pheno}
%%%%%%%%%%%%%%%%%%%%%%%%%%%%%%%%%%%%%%%%%%%%%%%%%%%%%%%%%%%%%%%%%%%%%%%%%%%%%
\subsection{Relaxion mass and mixing with the Higgs}
\label{sec:relaxion mixing}

\begin{figure}
\centering
\includegraphics[width=.4\textwidth]{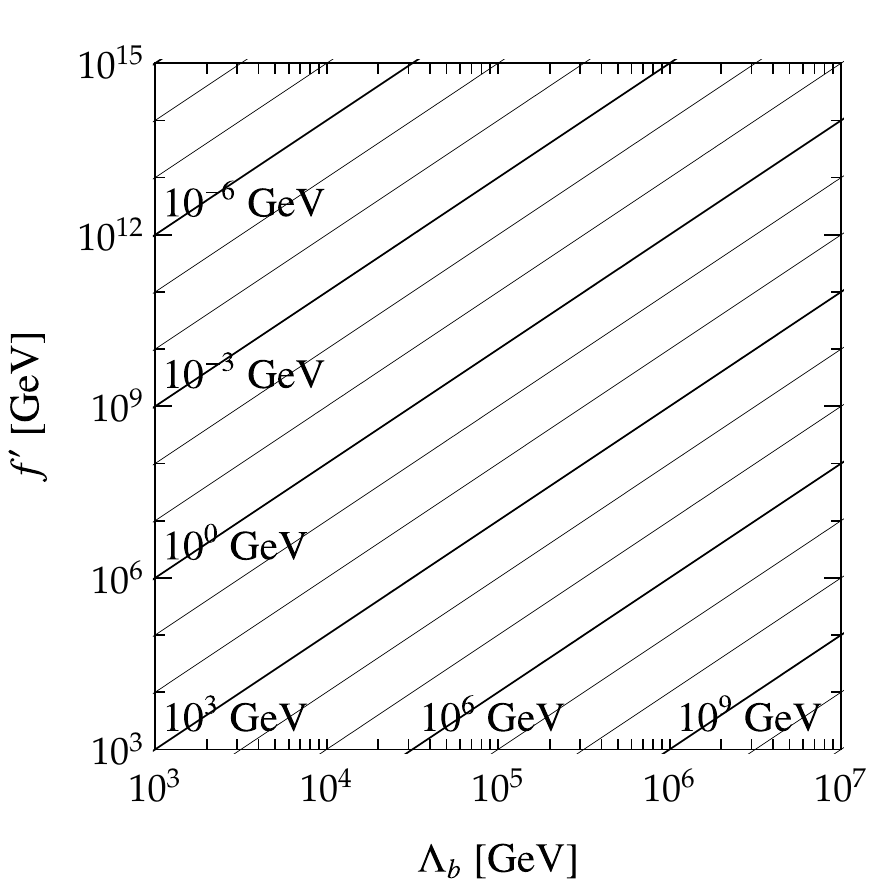}
\caption{Contours of $m_\phi$ in the $f'$--$\Lambda_b$ plane. The thick lines correspond to the indicated labels.}
\label{fig:phi mass}
\end{figure}

After the field $\phi$ has relaxed at one of the potential's minima, its mixing angle with the Higgs is obtained by expanding around the minimum of the potential $\phi_0$
\begin{equation}
\label{eq:potential after pp}
V(\phi,h) \supset -g\Lambda^3\phi + \frac{1}{2}[-\Lambda^2 + g' \Lambda(\phi_0+ \phi)]h^2 + \frac{\lambda}{4}h^4 + \frac{1}{2}m_\phi^2\phi^2 \,,
\end{equation}
where we now call $\phi$ the field displacement which is smaller than $f'$. The mass of the relaxion is
\begin{equation}
m_\phi\sim\Lambda_b^2/f'
\end{equation}
and is obtained by expanding the cosine potential. A contour plot of $m_\phi$ is shown in Fig.~\ref{fig:phi mass}.
The allowed mass range depends, given the values of $\Lambda$ and $g'$, on the intersection of the conditions displayed above. In particular, the maximal $m_\phi$ is obtained from Eqs.~(\ref{eq:2}),~(\ref{eq:8}) and~(\ref{eq:13}), which imply $m_\phi \lesssim \Lambda$, 
consistently with the fact that $\Lambda$ represents the cutoff of the theory.
The allowed mass ranges for the five benchmark points of Eq.~(\ref{eq:benchmarks}) are approximately given by 
\begin{align}
 %\nonumber \\
\text{scenario A:} \qquad & m_\phi \in [12\eV, \, 4\GeV]  \nonumber \\
\text{scenario B:} \qquad & m_\phi \in [58\MeV, \, 4\GeV] \nonumber \\ 
\text{scenario C:} \qquad & m_\phi \in [83\keV, \, 100 \GeV] \label{eq:mass range} \\
\text{scenario D:} \qquad & m_\phi \in [67\keV, \, 141\GeV] \nonumber \\
\text{scenario E:} \qquad & m_\phi \in [3\MeV, \, 178\GeV]. \nonumber
\end{align}

In our scenario, the barriers  do not depend on the Higgs VEV and the scale where this potential is generated can be as high as the cutoff ($\Lambda_b\lesssim \Lambda$). This implies that
 the upper bounds for the mass ranges in (\ref{eq:mass range})  can be much higher than the ones obtained in the relaxion models with a Higgs-dependent barrier  (see e.g. \cite{Espinosa:2015eda, Choi:2016luu, Flacke:2016szy, Beauchesne:2017ukw}).
Additionally, as we shall see in \Sec{sec:cosmology},  most of the lower bounds above are going to be shifted to higher masses after we consider the cosmological and astrophysical constraints.  

\begin{figure}
\centering
\includegraphics[width=.45\textwidth]{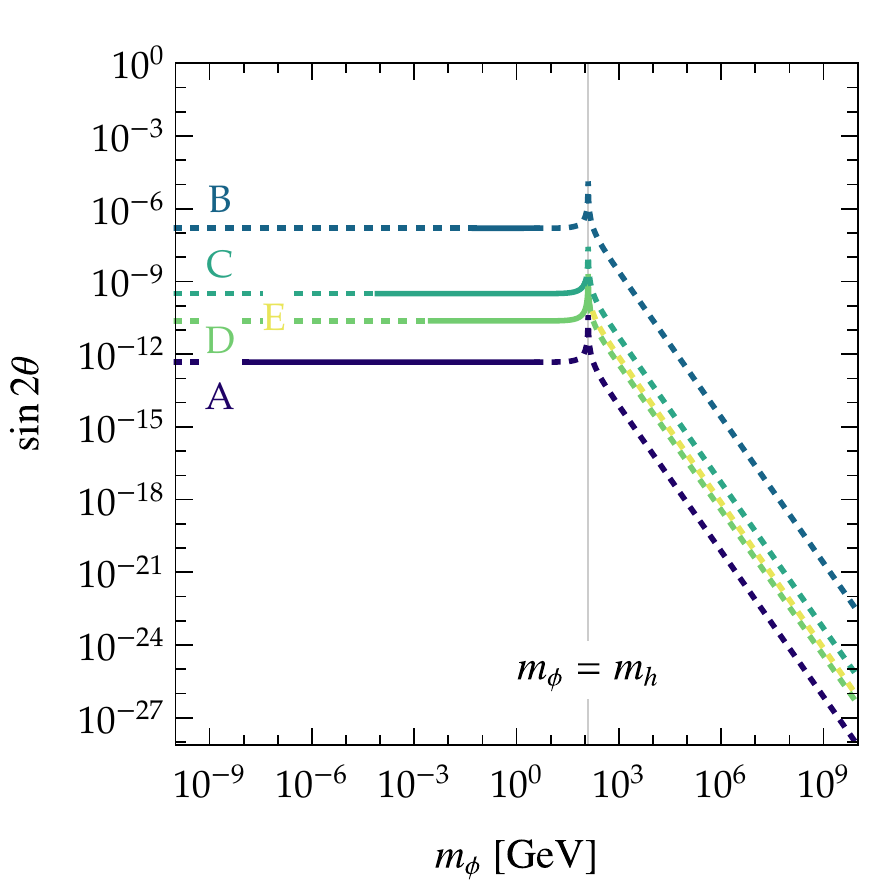}
\includegraphics[width=.45\textwidth]{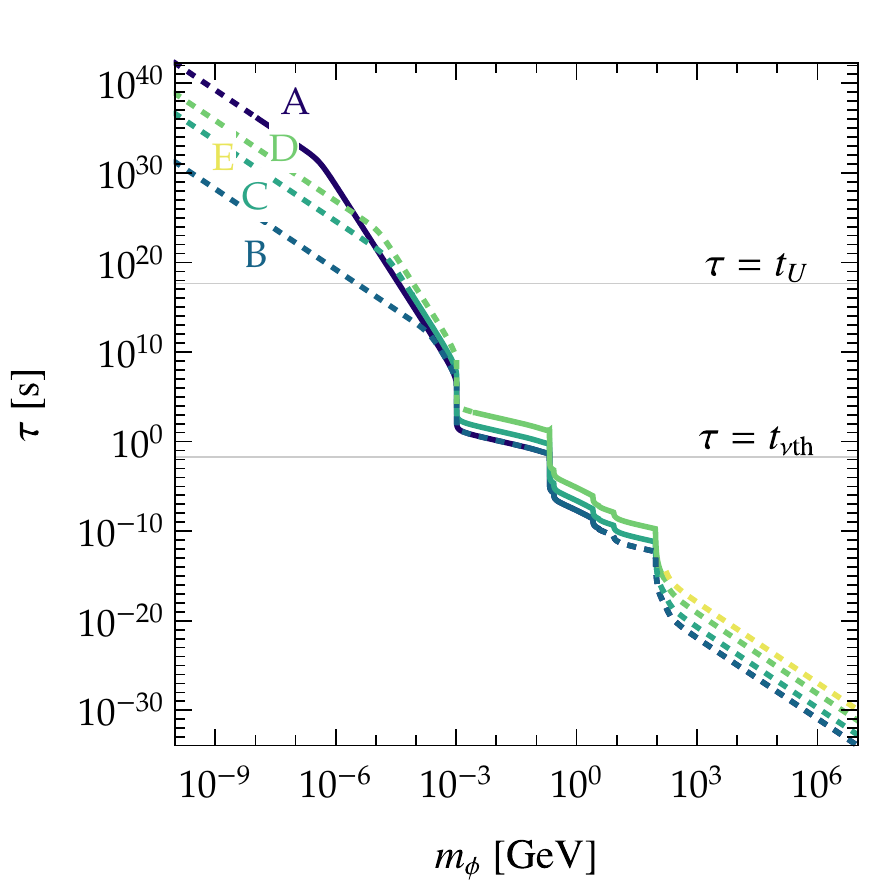}
\caption{
Relaxion--Higgs mixing angle (left) and relaxion lifetime (right), for scenarios A-E defined in \Eq{eq:benchmarks}. The solid lines correspond to the mass range allowed in our model once the constraints discussed previously are applied, see \Eq{eq:mass range}.
The two horizontal lines in the right panel correspond to the age of the universe $t_U$  and to $t_{\nu \rm{th}}$, which is the bound given by the neutrino thermalisation  (see Sec.\,\ref{sec:BBN}).}
\label{fig:mixing and lifetime}
\end{figure}

Given the potential in \Eq{eq:potential after pp}, one can compute the mixing angle of the relaxion with the Higgs as
\begin{equation}
\sin{2 \theta} = \frac{2  g' \Lambda  v_\ew}{\sqrt{4 g'^2 \Lambda^2 v_\ew^2 + (m_h^2 - m_\phi^2)^2}} \,.
\end{equation}
The mixing angle $\theta$ has a maximum of $\pi/4$ for $m_\phi=m_h$, which is a condition that can be realized in our model. Apart from a very narrow region (typically fractions of $\MeV$) around this point, the mixing angle is $\theta\ll 1$, and we can safely approximate $\sin2\theta$ with $2\theta$.
Let us briefly notice here that the VEV of the relaxion field is automatically smaller than $f'$ once condition~(\ref{eq:stopping}) is taken into account. 
Figure~\ref{fig:mixing and lifetime} shows a plot of the angle $\sin{2 \theta}$.

Additional contribution to the Higgs-relaxion mixing can be obtained by considering a diagram in which the bosonic lines of the $\phi V\widetilde{V}$ and $hZ_\mu Z^\mu$ vertices are connected via a fermion box. The corresponding mixing term can be estimated as $J v_\ew^4 f^{-2}(4\pi)^{-6}$, with $J=  \textrm{Im}[V_{ij}V_{kl}V^*_{il}V^*_{kj}]\sim 10^{-5}$ being the Jarlskog invariant  \cite{Patrignani:2016xqp}, which is suppressed compared to the above contribution in all the relevant parameter space.

\medskip

\begin{figure}
\centering
\includegraphics[width=.4\textwidth]{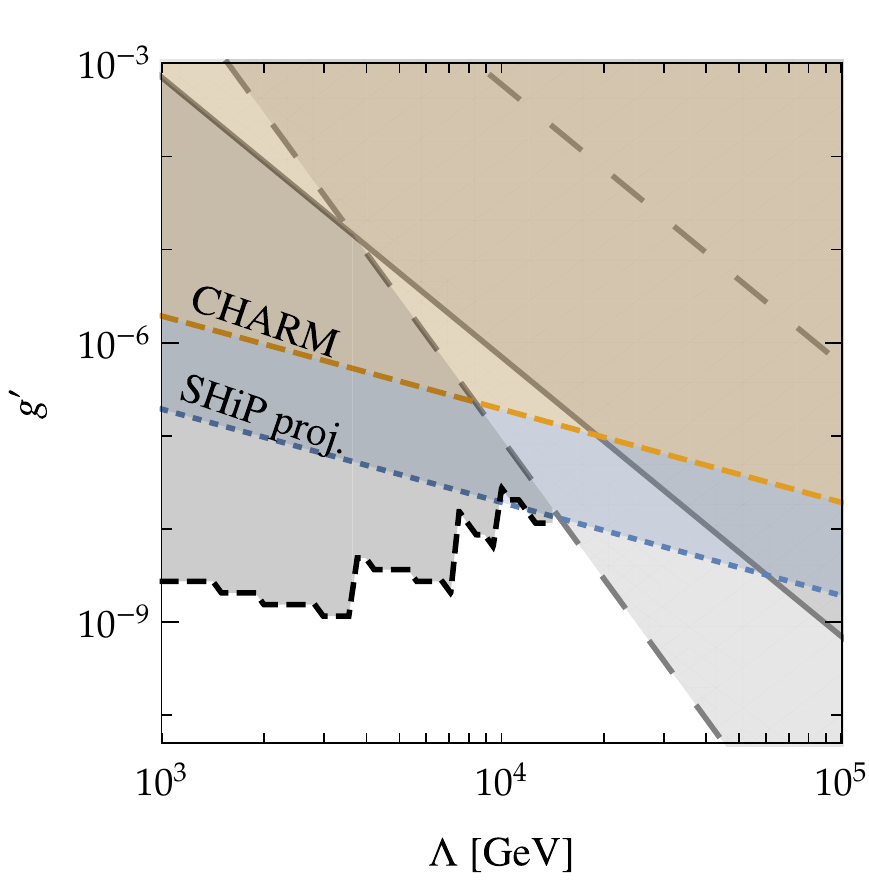}
\caption{  Zoom in the left-up corner of \Fig{fig:gpLambda} where we show the region where CHARM ($m_\phi \in [0.20 \GeV, 0.35 \GeV]$) and SHiP ($m_\phi \in [0.20 \GeV, 4 \GeV]$) can   probe part of the $f'$--$\Lambda_b$ parameter space. Such region is constrained by the conditions to obtain a successful  relaxation through particle production as indicated by the gray area.}

\label{fig:charmandship}
\end{figure}

The smallness of the mixing angle and the fact that $f$ is large make collider searches essentially harmless in our model.
As an example, we considered the production of $\phi$ particles at the LHC through its mixing with the Higgs, and its subsequent decay in the $Z\gamma$ channel. The predicted cross section turns out to be roughly $10-15$ orders of magnitude smaller than the limit from~\cite{CMS:2017fge}.
Even taking into account the couplings of Eq.~(\ref{eq:coupling fermions photons}), collider limits do not constrain the model. Indeed, for the relevant parameter space $\Lambda \gtrsim 10^4 \GeV$, the scale $f = \Lambda^2/(2 m_Z) \gtrsim 5.5\times 10^5 \GeV$. Using the results for photophobic axion-like particles discussed in~\cite{Craig:2018kne}, which can be directly applied to our relaxion scenario, we checked that the scale $f$ is high enough to avoid the constraints from LEP and LHC (similar bounds appeared previously in~\cite{Brivio:2017ije, Bauer:2017ris}). On the other hand, astrophysical probes can be relevant as we discuss in Sec.\,\ref{sec:astro}.

Furthermore, the relaxion  can be produced in rare decays of kaons and $B$-mesons, which are constrained by flavor and beam dump experiments.
We consider bounds on the relaxion-Higgs mixing from the CHARM experiment  given in \cite{Alekhin:2015byh}  which can probe relaxion masses  in the range $m_\phi \in [0.20 \GeV, 0.35 \GeV]$. Similarly, the future SHiP experiment can probe  the relaxion in the range $m_\phi \in [0.20 \GeV, 4 \GeV]$, as shown in Fig.\,\ref{fig:charmandship}.  This region is however excluded by other requirements (Eqs.\,(\ref{eq:1}), (\ref{eq:photons}), and (\ref{eq:10})) to get a  successful relaxation mechanism of the EW scale using particle production as it is shown in the gray area in Fig.\,\ref{fig:charmandship}.
In addition, we checked that  flavor constraints for the mixing for relaxion masses  from MeV to $5$ GeV cannot constrain our parameter space as our mixing angle is below the current bounds \cite{Flacke:2016szy}.
Similar bounds can be set by considering the production of relaxion particles through the couplings of Eq.~(\ref{eq:coupling fermions photons}). Again, the bound on $f$ is too weak to constrain the region of interest for this model~\cite{Craig:2018kne}.

\subsection{Relaxion lifetime}

Bounds on the properties of the relaxion field can come from its decay to SM particles after the relaxation mechanism took place. If kinematically allowed, the decay can proceed  through the mixing of $\phi$ with the Higgs boson (Sec.\,\ref{sec:relaxion mixing}), by the relaxion  $V\widetilde V$ leading interaction with the SM gauge bosons  (Sec.\,\ref{sec:gauge}), or through the induced  couplings to photons and SM fermions (Sec.\,\ref{sec:photons}), see App.~\ref{app:widths}.  
For the decays through the mixing with the Higgs, given by $\Gamma_\phi  = \theta^2 \Gamma_h(m_\phi)$, where $\Gamma_h(m_\phi)$ is the decay width of a Higgs boson with mass $m_\phi$,  we used the results of~\cite{Bezrukov:2009yw}.

Above $m_Z$, the $Z\gamma$ channel opens, followed by the $WW$ and $ZZ$ ones. These decays proceed mostly through the $V\widetilde V$ coupling, while at lower masses the decay proceeds through loop induced processe or it is suppressed by the small mixing angle.
The relaxion lifetime is shown in Fig.~\ref{fig:mixing and lifetime}, for the five benchmark scenarios.

%%%%%%%%%%%%%%%%%%%%%%%%%%%%%%%%%%%%%%%%%%%%%%%%%%%%%%%%%%%%%%%
\section{Cosmological constraints}\label{sec:cosmology}
%%%%%%%%%%%%%%%%%%%%%%%%%%%%%%%%%%%%%%%%%%%%%%%%%%%%%%%%%%%%%%%

After the relaxation process has ended, a population of $\phi$ particles is left in the universe that may lead to important cosmological observations. 
Depending on the lifetime of the relaxion, it may either lead to overabundance of dark matter in  the universe or ruin the predictions of Big Bang nucleosynthesis.
In this section, we  derive the corresponding bounds on the relaxion parameter space.

\subsection{Relaxion abundance in the early universe}

The population of $\phi$ particles is generated through two main mechanisms: vacuum misalignment and thermal production via the Primakoff process.

\subsubsection{Vacuum misalignment}
\label{sec:misalignment}

\begin{figure}
\centering
\includegraphics[width=.4\textwidth]{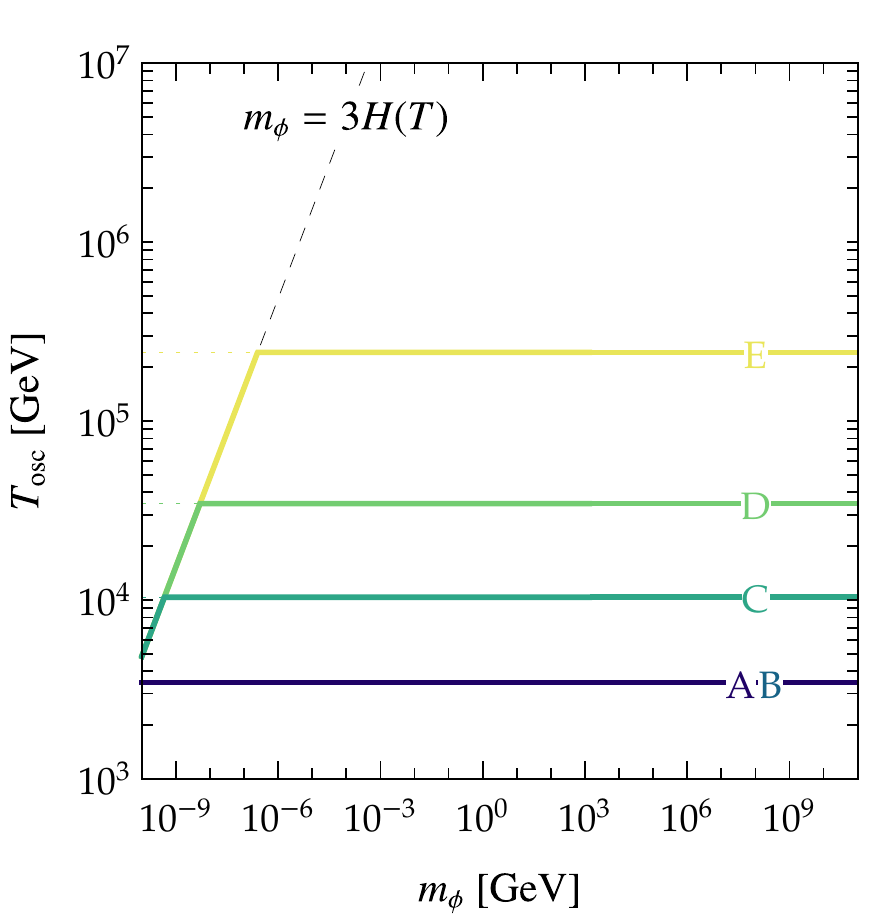}
\caption{Temperature at the onset of oscillations for the different scenarios. The black line shows the condition $m_\phi = 3H(T)$, that determines the initial temperature only at very low masses.}
\label{fig:Tosc}
\end{figure}

After the relaxion has been trapped in one of the minima of the potential, it will start oscillating with an amplitude that decreases with time due to the cosmological expansion and to the further production of gauge bosons, at least as far as its velocity is large enough. When the amplitude falls below a value that we are going to estimate in the following, the velocity of the field never becomes large enough to ignite particle production, and the oscillations are damped only by cosmic expansion.
Similarly to the QCD axion and other generic axion-like particles, during the oscillatory phase, the relaxion will contribute to the energy density of the universe as a cold dark matter component.
The abundance at the onset of the oscillatory phase is given by~\cite{Kolb:1990vq}
\begin{equation}
\Yphimis = \frac{1}{m_\phi}\frac{\rho_\phi}{s} = \frac{m_\phi \phi_i^2 /2}{2\pi^2 g_* T_\textrm{osc}^3 /45}
\end{equation}
where $\phi_i$ is defined as the displacement from the minimum of the periodic  potential in which the field has been trapped and $T_\textrm{osc}$ is the temperature when oscillations start as  the Hubble rate drops below the value of the thermal mass: $m_a(T_\textrm{osc}) = 3H(T_\textrm{osc})$. In our case, the maximum temperature of the plasma is $\Tpp\sim\Lambda$, that is lower than the value set by the overdamping of the oscillations except for very low masses, as shown in Fig.~\ref{fig:Tosc}. In most of our parameter space we can therefore assume an initial temperature $T_\textrm{osc}=\Tpp\sim\Lambda$, up to the numerical factors of Eq.~\ref{eq:Temperature}.

The initial amplitude of the oscillations $\phi_i$ can be simply estimated as follows.
In the initial stage of oscillations, if the amplitude is large, particle production will efficiently damp it down until the  velocity of the field $\phi$ is too low to induce the tachyonic growth. This happens when $\dot{\phi}_c \sim 2m_Z f$ (see \Eq{eq:Atachyion}), where the mass of the gauge boson is now given by the measured value $m_Z \approx 90\GeV$.
From this moment $\phi$ will oscillate freely, with an initial amplitude which is obtained by using energy conservation:
\begin{equation}
\Delta V \approx \frac{1}{2}\frac{\Lambda_b^4}{f'^2}\phi_i^2 = \frac{1}{2}\dot\phi_c^2 \lesssim 2 f^2m_Z^2 \quad \Longrightarrow \quad
\phi_i \lesssim 2\frac{f'f m_Z}{\Lambda_b^2} \,.
\end{equation}
Using  Eq.~(\ref{eq:mZprediction}), $f\sim \Lambda^2/(2v_\ew)$, gives
\begin{equation}
\phi_i \lesssim \frac{\Lambda^2}{\Lambda_b^2}\frac{m_Z}{v_\ew}f'\,.
\label{eq:phi initial}
\end{equation}
A similar bound is obtained by requiring that $\phi_i\lesssim\pi f'$. The two conditions are numerically similar in the region where $\Lambda$ and $\Lambda_b$ are of the same order.

In principle, this would imply that a very large relaxion population is produced from  vacuum misalignment. On the other hand, this estimate for $\phi_i$ is  very na\"ive. As discussed in Sec.~\ref{sec:gauge}, after the particle production turns on, the Higgs gets a large thermal mass and then the electroweak symmetry is restored, making $m_Z=0$. In this way, particle production is active also for low velocities. One can then expect an initial displacement angle much smaller than the one estimated above, which can considerably suppress the misalignment contribution to the relaxion abundance.  
A precise computation of the initial displacement requires a careful treatment of the thermalisation process which goes beyond the scope of this paper. In any case, in the next section, we will show that the thermal contribution to the relaxion abundance is  large. Therefore, independently of the misalignment contribution,  the relaxion is overabundant. 
This implies that in the absence of a dilution mechanism the relaxion cannot be dark matter and it must decay way before BBN to avoid bounds from the primordial abundances of light elements. These constraints are going to be discussed in the following sections.

\begin{figure}[h!]
\centering
\includegraphics[width=.42\textwidth]{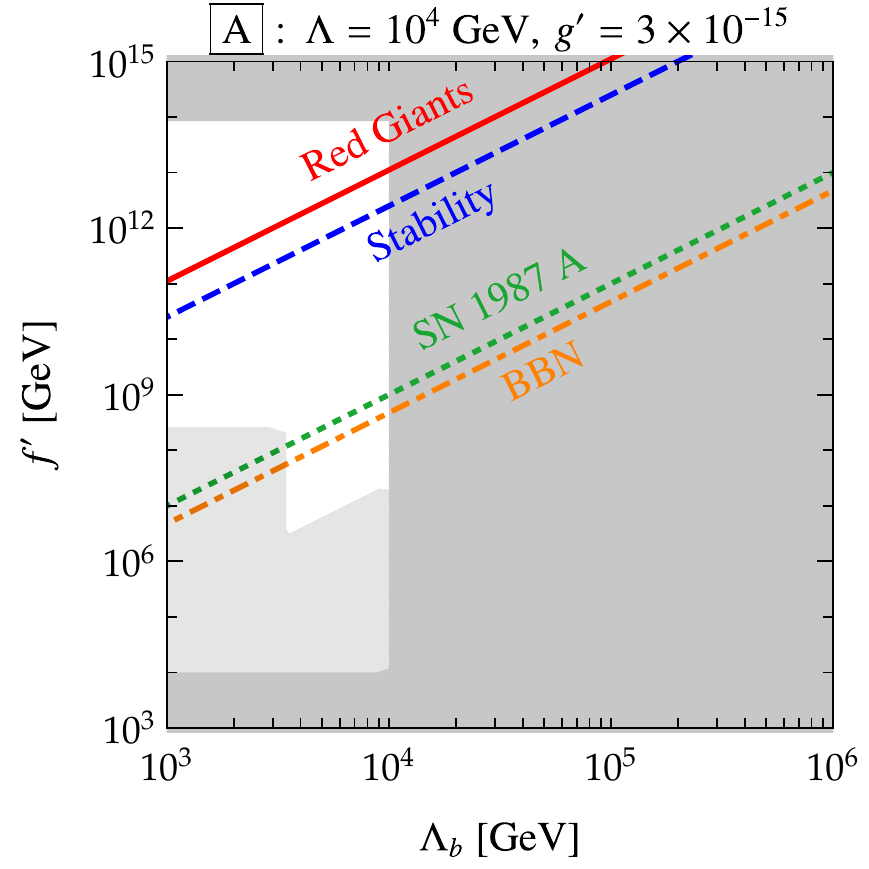}
\includegraphics[width=.42\textwidth]{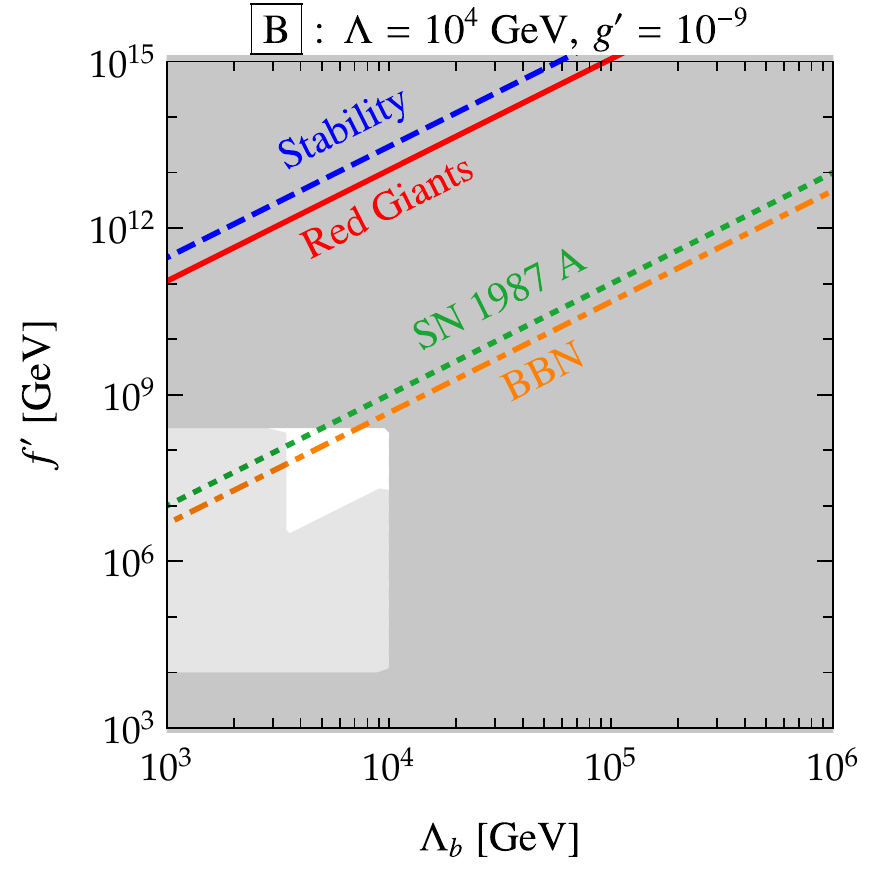}
\includegraphics[width=.42\textwidth]{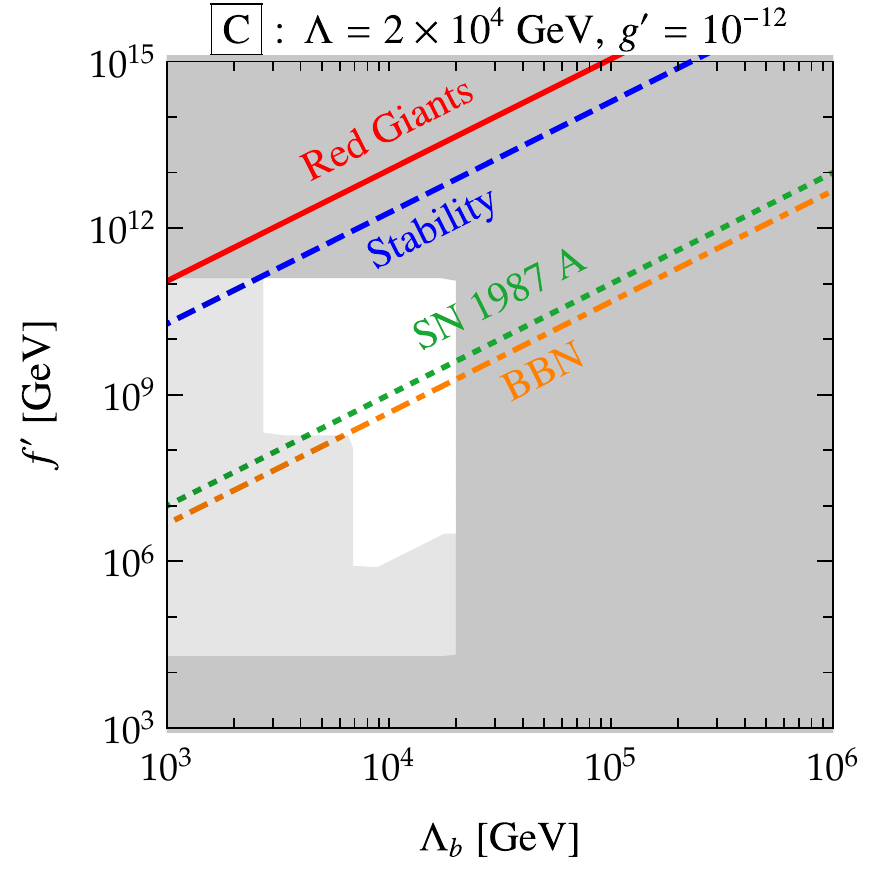}
\includegraphics[width=.42\textwidth]{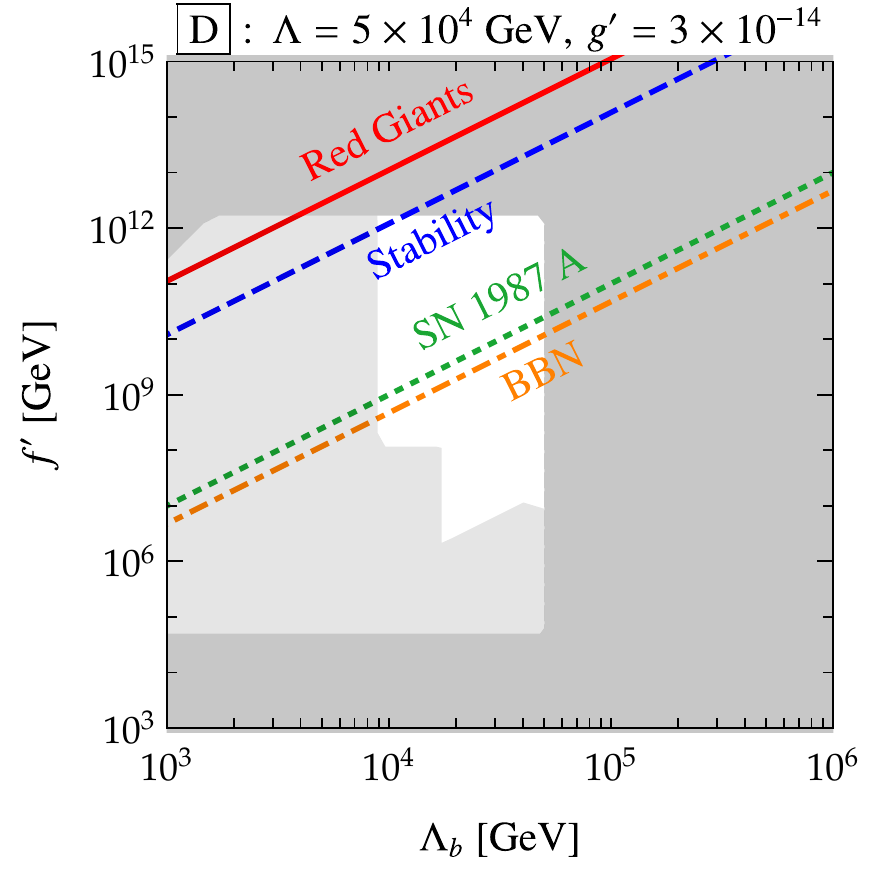}
\begin{minipage}{.42\textwidth}
\includegraphics[width=\textwidth]{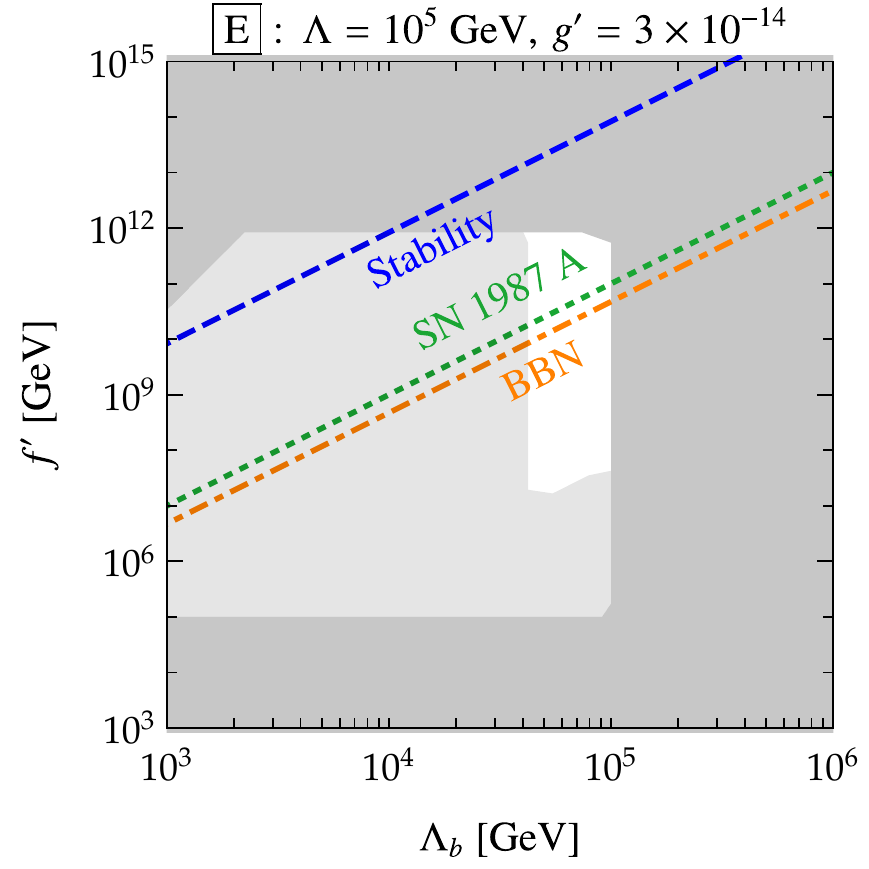}
\end{minipage}
\caption{Combination of all the cosmological and astrophysical constraints for each of the five benchmark points in the plane $[\Lambda_b, f']$
defined in \Eq{eq:benchmarks} and shown in red in Figure \ref{fig:gpLambda}.  The white area denotes the allowed region shown in Fig. \ref{fig:fpLambdac}. In the gray region the conditions necessary for particle production are not satisfied (the light gray area refers to the exclusion bounds which rely on   assumptions about the thermalisation process).  The astrophysics bounds (red giants and SN 1987 A) (see Sec.\,\ref{sec:astro}), the relaxion abundance (see Sec.\,\ref{sec:DM}) and distortions on the abundance of light elements (see Sec.\,\ref{sec:BBN}) set an upper bound on the scale $f'$ for a given $\Lambda_b$.
\label{fig:cosmo bounds}}
\end{figure}

\subsubsection{Thermal production}

A hot population of $\phi$ particles is produced via the process $Z/\gamma + q/\ell \to \phi+ q/\ell$. This is mediated by the $\phi V\widetilde V$ coupling, analogously to the standard QCD axion coupled to photons, and the role of the photon is played by the massless hypercharge gauge boson.
The abundance due to this process is given by~\cite{Kolb:1990vq}
\begin{equation}
\Yphith = Y_\textrm{eq}\left[ 1 - \exp\left(-\int_1^x \frac{\Gamma}{x'H}dx' \right)\right] \,,
\end{equation}
where $x = \Tpp/T$, $Y_\textrm{eq}=45\zeta(3)/(2\pi^4g_*^S)$ is the equilibrium abundance, and $\Gamma$ is the interaction rate~\cite{Cadamuro:2011fd}
\begin{equation}
\Gamma = \frac{1}{9 \pi}g_1^2 \left(2 \log(3/g_1)+0.82\right) \frac{T^3}{f_B^2}\,,
\end{equation}
where $f_B$ is the scale in $\mathcal{L} \supset \frac{\phi}{f_B} \epsilon^{\mu \nu \rho \sigma} \partial_\mu B_{\nu} \partial_\rho B_\sigma $, which can be written as $f_B = \frac{(g_2^2 -g_1^2)}{g_1^2} f$. Depending on whether $m_\phi$ is smaller or larger than the EW scale, the integral above has a natural cutoff at $T\sim v_\ew$ or $T\sim m_\phi$, respectively. In the first case, in the broken electroweak phase, the rate $\Gamma$ is suppressed by powers of $T/m_Z$, while in the second case, at $T\lesssim m_\phi$ the relaxion is Boltzmann suppressed. We checked numerically that, in our parameter space, the integral is always large enough to suppress the negative exponential, resulting in a final abundance  equal to the thermal value $Y_\textrm{eq}$. 

Below $T\sim v_\ew$, the relaxion is kept in thermal equilibrium by processes involving its couplings to photons and light fermions.

In addition to the process considered above, the relaxion can also be thermally produced in processes involving the $\phi gg$ and $\phi \bar q q$ couplings that are obtained through the relaxion--Higgs mixing~\cite{Flacke:2016szy}. In our case, the insertion of the mixing angle makes this contribution negligible. This is another crucial difference compared to the standard relaxion scenarios, where the main contribution to the thermal abundance is proportional to the relaxion--Higgs mixing~\cite{Flacke:2016szy}, making the total thermal production very suppressed.

\subsection{Dark matter relic abundance}
\label{sec:DM}

If the relaxion is stable on cosmological timescales, its relic abundance can contribute to the present dark matter density, and must therefore not exceed the measured value of $\Omega_\textrm{cdm}h^2\simeq 0.12$.
The present abundance is given by
\begin{equation}
\Omega_\phi = \frac{s_0 m_\phi}{\rho_\textrm{crit}}  Y_\phi\,,
\end{equation}
where $\rho_\textrm{crit}\approx 4\times 10^{-47}\GeV^4$ is the critical density of the universe and $s_0\approx 2\times 10^{-38}\GeV^3$ is the present entropy density.
In the region of $m_\phi$ where the relaxion is cosmologically stable, its relic abundance is too large in our model. 
The corresponding stability bound is shown in blue in Fig.~\ref{fig:cosmo bounds}. One interesting possibility is that  the relaxion could constitute the DM. From the results of Fig.~\ref{fig:cosmo bounds}, this possibility could be realized only in scenario A, which is the only one in which the stability of $\phi$ is allowed by the constraints on particle production. Nevertheless, this possibility is not viable for a number of reasons. First, the stability region is excluded by the SN 1987A bound and, partly, by red giants. Secondly, even assuming some mechanism to evade the SN 1987A bound, and if the relic abundance is diluted to the correct value, then one should take into account the bounds on IR radiation, which excludes lifetime orders of magnitude larger than the age of the Universe \cite{Essig:2013goa}, making the DM hypothesis viable only well inside the region excluded by red giants.

\subsection{Primordial abundances of light elements} \label{sec:BBN}

Relaxions decaying after the Big Bang Nuclesynthesis (BBN) epoch are strongly constrained by  bounds  on the primordial abundances of light elements.
A long-lived unstable particle which decays into electromagnetic or hadronic  particles can affect the light element abundances by photodissociation, hadrodissociation, and $p\leftrightarrow n$ conversion processes~\cite{Kawasaki:2017bqm}.
In our case, the decaying particle dominates the energy density, thus heavily affecting neutrino thermalisation, which constrains the decay to happen well before the BBN.
Therefore, the abundance of light elements exclude the scenario where the relaxion decays after $\tau_\phi \lesssim 2\times10^{-2}\,s$~\cite{Kawasaki:2000en}. The constrained region is indicated by the orange line in \Fig{fig:cosmo bounds}.

\subsection{Astrophysical constraints}
\label{sec:astro}

In the range of interest, $\Lambda \gtrsim 10^4$ GeV,   red giants  and   Supernova 1987A  observations impose relevant constraints to our parameter space. 
Here we consider the study of photophobic axion-like particles of \Ref{Craig:2018kne}  which is    applied to the relaxion case (see also \cite{Raffelt:1994ry, Raffelt:2006cw}). 
The  relaxion coupling to electrons (see Sec.\,\ref{sec:photons}) is constrained by the cooling  of red giant stars due to presence of additional processes  like  bremsstrahlung emission  that  can delay the helium ignition, implying that the core can grow to higher masses  before the helium ignites. For $f\lesssim 3 \times 10^7\GeV$, red giant observations exclude  relaxion masses below  $m_\phi \lesssim 10^{-5}\GeV$ \cite{Craig:2018kne}, which constrains scenarios A, B, C, and D as shown in \Fig{fig:cosmo bounds}.

Additionally, using the neutrinos observed from SN 1987A one can constrain  weakly interacting particles as the emission of such new states would be an efficient energy-loss channel. The relevant cooling channel in the relaxion case is through the nucleon bremsstrahlung process ($N + N \rightarrow N + N +\phi$). For $f \lesssim  10^8\GeV$, the SN 1987A energy-loss bound rules out relaxion masses below $m_\phi  \lesssim 0.1~\GeV$~\cite{Craig:2018kne}.
This constrains scenarios A, B, C, D, and E as indicated in \Fig{fig:cosmo bounds}. Scenario E is not constrained by the astrophysical probes discussed here red giants as in this case the scale $f\sim 5\times 10^7$ GeV  is large enough to evade this bound.

\subsection{Overview}

The combination of constraints is shown in \Fig{fig:cosmo bounds}.
In the region where the relaxion is unstable, the  strongest bound comes from primordial abundances of light elements, which is shown in orange in \Fig{fig:cosmo bounds}. In our scenario, the relaxion relic abundance is too large, so the quantity  $m_{\phi} Y_{\phi}$ is always   above the current bound  in the region of the parameter space where the relaxion decays between $\tau_\phi \sim 10^{-2}\s$ and $\tau_{\phi} \sim 10^{17}\s$. 
 If instead the relaxion is cosmologically stable, it is overabundant, and therefore we impose $\tau_\phi\lesssim 10^{17}\s$. This corresponds to the exclusion region in blue in \Fig{fig:cosmo bounds}.
 
\clearpage

Taking into account all the constraints in this section, the allowed mass ranges for our five benchmarks  are approximately given by
\begin{align} \nonumber
\text{scenario A:} \qquad & m_\phi \in [2 m_\mu, \, 4\GeV] \\ \nonumber
\text{scenario B:} \qquad & m_\phi \in [2 m_\mu, \, 4\GeV] \\ \label{eq:mass range final}
\text{scenario C:} \qquad & m_\phi \in [2 m_\mu, \, 100\GeV] \\ \nonumber
\text{scenario D:} \qquad & m_\phi \in [2 m_\mu ,\, 141\GeV] \\ \nonumber
\text{scenario E:} \qquad & m_\phi \in [2 m_\mu , \, 178\GeV],
\end{align}
where $m_\mu \approx 106 \MeV$ is the mass of the muon.
Comparing the  resulting mass ranges above with the ones in (\ref{eq:mass range}), we see that all the lower bounds are shifted to higher masses. The constraints from astrophysical probes (SN 1987A and red giant starts observations) are also shown in \Fig{fig:cosmo bounds}.

%%%%%%%%%%%%%%%%%%%%%%%%%%%%%%%%%%%%%%%%%%%%%%%%%%%%%%%%%%%%%%%%%%%%%%%%%%%%%

\section{Summary and Conclusions}
\label{sec:conclusion}

In summary, we have investigated in detail the viability of the cosmological relaxation mechanism of the electroweak scale taking place after inflation \cite{Hook:2016mqo}. In this case,  the friction needed to stop and prevent the relaxion from running away down its potential, comes from particle production instead of exponential Hubble expansion.
We showed that Higgs particle production cannot be used for this purpose. Instead, particle production is  sourced by a relaxion coupling to a $U(1)$ electroweak gauge field of the type 
\begin{equation}
 - \frac{\phi}{4 \mathcal{F}} \left(g_2^2 W \widetilde{W} - g_1^2 B \widetilde{B} \right) 
 \label{eq:keycoupling}
\end{equation}
 where $g_1$ and $g_2$ are respectively the couplings of $U(1)_Y$ and $SU(2)_W$.
 This particular combination is crucial since it does not contain the photon.
 It arises naturally in some UV completions, for instance PNGBs may inherit such anomalous coupling structure.
 
Such coupling induces exponential particle production only when the Higgs VEV approaches zero and the $U(1)$ gauge field (\ref{eq:keycoupling}) becomes nearly massless. 
Particle production comes at the expense of kinetic energy of the relaxion. Being slowed down, it can no longer overcome the large (Higgs-independent) barriers.
This stops very efficiently the relaxion when the Higgs mass parameter approaches its critical value from above, as illustrated in Fig.~\ref{fig:numerics}. 
 In this realisation of the relaxion mechanism, the universe starts in the broken electroweak phase, with a Higgs VEV of the order of the cutoff scale $\Lambda$. 
 The universe is initially reheated in a hidden sector, such that the Standard Model is not thermalised and the Higgs potential receives negligible thermal corrections.
This is  the only  non-trivial assumption for a successful implementation of this scenario. Reheating of the Standard Model sector takes place at the end of the relaxation mechanism and is induced by the same relaxion coupling (\ref{eq:keycoupling}) which is responsible for the stopping mechanism through gauge boson production. 
As the relaxion potential energy is transferred to the Standard Model thermal bath, the reheat temperature is expected to be close to the cutoff scale $\Lambda$. 
Interestingly, the more minimal scenario in which the Standard Model is reheated just after inflation and the relaxation phase starts after the temperature has been redshifted below the scale $\Lambda$, is constrained but still viable, see Fig.~\ref{fig:reheating SM allowed g}.
 
The relaxion initial velocity can be obtained either from a coupling with the hidden sector or through an interaction with the inflation sector.  
For instance,   a coupling of the relaxion to the inflaton can provide an effective slope, which results in a large initial velocity.
 
We determined the parameter space in which such mechanism works and satisfies all cosmological constraints. Free parameters are the cutoff scale $\Lambda$, the Higgs-relaxion coupling $g'$, the height of the barriers $\Lambda_b$ and the frequency of the periodic potential $f'$.
Our results are summarised in Figs. \ref{fig:gpLambda} and \ref{fig:cosmo bounds}.
Fig.~\ref{fig:gpLambda} shows the allowed region in the plane $[g', \Lambda]$ while 
Fig.~\ref{fig:cosmo bounds} shows the open region in the plane $[ f',\Lambda_b]$ for the five benchmark points of Fig.~\ref{fig:gpLambda}.
The cutoff scale $\Lambda$ can be as large as $\sim$ 100 TeV.
Large couplings $g'\gtrsim 10^{-3}$ are incompatible with the condition that the Higgs field tracks the minimum of its potential during the cosmological evolution. This condition is absent in the relaxion proposal relying on inflation and Higgs-dependent barriers, although  an effectively comparable condition  comes from preventing large quantum corrections induced by the coupling generating the Higgs-dependent barrier.  Additionally, the region with $g'\gtrsim 10^{-9}$ is excluded by the loop-induced coupling with photons.
Coupling values smaller than $10^{-16}$ are forbidden in our framework as they would lead to slow-rolling of the relaxion and therefore inflation induced by the relaxion field. 
In comparison, the relaxion-associated-with-inflation proposal typically has couplings which are smaller by many orders of magnitude (see comparison in right plot of Fig.~\ref{fig:gpLambda}), and  are only bounded by the condition that quantum displacements of the relaxion do not dominate its classical motion.

This mechanism is very difficult to test experimentally, despite the relatively low cutoff scale and rather large $g'$ values that we are pointing to, as the relaxion manifests itself either via its mixing with the Higgs, via its coupling to the $Z$ and $W$ gauge bosons through (\ref{eq:keycoupling}) or via the induced couplings with photons and fermions.  
The relaxion is heavy compared to the original relaxion proposal. Its mass ranges values from $\mathcal{O}(100)\MeV$ up to the EW scale. It cannot be cosmologically stable, as otherwise it would overclose the universe. The relaxion cannot be dark matter in this scenario unless invoking an additional dilution mechanism. Thus it has to decay before BBN.

One further step to probe this mechanism will be to determine in more detail cosmological implications of the stopping mechanism, which involves out-of-equilibrium conditions that may lead to observable imprints. The reheating process in itself deserves further investigation.
The coupling  (\ref{eq:keycoupling}) has been used in the context of inflation as a source of gravitational waves,  CMB non-gaussianities or magnetogenesis, and its effects in the context of the lower scale relaxion mechanism should be studied.

Another important insight will come from understanding how  
the baryon asymmetry can be explained in this set up. 
Baryogenesis cannot take place before relaxation as it would be diluted away by the entropy injection during reheating coming from the relaxion decay.
As the philosophy of the relaxion mechanism is that no new physics occurs at the EW scale, we expect the EW phase transition that occurs after reheating and EW symmetry  restoration to be standard-like.
This forbids the possibility of standard EW baryogenesis as the requirement of a first-order EW phase transition typically requires an additional weak scale scalar field. An alternative baryogenesis mechanism has to be found. 
For concrete progress to be made in this direction, the determination of the reheat temperature is required, and this work strongly motivates such a dedicated study. 

In conclusion, cosmological relaxation of the EW scale rather independently from inflation is a viable option that opens interesting opportunities and deserves further investigation.

%%%%%%%%%%%%%%%%%%%%%%%%%%%%%%%%%%%%%%%%%%%%%%%%%%%%%%%%%%%%%%%%%%%%%%%%%%%%%

\acknowledgments

 We would like to thank  Francesco Cicciarella, Mafalda Dias, Jonathan Frazer, David E. Kaplan, Kazunori Kohri, Gustavo Marques-Tavares, Mauro Pieroni,  Antonio Riotto, Lorenzo Ubaldi, Augustin Vanrietvelde, and Sebastian Wild for useful discussions. In particular,  we are grateful to Oleksii Matsedonskyi and  Alexander Westphal for discussions regarding the initial conditions, to Ben Stefanek for discussions about the relaxion relic abundance, and to  Valerie Domcke for innumerable discussions and comments. NF and EM thank  the organizers and participants of the  
Cosmological probes of BSM Workshop in Benasque for interesting discussions during the completion of this work.   

\appendix

\newpage

\section{Equations of Motion}
\label{sec:appendix}

\subsection{Higgs case} \label{app:Higgs_pp}

Here we consider the equations of motions for the model given in \Eq{eq:relaxion_potential},  the Higgs field $h$ is decomposed as a classical field  and quantum fluctuation,
\begin{equation}
h = h_0 + \chi.
\end{equation}
The $\chi$ field can be expanded in Fourier modes as  \begin{equation}
\chi(t, \vec{x}) = \int\frac{d^3k}{(2 \pi)^{3/2}}~ \left(a_{\vec{k}} \,\chi_{\vec{k}}(t)e^{i \vec{k} \cdot \vec{x}} + a_{\vec{k}}^\dagger\, \chi_{\vec{k}}^*(t)e^{-i \vec{k} \cdot \vec{x}}\right),
\end{equation}
where  the creation and annihilation operator $a_{\vec{k}}^\dagger, a_{\vec{k}}$ satisfy the usual commutation relations
\begin{equation}
\left[a_{\vec{k}}, a_{\vec{k'}}\right] = \left[a_{\vec{k}}^\dagger, a_{\vec{k'}}^\dagger\right] =0, ~~~~ \left[a_{\vec{k}}, a_{\vec{k'}}^\dagger\right] = (2 \pi)^3 \delta^3(\vec{k} -\vec{k'}),
\end{equation}
and the normalized $\chi_k$  wave functions satisfy
\begin{equation}
\dot{\chi}^*_{\vec{k}} \chi_{\vec{k}} - \chi_{\vec{k}}^* \dot{\chi}_{\vec{k}} = i\,.
\end{equation}
As in our scenario the relaxation dynamics happens  after inflation,  for simplicity,  here we  consider flat space so the equations of motion can be written as
 \begin{align}
 \label{eq:brokenphi}
\ddot{\phi} - g \Lambda^3 + \frac{1}{2} g' \Lambda h_0^2  +\frac{1}{2} g' \Lambda  \int\frac{d^3k}{(2\pi)^3} \left( |\chi_{\vec{k}}|^2 - \frac{1}{2\omega_k}\right) +\frac{\Lambda_b^4}{f'} \sin{\left(\frac{\phi}{f'}\right)}  &=0 \\
\label{eq:brokenhiggs}
\ddot{h} + \left [ g' \Lambda \phi -\Lambda^2   + 3\lambda  \int\frac{d^3k}{(2\pi)^3}\left( |\chi_{\vec{k}}|^2- \frac{1}{2\omega_k}\right) \right]h_0  + \lambda\, h_0^3 &=0 \\ 
%\label{eq:brokenchi}
\left(\partial^2 + k^2 + g' \Lambda \phi - \Lambda^2 + 3\lambda \,h_0^2 \right)\chi_{\vec{k}} &=0,
\end{align}
where  $\omega_k = \sqrt{k^2 + (g' \phi \Lambda -\Lambda^2 +3\,\lambda\, h_0^2)}$ is the frequency of the quantum field $\chi_{\vec{k}}$,
 with $ k^2 \equiv \vec{k} \cdot \vec{k} $. Note that we consider the linearized equation of motion for $\chi_{\vec{k}}$  as higher order terms  are sub-dominant.    The subtraction in the parentheses in Eqs.\,(\ref{eq:brokenphi}) and (\ref{eq:brokenhiggs}) refers to the first order WKB  mode functions $\chi_{\vec{k}}^{\textrm{WKB}}= \exp(i\omega_k)/\sqrt{2\omega_k}$, which is needed to cancel a divergence in the effective potential  (see e.g. \cite{Kofman:2004yc, Enomoto:2013mla}).   In addition,
  we assume that the relaxion field is homogeneous in space, i.e.  $\phi(t,\vec{x})=\phi(t)$.

\subsection{Gauge bosons case}
\label{app:gaugebosons}

The aim of this appendix is to derive equations (\ref{eq:phieomvector}), 
(\ref{eq:heomvector}), (\ref{eq:Aeomvector}) and (\ref{eq:fftilde2}) and display numerical solutions.
We start from the gauge invariant Lagrangian
\begin{equation}
\mathcal{L} = -\frac{1}{4}F_{\mu\nu}F^{\mu\nu} + \frac{1}{2}\partial_\mu \phi \partial^\mu \phi +
(D_\mu \HD)^\dagger D^\mu \HD - \frac{\phi}{4f} \epsilon^{\mu\nu\rho\sigma} V_{\mu\nu}V_{\rho\sigma} -V(\phi,\HD^\dagger\HD)
\end{equation}
where $\HD$ is the Higgs doublet and the potential $V(\phi,\HD^\dagger\HD)$ is given in \Eq{eq:relaxion_potential}.
After gauge symmetry breaking, the Goldstone bosons can be reabsorbed in the field $V_\mu$, and the Lagrangian reads
\begin{equation}
\mathcal{L} = -\frac{1}{4}F_{\mu\nu} F^{\mu\nu} \frac{1}{2}\partial_\mu \phi \partial^\mu \phi + \frac{1}{2}\partial_\mu h\partial^\mu h
+\frac{1}{2}g_V^2h^2V_\mu V^\mu - \frac{\phi}{4f} \epsilon^{\mu\nu\rho\sigma} V_{\mu\nu}V_{\rho\sigma} - V(\phi,h)\,.
\end{equation}
The equations of motion for the $h$ and for $\phi$ fields are trivial:
\begin{align}
& \Box \phi + \frac{\partial V(\phi,h)}{\partial \phi} - \frac{1}{4f}V\widetilde V = 0\\
& \Box h + \frac{\partial V(\phi,h)}{\partial h} - g_V^2 V_\mu V^\mu h = 0
\end{align}
For the field $V_\mu$ we get instead
\begin{equation}
\Box V^\mu - \partial^\mu \partial_\nu V^\nu - 2 \epsilon^{\alpha\beta\gamma\mu}\frac{\partial_\gamma\phi}{f}\partial_\alpha V_\beta + g_V^2 h^2 V^\mu =0 \,.
\label{eq:eomBmu}
\end{equation}
If we take the divergence of this relation we obtain
\begin{equation}
\partial_\mu (h^2 V^\mu) = 0\,, \quad \textit{i.e.} \quad \partial_\mu V^\mu = -2\frac{\partial_\mu h}{h} V^\mu
\label{eq:demuBmu}
\end{equation}
which shows that the usual assumption $\partial_\mu V^\mu = 0$ is not consistent in this case.

Assuming for simplicity that the fields $\phi$ and $h$ are spatially uniform, the equations of motion simplify to
\begin{equation}
\left\{
\begin{aligned}
&\ddot \phi + \frac{\partial V(\phi,h)}{\partial \phi} - \frac{1}{4f}V\widetilde V = 0\\
&\ddot h + \frac{\partial V(\phi,h)}{\partial h} - g_V^2 V_\mu V^\mu h = 0 \\
&\ddot V^j + \partial_i\partial^i V^j + 2 \partial^j \frac{\dot h}{h} V^0 - 2 \epsilon^{j k l}\frac{\dot \phi}{f}\partial_k V_l + g_V^2 h^2 V^j = 0 \\
&(\partial_i\partial^i + g_V^2h^2) V^0 - \partial_i \dot V^i = 0
\end{aligned}
\right.
\label{eq:eomsystem}
\end{equation}
The first two equations in (\ref{eq:eomsystem}) reproduce the equations (\ref{eq:phieomvector}) and (\ref{eq:heomvector}).
After a Fourier transform, the equation for the $V^0$ component becomes a simple algebraic equation, and $V_0$ can be expressed in terms of the other components. Projecting the spatial part onto helicity components we obtain
\begin{equation}
\left\{
\begin{aligned}
&\ddot V_\pm + (k^2 + g_V^2 h^2)V_\pm \pm \frac{\dot\phi}{f}k V_\pm = 0 \\
&\ddot V_L + 2\frac{\dot h}{h} \frac{k^2}{k^2+g_V^2h^2} \dot V_L + (k^2+g_V^2h^2)V_L = 0 \\
& V_0 = -i \frac{k \dot V_L}{k^2 + g_V^2 h^2}
\end{aligned}
\right.
\end{equation}
The first equation above  reproduces (\ref{eq:Aeomvector}).

\begin{figure}
\centering
\includegraphics[width=.32\textwidth]{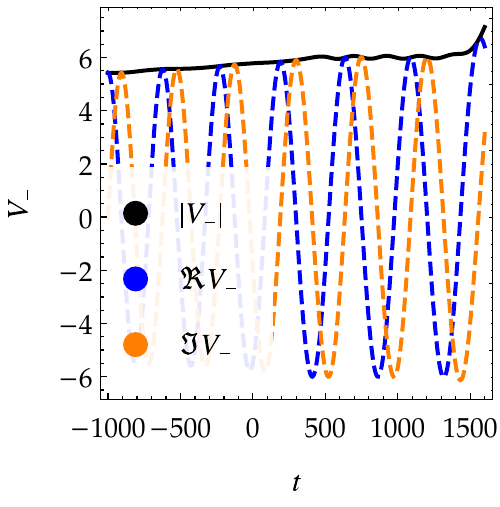}
\includegraphics[width=.32\textwidth]{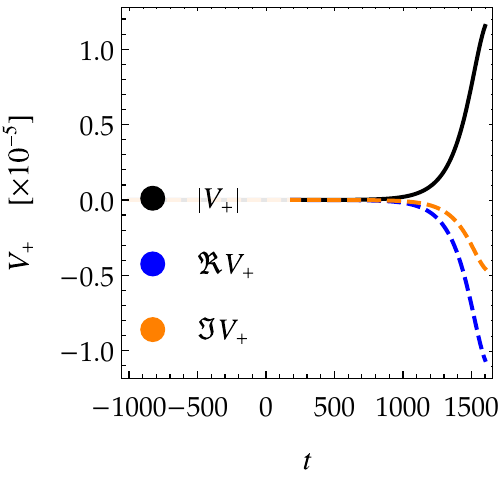}
\includegraphics[width=.32\textwidth]{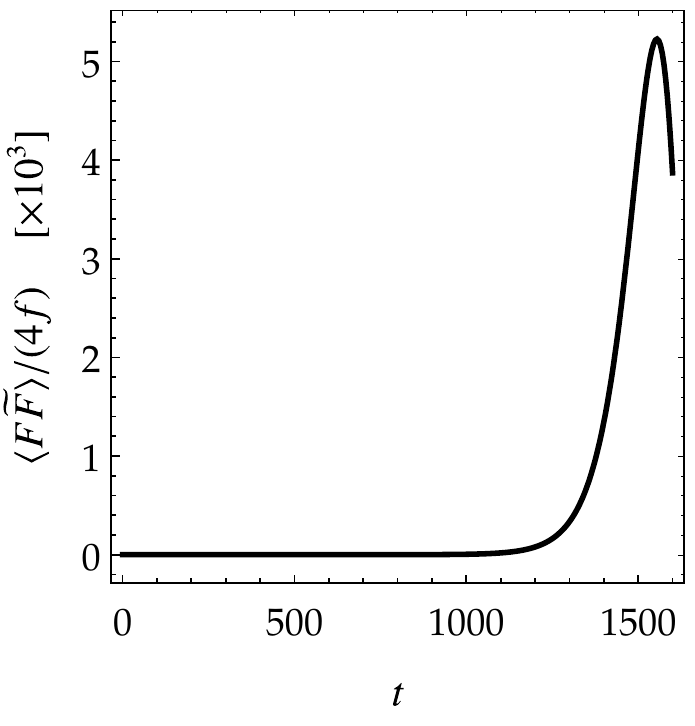}
\includegraphics[width=.32\textwidth]{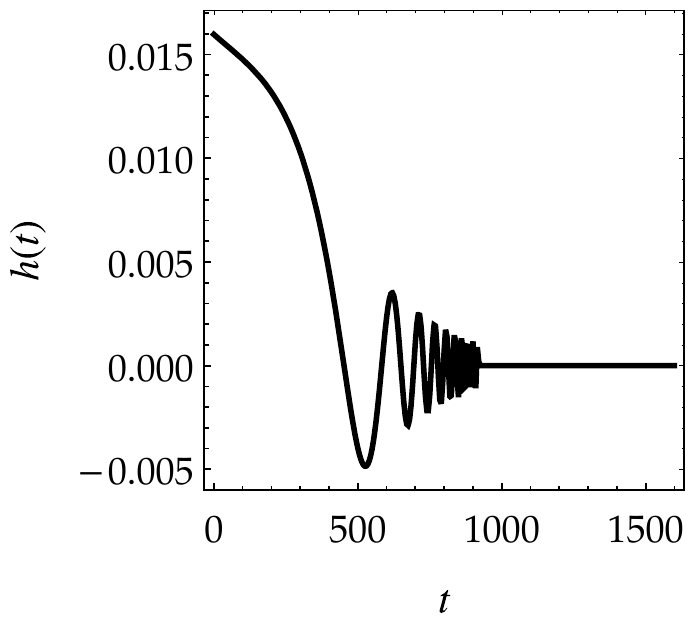}
\includegraphics[width=.32\textwidth]{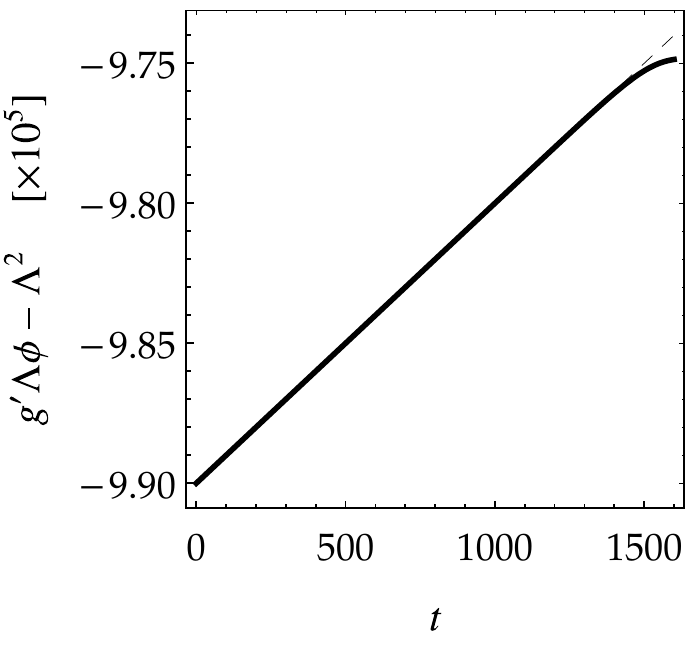}
\includegraphics[width=.32\textwidth]{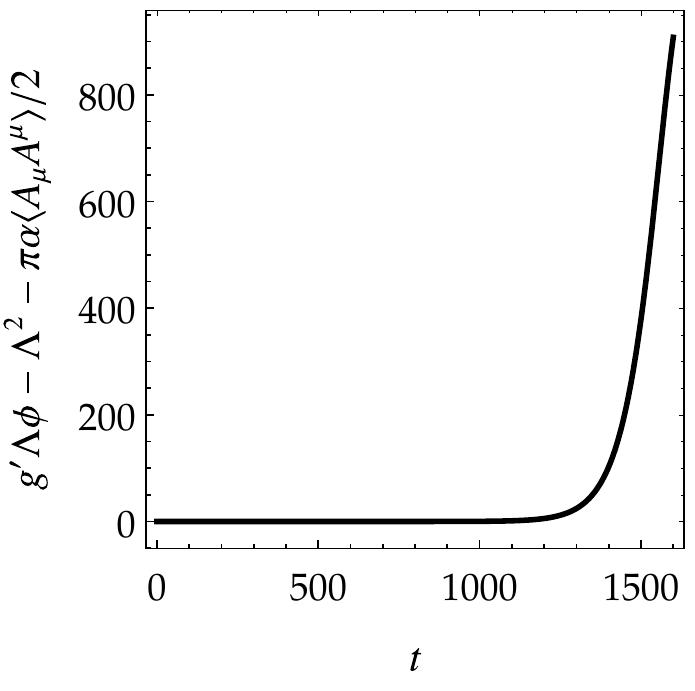}
\includegraphics[width=.32\textwidth]{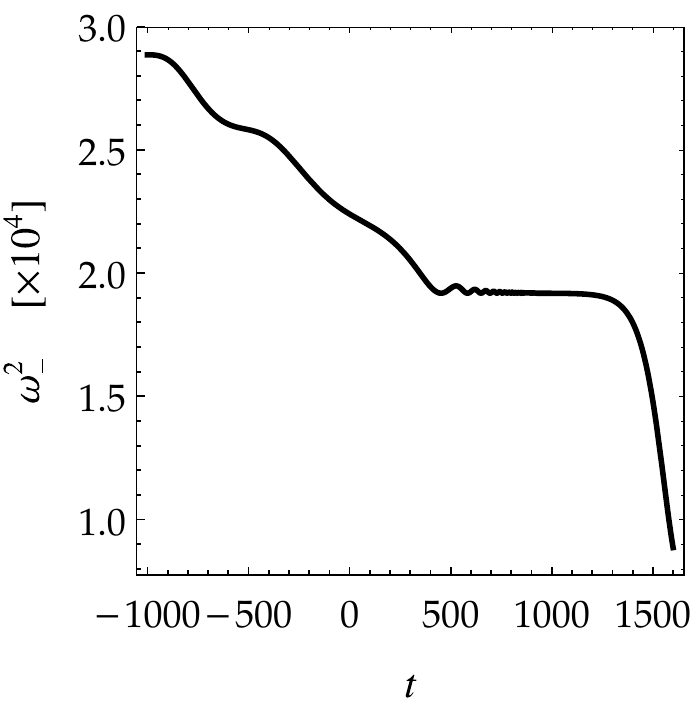}
\includegraphics[width=.32\textwidth]{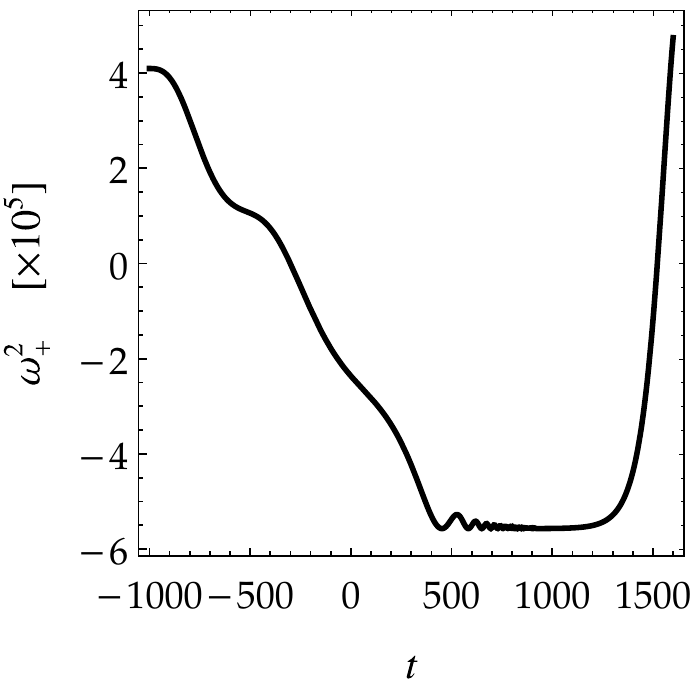}
\includegraphics[width=.32\textwidth]{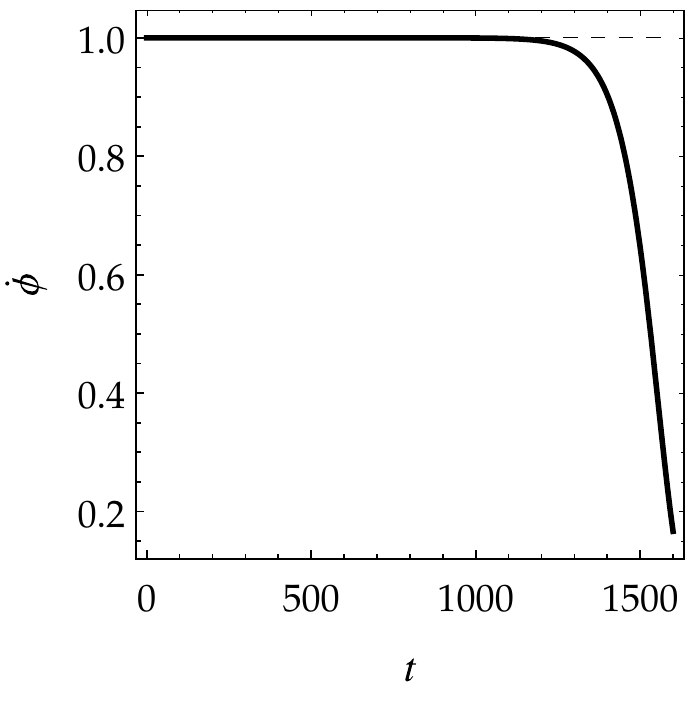}
\caption{Solutions of the system of equations (\ref{eq:phieomvector},
\ref{eq:heomvector}, \ref{eq:Aeomvector}) (which does not include thermalisation effects). While time scales will be different, we expect similar qualitative features when the effect of temperature is taken into account. All quantities are expressed in units of $\Lambda$. The top panel shows that the transverse polarisation $V_-$  does not feature any tachyonic instability and is nearly constant, while the other transverse polarization $V_+$ exhibits a tachyonic growth. The operator $\langle V \tilde{V}\rangle$ grows accordingly. The relaxion velocity $\dot{\phi}$ drops. 
The Higgs mass parameter $g' \Lambda \phi - \Lambda^2$ stabilises. The contribution to the Higgs mass parameter from the gauge field quickly grows and restores the EW symmetry.  The Higgs vev stabilizes to a vanishing value (and, for simplicity, we set it to zero after a few oscillations). The left  plot of the bottom panel shows that the time when the frequency squared of one of the modes  $\omega^2_+ $ becomes negative coincides with gauge field particle production.}
\label{fig:numerics}
\end{figure}

The terms $V_{\mu\nu}\widetilde{V}^{\mu\nu}$ and $V_\mu V^\mu$ in the equations of motion for $\phi$ and $h$ must be interpreted as the expectation values of the quantum operators on the in-vacuum state.
Introducing the creation and annihilation operators we write
\begin{equation}
V_\mu (t,\vec x) = \int \frac{d^3k}{(2\pi)^3} \sum_\lambda V_\mu^{(\lambda)}(t,\vec k) e^{i\vec k\cdot\vec x} \left(a_{\vec k}^\lambda + (a_{-\vec k}^\lambda)^\dagger\right)
\end{equation}
where $V_\mu^{(\lambda)}(t,-\vec k) = V_\mu^{(\lambda)}(t,\vec k)^*$ for the reality of the $V_\mu(t,\vec x)$  field, and the creation/annihilation operators satisfy the usual commutation relations
\begin{equation}
[a_{\vec k}^\lambda, a_{\vec k'}^{\lambda'} ] = (2\pi)^3 \delta^{\lambda\lambda'} \delta^3(\vec k-\vec k')\,, \qquad
[a,a]=[a^\dagger, a^\dagger]=0 \,.
\end{equation}
With this we get
\begin{align}\label{eq:BB}
\langle V_\mu V^\mu \rangle & = \int \frac{d^3k}{(2\pi)^3} \sum_\lambda \left( |V_0|^2 - |V_+|^2 - |V_-|^2 - |V_L|^2\right) \nonumber\\
&  \approx - \int \frac{d^3k}{(2\pi)^3} \sum_\lambda \left( |V_+|^2 + |V_-|^2 \right) \,.
\end{align}
where in the last line we have neglected the contribution of the longitudinal and time-like components.
Similarly, for $V\widetilde V$ we get
\begin{equation}\label{eq:fftilde3app}
\frac{1}{4}\langle V\widetilde V\rangle = \langle \epsilon^{\mu\nu\rho\sigma}\partial_\mu V_\nu \partial_\rho V_\sigma\rangle = \int \frac{d^3k}{(2\pi)^3} \, k \, \frac{\partial}{\partial t} \left( |V_+|^2 - |V_-|^2 \right) \,.
\end{equation}
which is equation (\ref{eq:fftilde2}). Equations (\ref{eq:BB}) and (\ref{eq:fftilde3app})  must be renormalized by subtracting the same quantities computed on the first order WKB  mode functions $V_\pm^{\textrm{WKB}} = \exp(i\omega_{k\pm})/\sqrt{2\omega_{k\pm}}$. In Fig.~\ref{fig:numerics}, we show numerical solutions of these equations, as commented in Section~\ref{sec:gauge}.

\clearpage

\section{Relaxion Decay Widths}
\label{app:widths}

In this appendix, we list the  relaxion decay widths through the  $V\widetilde V$ leading interaction with the SM gauge bosons  (Sec.\,\ref{sec:gauge}) and by the loop-induced  couplings to photons and SM fermions (Sec.\,\ref{sec:photons}).  For the decay through the mixing of $\phi$ with the Higgs boson  we used the results of~\cite{Bezrukov:2009yw}. In our analysis we only consider 2-body decays. When kinematically allowed, the decay can proceed  through (see e.g. \cite{Craig:2018kne}): 
\begin{eqnarray}
\Gamma_{\phi\rightarrow\gamma \gamma} &=& \frac{1}{64 \pi}  \frac{m_\phi^3}{f_\gamma^2}, \\
\Gamma_{\phi\rightarrow l\bar{l}} &=& \frac{1}{2 \pi} \frac{m_\phi m_l^2}{f_l^2} \left(1 -\frac{4 m_l^2}{m_\phi^2}\right)^{1/2},\\
\Gamma_{\phi\rightarrow Q\bar{Q}} &=& \frac{3}{2 \pi} \frac{m_\phi m_Q^2}{f_Q^2} \left(1 -\frac{4 m_Q^2}{m_\phi^2}\right)^{1/2},\\
\Gamma_{\phi\rightarrow \textrm{hadrons}} &=& \frac{1}{8 \pi^3} \alpha_s^2 m_\phi^3 \left(1 + \frac{83 \alpha_s}{4 \pi}  \right) \left| \sum_{q=u,d,s} \frac{1}{f_q} \right|^2\\
\Gamma_\phi & =& \theta^2 \Gamma_h(m_\phi),\\
\Gamma_{\phi\rightarrow ZZ} &=& \frac{1}{8 \pi f^2} \left(m_\phi^2- 4m_Z^2 \right)^{3/2},\\
\Gamma_{\phi\rightarrow WW} &=&\frac{1}{4 \pi f^2} \frac{g_2^4}{(g_2^2-g_1^2)^2}\left(m_\phi^2- 4m_W^2 \right)^{3/2}, \\
\Gamma_{\phi\rightarrow Z\gamma} &=&\frac{1}{4 \pi f^2} \frac{g_1^2 g_2^2}{(g_2^2-g_1^2)^2}\frac{\left(m_\phi^2- m_Z^2 \right)^{3}}{m_Z^2 m_\phi + m_\phi^3},
\end{eqnarray}
where $f_\gamma$ for the photons and $f_F$ for the fermions are given in Sec.\,\ref{sec:photons}. The index $l$ refers to the SM charged leptons,  $Q=c,b,t$ and $q=u,d,s$ are respectively the heavy and light SM quarks.  Note that the decay constants above follow the convention in Eq.~(\ref{eq:f convention}) where the gauge coupling is absorbed in the definition of $f$.

%\clearpage

\bibliography{PPRelBib}{} \bibliographystyle{JHEP}

\end{document}